\documentclass[numberedappendix,12pt,preprint]{emulateapj} 

\usepackage[]{amsmath}

\newcommand{\simgt}{\,\rlap{\lower 3.5 pt \hbox{$\mathchar \sim$}} \raise
1pt \hbox {$>$}\,}
\newcommand{\simlt}{\,\rlap{\lower 3.5 pt \hbox{$\mathchar \sim$}} \raise
1pt \hbox {$<$}\,}

\newcommand{\QHII}{Q}
\newcommand{\nh}{\langle n_{\rm H}\rangle}
\newcommand{\nion}{n_{\rm ion}}
\newcommand{\trec}{t_{\rm rec}}

\newcommand{\be}{\begin{equation}}
\newcommand{\ee}{\end{equation}}
\newcommand{\lya}{Ly$\alpha$}

\shorttitle{The BoRG13 Luminosity Function}
\shortauthors{Schmidt et al. (2014)}

\usepackage{color}			      
\definecolor{midgray}{gray}{0.4}		
\definecolor{orange}{rgb}{1,0.5,0}    

\usepackage[pdftex,colorlinks=true,citecolor=midgray,linkcolor=midgray]{hyperref}        

\newcommand{\BE}{\begin{equation}}
\newcommand{\EE}{\end{equation}}
\newcommand{\BEA}{\begin{eqnarray}}
\newcommand{\EEA}{\end{eqnarray}}
\newcommand{\Eq}[1]{Equation~(\ref{#1})}

\newcommand{\Eqs}[2]{Equations~(\ref{#1})~and~(\ref{#2})}

\begin{document}


\title{The Luminosity Function at $z\sim8$ from 97 Y-band dropouts: Inferences About Reionization}


\author{
Kasper B. Schmidt$^{1}$,
Tommaso Treu$^{1}$,
Michele Trenti$^{2}$,
Larry D. Bradley$^{3}$,
Brandon C. Kelly$^{1}$,
Pascal A. Oesch$^{4}$,
Benne W. Holwerda$^{5}$,
J. Michael Shull$^{6}$, and
Massimo Stiavelli$^{3}$
 }
\affil{$^{1}$ Department of Physics, University of California, Santa Barbara, CA, 93106-9530, USA}
\affil{$^{2}$ Kavli Institute for Cosmology and Institute of Astronomy, University of Cambridge, Madingley Road, Cambridge, CB3 0HA, United Kingdom} 
\affil{$^{3}$ Space Telescope Science Institute, 3700 San Martin Drive, Baltimore, MD, 21218, USA}
\affil{$^{4}$ UCO/Lick Observatory, University of California, Santa Cruz, CA, 95064, USA}
\affil{$^{5}$ Leiden Observatory, Leiden University, NL-2300 RA Leiden, Netherlands}
\affil{$^{6}$ CASA, Department of Astrophysical and Planetary Science, University of Colorado, Center for Astrophysics and Space Astronomy, 389-UCB, Boulder, CO 80309, USA}

\email{kschmidt@physics.ucsb.edu}




\begin{abstract}
  We present the largest search to date for Y-band dropout galaxies
  ($z\sim8$ Lyman break galaxies, LBGs) based on 350 arcmin$^2$ of HST
  observations in the V-, Y-, J- and H-bands from the Brightest of
  Reionizing Galaxies (BoRG) survey.  In addition to previously
  published data, the BoRG13 dataset presented here includes
  approximately 50 arcmin$^2$ of new data and deeper observations of
  two previous BoRG pointings, from which we present 9 new $z\sim8$
  LBG candidates, bringing the total number of BoRG Y-band dropouts to 38 with
  $25.5\leqslant m_{J} \leqslant 27.6$ (AB system).  We introduce a
  new Bayesian formalism for estimating the galaxy luminosity function, 
  which does not require binning (and thus smearing) of the data
  and includes a likelihood based on the formally correct binomial distribution
  as opposed to the often used approximate Poisson distribution. We
  demonstrate the utility of the new method on a sample of $97$
  Y-band dropouts that combines the bright BoRG galaxies with the fainter
  sources published in \cite{Bouwens:2011p8082} from the Hubble
  Ultradeep Field (HUDF) and Early Release Science (ERS) programs.  We
  show that the $z\sim8$ luminosity function is well described by a
  Schechter function over its full dynamic range with a
  characteristic magnitude $M^\star = -20.15^{+0.29}_{-0.38}$, a
  faint-end slope of $\alpha = -1.87^{+0.26}_{-0.26}$, and a number
  density of $\log_{10} \phi^\star [\textrm{Mpc}^{-3}] =
  -3.24^{+0.25}_{-0.24}$.  Integrated down to $M=-17.7$ this
  luminosity function yields a luminosity density, $\log_{10} \epsilon
  [\textrm{erg}/\textrm{s/Hz/Mpc}^{3}] = 25.52^{+0.05}_{-0.05}$. Our
  luminosity function analysis is consistent with previously published
  determinations within 1$\sigma$. The error analysis
  suggests that uncertainties on the faint-end slope are still too
  large to draw firm conclusion about its evolution with redshift.  We
  use our statistical framework to discuss the implication of our
  study for the physics of reionization.  By assuming theoretically
  motivated priors on the clumping factor and the photon escape
  fraction we show that the UV luminosity density from galaxy samples down to
  $M=-17.7$ can ionize only 10-50\% of the neutral hydrogen at
  $z\sim8$. Full reionization would require extending the luminosity function down to
  $M=-15$. The data are consistent with a substantial fraction of
  neutral hydrogen at $z>7$, in agreement with recent suggestions
  based on deep spectroscopy of redshift 8 LBGs.
\end{abstract}


\section{Introduction}\label{sec:intro}

Characterizing the epoch of reionization, i.e., the epoch at which the
first stars and galaxies in the Universe reionized the vast majority
of the neutral hydrogen, is a key outstanding issue in the
continued effort to map the formation and early evolution of
galaxies. For almost four years, since the installment of the Wide
Field Camera 3 (WFC3) onboard the Hubble Space Telescope (HST), the
frontier of this knowledge has been pushed further and further back
towards the dawn of cosmic reionization.

At present, several hundred high redshift galaxy candidates have been
found at redshift $z\sim6$
\citep[e.g.,][]{Stark:2010p27668,Bouwens:2007p29848,Bouwens:2012p31891,Ouchi:2010p27936,Bradley:2013p32053},
and with the improved near-IR efficiency of WFC3 the search for
candidates has been pushed to $z\gtrsim 8$ using the Lyman Break
technique. In particular the Hubble Ultra Deep field efforts in
2009
\citep{Oesch:2010p30140,Oesch:2010p30134,Lorenzoni:2011p12992,Bouwens:2010p30142,Bouwens:2011p8082,McLure:2010p31124}
and 2012
\citep{Koekemoer:2012p26719,Ellis:2012p26700,Dunlop:2012p26717,Schenker:2013p26914,McLure:2013p27183,Ono:2012p26883,Oesch:2013p27877,Illingworth:2013p31165}
have revealed the faintest samples of galaxy candidates at
$z\gtrsim8$. Simultaneously larger area observations are targeting
brighter and rarer candidates, either in legacy fields, such as
GOODS/CANDELS \citep{Grogin:2011p9457,Koekemoer:2011p9456}, or in
pure-parallel random pointings, like those of our \emph{Brightest of
Reionizing Galaxies
Survey}\footnote{\url{https://wolf359.colorado.edu}}
\citep[hereafter BoRG,][]{Trenti:2011p12656,Trenti:2012p13020,Bradley:2012p23263}.

Our ongoing BoRG survey has two key goals. The first goal is to
provide bright targets that can potentially yield spectroscopic
confirmation of $z\sim8$ galaxies by follow-up observations
\citep{Treu:2012p12658,Treu:2013p32132}. In fact, while $z\sim6$ dropout
samples have extensive spectroscopic redshifts,
\citep[e.g.,][]{Vanzella:2009p29479,Stark:2010p27668,Stark:2013p28962},
only a handful of $z\gtrsim 7$ galaxies \citep[e.g.,][]{Ono:2012p27651}
have currently confirmed redshifts with the highest being at $z=7.5$
\citep{Finkelstein:2013p32467}. 
So far no Y-band dropout at $z\sim8$ has been spectroscopically confirmed.  
Only upper limits on Ly$\alpha$ flux have been provided to date
\citep{Caruana:2012p27502,Caruana:2013p32713,Capak:2013p31627,Treu:2013p32132,Faisst:2014p34184} and those leave
open the interpretation of whether the photometric selection technique
breaks down at $z\gtrsim 7$ (which would be surprising given the small
change in magnitudes and filters) versus the more
interesting physical explanation of an increase in the intergalactic medium (IGM) optical depth to
Ly$\alpha$ arising from a higher neutral hydrogen fraction at $z\sim8$
\citep{Treu:2013p32132} with respect to  redshift 7 
\citep{Fontana:2010p29506} and 6 \citep{Stark:2010p27668,Stark:2013p28962}.

The second goal of the BoRG survey is to improve the determination of
the $z\sim8$ luminosity function, by identifying rare and bright
dropouts to extend the dynamic range of observations in smaller area
deep fields, which are dominated by fainter sources. An accurate
measure of the luminosity function is necessary not only to study how galaxies evolve
across time, but also to quantify the photon budget available for
hydrogen ionization
\citep{Trenti:2010p29335,Zaroubi:2013p24088,Dunlop:2013p23759}. At
lower redshift, it is well established \citep{Bouwens:2007p29848} that
the luminosity function is accurately described by a Schechter function
\citep{Schechter:1976p29330} so it is natural to expect a
similar form at higher redshift. However, data covering a wide
dynamic range are needed to establish that this is indeed the case,
and to resolve the degeneracy between the Schechter function parameters in the luminosity function fit
\citep{Bradley:2012p23263,Oesch:2012p30149}.

In this paper we have two goals. The first is to present the complete
sample of Y-band dropouts from the BoRG cycle-19 data, and to use these in combination
with the literature to determine the galaxy luminosity function at $z\sim8$. The second goal is
to study the consequences for cosmic reionization the inferred luminosity function has.
To determine the luminosity function we develop and implement a rigorous
statistical Bayesian method to infer the posterior distribution
function of the parameters of the luminosity function from the
data. The method supersedes those commonly adopted in this field
\cite[e.g][]{Bradley:2012p23263,McLure:2013p27183,Schenker:2013p26914,Oesch:2012p30149}
in several ways: the data are not binned, thus avoiding smearing the
luminosity function
\citep{Trenti:2008p32309}; the flux uncertainties are correctly
taken into account; the counts are modeled using the formally correct binomial
distribution \citep{Kelly:2008p29070}, instead of the Poisson
approximation; the full posterior probability distribution function is computed using Markov Chain
Monte Carlo methods instead of relying on maximum likelihood estimators
based on the 
asymptotic covariance matrix from the observed Fisher information
for uncertainties.  
By applying this framework to a large sample of $z\sim 8$ galaxies,
consisting of $N=97$ objects both bright (from BoRG) and faint (from
the Hubble UDF/ERS fields), we show that the credible intervals
include previous best-fit estimates
By treating the problem in a self-consistent statistical manner we carry out an inference
about reionization by combining the inferred observational uncertainties with
various theoretical priors.

The paper is organized as follows. We start by briefly describing the
BoRG survey in Section~\ref{sec:borg}. The current sample of Y-band 
dropouts containing 9 new and 2 improved
$z\sim8$ galaxy candidates (BoRG13) with respect to those
previously published by our team \citep[BoRG09;
BoRG12][]{Trenti:2011p12656,Bradley:2012p23263} is described in
Section~\ref{sec:ydrop}. 
Two Appendixes (\ref{app:borg58} and \ref{app:noise}) take advantage of the
follow-up observations of one field and of the large number of BoRG
pointings to characterize and discuss the statistics of detections and
contaminants in dropout searches.
In Section~\ref{sec:LF} we apply our inference of the Schechter luminosity 
function parameters using our rigorous Bayesian
framework, discussed in detail in Appendix~\ref{sec:BF}.  The results are
presented and discussed in the context of
cosmic reionization in Section~\ref{sec:results}. 
A brief summary is given in Section~\ref{sec:conc}.

All magnitudes are AB magnitudes and a standard concordance
cosmology with $\Omega_m=0.3$, $\Omega_\Lambda=0.7$, and $h=0.7$ is assumed.

\section{The BoRG Survey}\label{sec:borg}

The $z\sim 8$ galaxy candidates from the latest data
obtained as part of the BoRG survey are described briefly below. We
refer the reader to \cite{Trenti:2011p12656} and
\cite{Bradley:2012p23263} for a more in-depth description of the survey.

The BoRG survey is a pure-parallel WFC3 imaging HST program. As of
April 2013 the survey has obtained $\sim$350~arcmin$^2$ of visual and
near-infrared HST photometry over 71 fields randomly located in
the sky. The pure-parallel nature of the survey implies that the
survey area is divided into 71 independent lines of sight on the
sky, reducing sample (or cosmic) variance below the level of
statistical noise
\citep{Trenti:2008p32309,Bradley:2012p23263}.
53 out of the 71 fields represent the core of the
BoRG survey and have been observed in the four WFC3/HST filters F606W, F098M,
F125W, and F160W. This was primarily done as part of programs GO/PAR 11700 and GO/PAR
12572 (PI: Trenti) complemented by a small number of COS-GTO
coordinated parallels. One of these 53 fields furthermore has data in
F105W from a recent follow-up campaign (described in
Appendix~\ref{app:borg58}).
The core of BoRG is complemented by other archival data consisting of
8 fields from GO/PAR 11702 \citep[PI: Yan,][]{Yan:2011p27097} and 10
COS-GTO fields, where instead of the F606W-band the F600LP-band was
used. For a discussion of the benefits of using F606W, as in the BoRG
core, instead of F600LP see \cite{Bradley:2012p23263}.

In this work we will refer to F606W, F098M, F125W, and
F160W as V-, Y-, J-, and H-band observations unless mentioned
otherwise. Note that the data added in BoRG13 only has F606W V-band
observations as opposed to BoRG09 and BoRG12 which also contained  
F600LP V-band data.

\section{Y-band dropout sample}\label{sec:ydrop}

The dropout technique
\citep{Steidel:1996p23265,Steidel:2000p25194,Madau:1996p24662} was applied to the BoRG data to identify $z\sim
8$ galaxy candidates as detailed in Section~\ref{sec:zsel}

The first fields of the BoRG survey (referred to as BoRG09 and BoRG12)
were analyzed by \citet[][29 fields]{Trenti:2011p12656} and
\citet[][29+30=59]{Bradley:2012p23263}. Based on this data, \citet{Bradley:2012p23263}
presented a sample of 33 Lyman break galaxy (LBG) candidates at
$z\sim8$ and estimated the corresponding high-redshift luminosity
function.

In this work we augment this sample by analyzing 13 additional fields,
taken in HST Cycle 19 (GO/PAR 12572, PI: Trenti).  Note that the field
BoRG\_1510+1115 also appeared in the study by \cite{Bradley:2012p23263} based on
partial data. This field is therefore included in the present
analysis and supersedes the previous release. Results of the analysis
of the 13 new fields are presented in the next
Section~\ref{sec:C19}. We also present an updated analysis of
BoRG\_1437+5043 based on deeper and wider-field observations obtained
in November 2012 (GO 12905, PI: Trenti). We describe these
observations in Appendix~\ref{app:borg58}.

\subsection{The Latest Survey Extension: BoRG13}\label{sec:C19}

\begin{figure}
\includegraphics[width=0.49\textwidth]{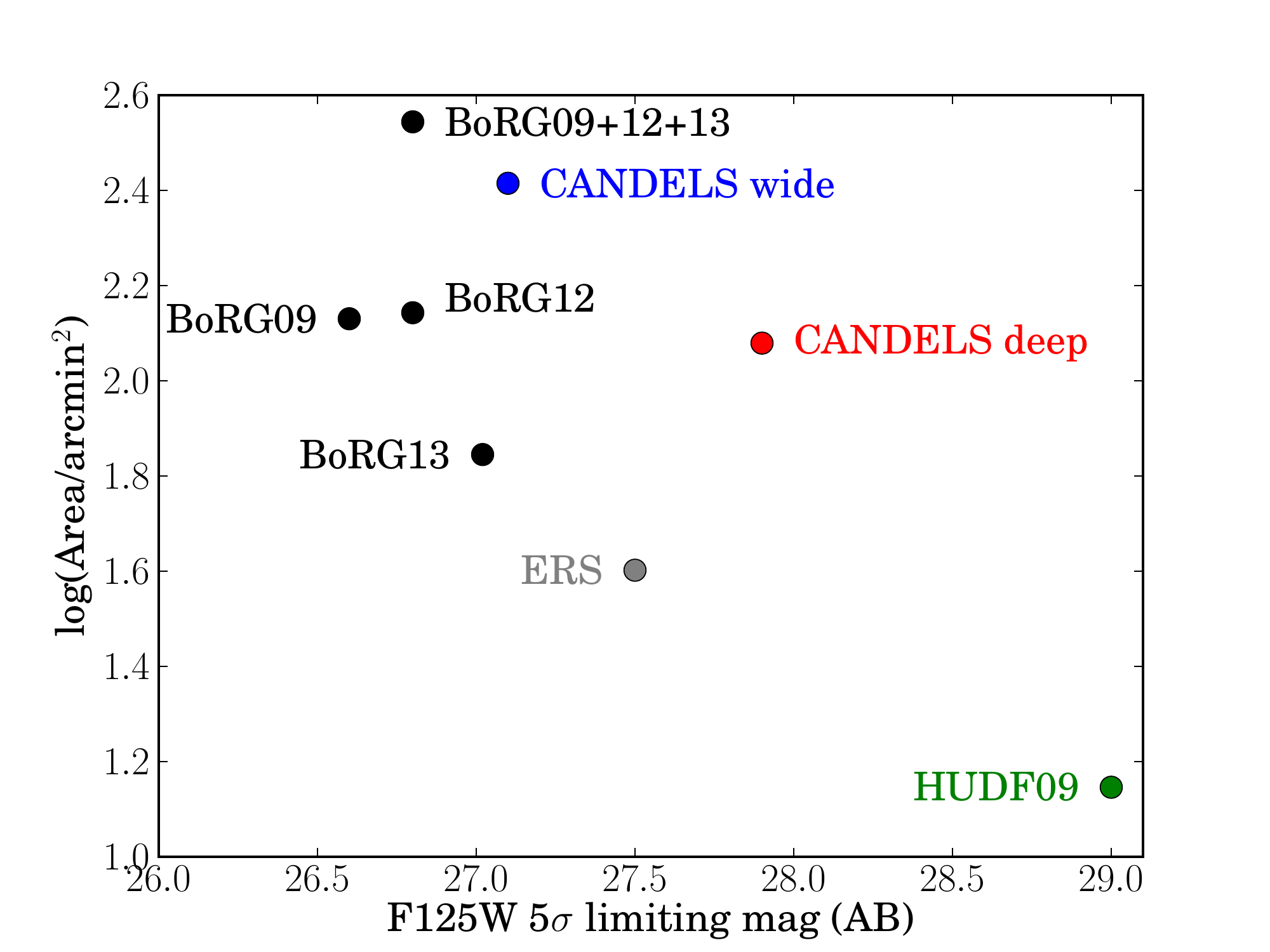}
\caption{The area and depth of surveys where $z\sim8$ Lyman Break Galaxy (LBG) dropout selection is currently possible. It is clear that BoRG is by far the largest area with the needed band coverage (VYJH). The CANDELS wide area for instance has Y-band coverage in approximately 260 arcmin$^2$. The HUDF and ERS fields were used to obtain the 59 $z\sim8$ LBGs selected by \cite{Bouwens:2011p8082} used in the present study at the faint end of the luminosity function.}
\label{fig:AvsD}
\end{figure}

The 13 new Cycle 19 BoRG fields are summarized in
Table~\ref{tab:C19fields} together with the follow-up in
BoRG\_1437+5043. Together with Table~1 and Table~2 of
\cite{Bradley:2012p23263} this summarizes the current status of
BoRG09, BoRG12, and BoRG13.  This brings the total J-band area of the
BoRG survey to $\sim$350~arcmin$^2$. This makes BoRG the largest
existing area which can be searched for Y-band dropouts. In comparison, the
CANDELS \citep{Koekemoer:2011p9456,Grogin:2011p9457} survey has Y-band
coverage of approximately 260~arcmin$^2$ (wide) and 120~arcmin$^2$
(deep). Furthermore the depth reached by BoRG (J and H $\sim26$) at
$5\sigma$ has not been achieved from the ground over large areas.  The
depth and area of the BoRG campaigns are compared with those achieved
by other $z\sim8$ surveys in Figure~\ref{fig:AvsD}.

As noted in above and shown in \cite{Bradley:2012p23263} the effect of large scale structures, i.e. cosmic variance, is less important than the statistical noise for the BoRG12 sample. Extending BoRG12 by the 13 new randomly pointed fields  of BoRG13 makes cosmic variance even more negligible.

The BoRG13 fields were reduced using publicly available code. First,
the cosmic rays were removed in each exposure using the Laplacian
cosmic ray detection developed by \cite{vanDokkum:2001p31886}.
The individual exposures in each filter were then combined using
\verb+AstroDrizzle+ \citep[the replacement of MultiDrizzle as of June 2012][]{Koekemoer:2003p31861}. 
The images were drizzled to a final pixel scale of $0\farcs08$/pixel
using a `pixfrac' of 0.75 as in our previous analyses.
The correlated noise introduced by this drizzling \citep[e.g.,][]{Casertano:2000p30821}
is explicitly corrected for by normalizing the r.m.s. maps by the 
empirical noise as described by \cite{Trenti:2011p12656}.
The 5$\sigma$ limiting magnitudes ($r=0\farcs32$) and exposure times
obtained in each field are quoted in Table~\ref{tab:C19fields}.

Having created the reduced science images and r.m.s. maps we used
\verb+SExtractor+ \citep{Bertin:1996p12964} in dual-image mode with
the J-band image as detection image to create source catalogs in each of
the available photometric bands. These catalogs where used to search
for Y-band dropouts as described in Section~\ref{sec:zsel}. In this
process signal-to-noise (S/N) was estimated using isophotal apertures
(\verb+ISOMAG+) whereas total magnitudes are estimated in scalable
Kron apertures (\verb+AUTOMAG+).

This approach follows the well-established procedure adopted and
described by \cite{Bradley:2012p23263}, and we refer to this work for
further details.

\begin{table*}
\centering{
\caption[ ]{BoRG13 Survey Fields, Exposure Times, and $5\sigma$ Limiting Magnitudes$^a$}
\label{tab:C19fields}
\begin{tabular}[c]{lrrrrrrrrrrrrcc}
\hline
\hline
Field & $\alpha_{J2000}$  & $\delta_{J2000}$ & \multicolumn{2}{c}{F606W} & \multicolumn{2}{c}{F098M} & \multicolumn{2}{c}{F105W} & \multicolumn{2}{c}{F125W} & \multicolumn{2}{c}{F160W} & Area & E(B$-$V) \\
   & [deg] & [deg] & $t$ [s] & $m_\textrm{lim}$ & $t$ [s] & $m_\textrm{lim}$ & $t$ [s] & $m_\textrm{lim}$ & $t$ [s] & $m_\textrm{lim}$ & $t$ [s] & $m_{lim}$ & [$'^2$] & \\
\hline  
BoRG\_0456-2203 		& 73.9646 & -22.0489 	& 2647 & 26.58 & 3718 & 26.75 & $\dots$ & $\dots$ & 1809 & 26.74 & 1809 & 26.52 & 3.06 & 0.038 \\
BoRG\_0951+3304 		& 147.7003 & 33.0737 	& 2660 & 26.43 & 4518 & 26.43 & $\dots$ & $\dots$ & 2212 & 26.52 & 2212 & 26.20 & 1.82 & 0.013 \\
BoRG\_0952+5304 		& 147.9448 & 53.0714 	& 2506 & 26.78 & 3912 & 26.69 & $\dots$ & $\dots$ & 1806 & 26.77 & 1806 & 26.50 & 3.72 & 0.011 \\
BoRG\_1059+0519 		& 164.7039 & 5.3125 	& 2386 & 26.43 & 3812 & 26.72 & $\dots$ & $\dots$ & 1806 & 26.83 & 1806 & 26.38 & 2.15 & 0.028 \\
BoRG\_1118-1858 		& 169.4101 & -18.9726 	& 8514 & 26.94 & 13235 & 27.13 & $\dots$ & $\dots$ & 6276 & 27.17 & 6276 & 26.94 & 2.04 & 0.050 \\
BoRG\_1358+4326 		& 209.4754 & 43.4338 	& 2451 & 26.77 & 3812 & 26.77 & $\dots$ & $\dots$ & 1606 & 26.83 & 1606 & 26.47 & 3.86 & 0.008 \\
BoRG\_1358+4334 		& 209.4636 & 43.5610 	& 4866 & 27.02 & 7023 & 27.21 & $\dots$ & $\dots$ & 3812 & 27.32 & 3812 & 27.06 & 2.71 & 0.007 \\
BoRG\_1416+1638 		& 214.0048 & 16.6269 	& 3112 & 26.86 & 5271 & 26.84 & $\dots$ & $\dots$ & 2462 & 26.85 & 2462 & 26.55 & 3.64 & 0.020 \\
BoRG\_1429-0331 		& 217.3717 & -3.5185 	& 9164 & 26.96 & 13235 & 27.06 & $\dots$ & $\dots$ & 5726 & 27.00 & 5726 & 26.79 & 3.37 & 0.083 \\
BoRG\_1437+5043\_r1 	& 219.2153 & 50.7244 	& 13570 & 27.15 & 19720 & 26.98 & $\dots$ & $\dots$ & 11394 & 27.04 & 10691 & 26.66 & 1.66 & 0.013 \\
BoRG\_1437+5043\_r2$^b$ & 219.2153 & 50.7244 	& 13570 & 27.70 & 19720 & 27.65 & 8885 & 27.21 & 11394 & 27.74 & 10691 & 27.49 & 1.67 & 0.013 \\
BoRG\_1437+5043\_r3$^b$ & 219.2153 & 50.7244 	& 13570 & 27.54 & 19720 & 27.42 & 8885 & 27.14 & 11394 & 27.54 & 10691 & 27.39 & 1.52 & 0.013 \\
BoRG\_1459+7146 		& 224.7501 & 71.7638 	& 3724 & 26.61 & 6023 & 26.75 & $\dots$ & $\dots$ & 2812 & 27.04 & 2812 & 26.82 & 2.78 & 0.027 \\
BoRG\_1510+1115 		& 227.5371 & 11.2415 	& 13315 & 27.19 & 21059 & 27.43 & $\dots$ & $\dots$ & 9529 & 27.67 & 9529 & 27.16 & 1.78 & 0.046 \\
BoRG\_2132-1202 		& 322.9467 & -12.0397 	& 2656 & 26.20 & 3718 & 26.19 & $\dots$ & $\dots$ & 1809 & 26.23 & 1809 & 26.00 & 1.21 & 0.062 \\
BoRG\_2313-2243 		& 348.2326 & -22.7252 	& 8308 & 26.98 & 13335 & 27.11 & $\dots$ & $\dots$ & 6326 & 27.11 & 6326 & 26.91 & 3.26 & 0.026 \\
\hline
\multicolumn{15}{l}{\textsc{Note.} -- $^a$5$\sigma$ magnitude limits are for $r=0.32\arcsec$ apertures corrected for Galactic extinction. The total effective search area for Y-band}\\
\multicolumn{15}{l}{dropouts in BoRG13 is 40.26 arcmin$^2$. The combined effective search area for Y-band dropouts in BoRG09 + BoRG12 + BoRG13 is}\\
\multicolumn{15}{l}{$\sim$247 arcmin$^2$. $^b$Include follow-up observations from November 2012 (GO 12905, PI: Trenti) described in Appendix~\ref{app:borg58}.}
\end{tabular}}
\end{table*}

\subsection{Selecting $z~\sim8$ Galaxy Candidates in BoRG13}\label{sec:zsel}

The selection of the high redshift LBGs closely follows \cite{Trenti:2011p12656} and \cite{Bradley:2012p23263}. 
We use an identical Y-band dropout selection scheme, i.e., we require that
\BEA
\textrm{S/N}_\textrm{V-band}       & < & 1.5  \label{eqn:SNcut} \\
\textrm{S/N}_\textrm{J-band}       & > & 5.0  \label{eqn:JSNcut} \\
\textrm{S/N}_\textrm{H-band}       & > & 2.5  \label{eqn:HSNcut} \\
(\textrm{Y}-\textrm{J}) & > & 1.75 \\
(\textrm{J}-\textrm{H}) & < & 0.02 + 0.15 \; (\textrm{Y}-\textrm{J} - 1.75) \; , \label{eqn:JHcut} 
\EEA
in each of the fields from Table~\ref{tab:C19fields}. 
The colors are insensitive to whether isophotal, Kron or aperture magnitudes are used \citep{Finkelstein:2010p34137,Trenti:2012p13020}.
Candidates passing these selection criteria have been vetted by visual inspection to remove false positives such as hot pixels and diffraction spike features from bright objects.

As in our previous work, the \verb+SExtractor+ stellarity index was used as
an additional criterion to help reject stars as potential
contaminants. For the J-band 8$\sigma$ sources the stellarity index is quite
reliable and we required it to be $<0.85$. 
For sources with $5\sigma<\textrm{S/N}_\textrm{J}<8\sigma$,
the stellarity index is more noisy and therefore we
did not imply a strict cut, but we used it as a criterion in
combination with visual inspection. Three of the authors (KBS, TT, MT)
inspected all the dropouts independently with the aim of rejecting
spurious or star-like features. After the initial independent
classification each dropout was discussed by the three co-authors
resulting in the consensus list given in this paper.
Only a few objects did not initially get rejected by all authors. These were potential hot 
pixels being mistaken for real sources in the drizzled un-dithered BoRG data and potential 
stars (point-like sources). 
In the former case disagreement led to exclusion from the final sample whereas objects were 
revisited to obtain agreement in the latter case.
We note that the final list therefore includes potential real-source contaminants.
However, invoking a relatively high fiducial contamination fraction of the
dropout sample as described in Section~\ref{sec:contamination} this is 
accounted for when inferring the intrinsic luminosity function.

\begin{figure*}
\includegraphics[width=0.75\textwidth]{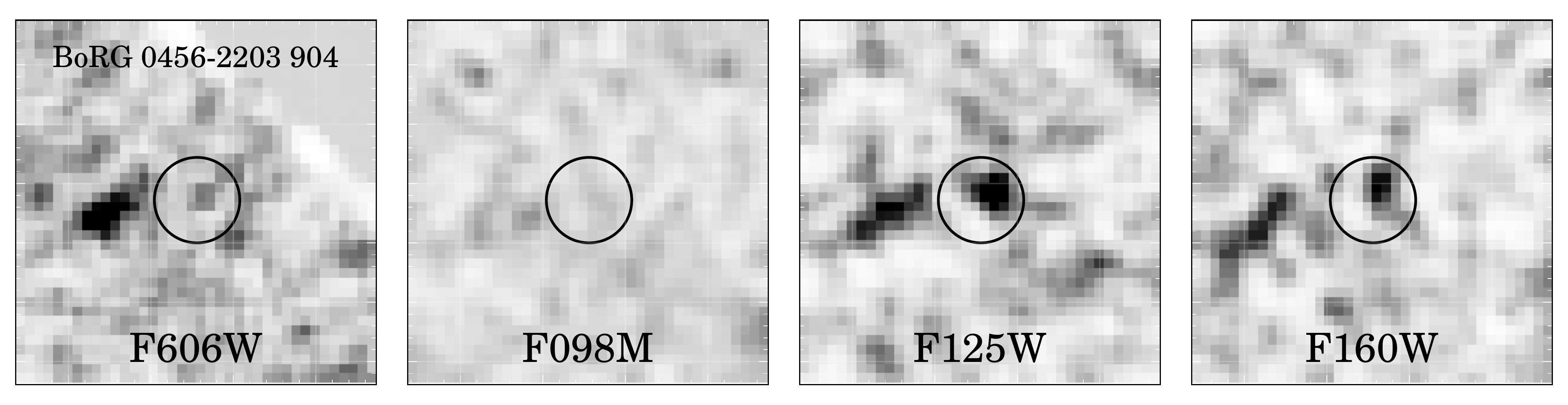}
\includegraphics[width=0.24\textwidth]{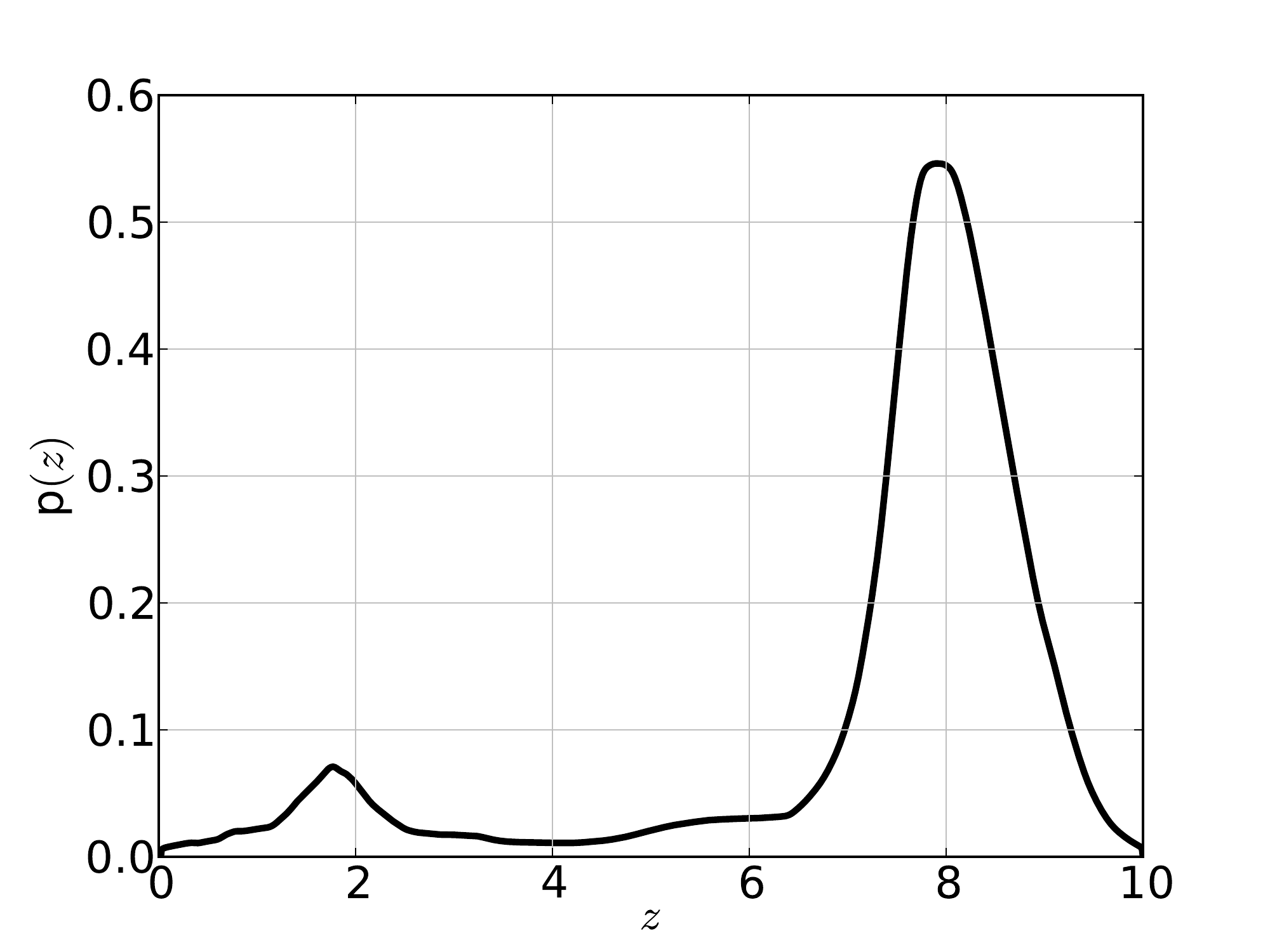}\\
\includegraphics[width=0.75\textwidth]{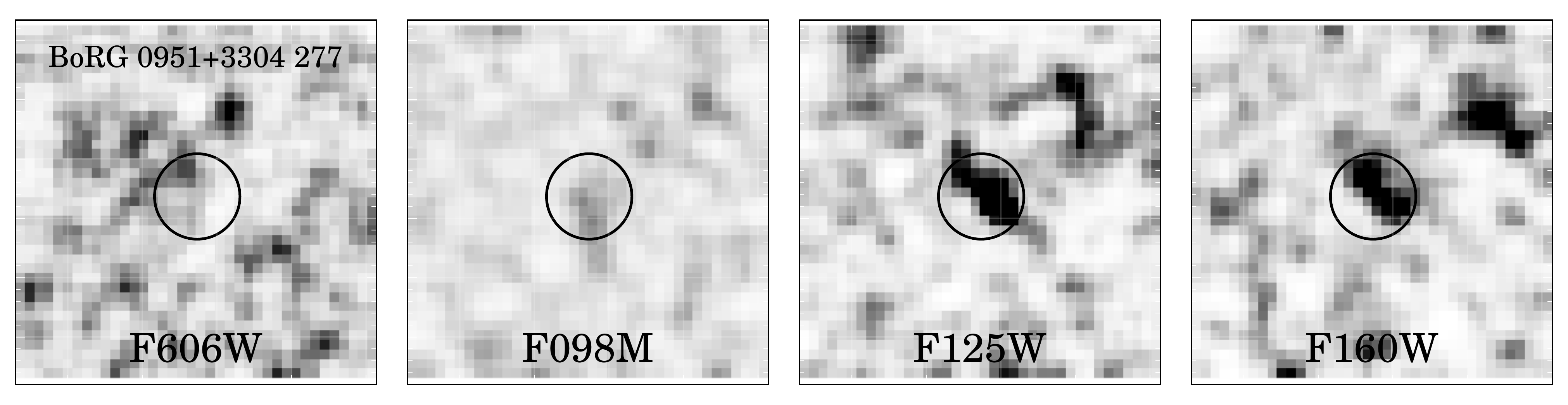}
\includegraphics[width=0.24\textwidth]{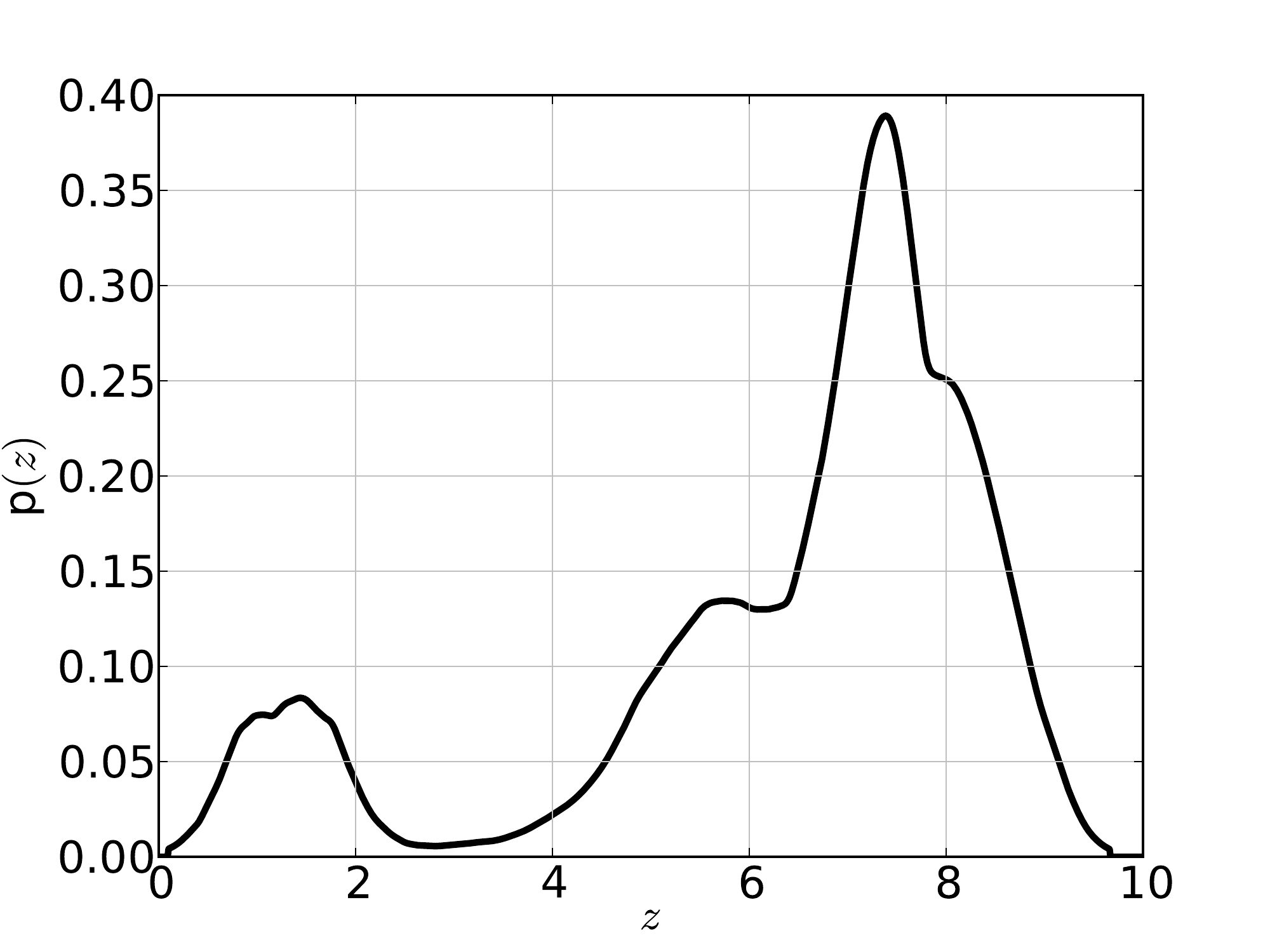}\\
\includegraphics[width=0.75\textwidth]{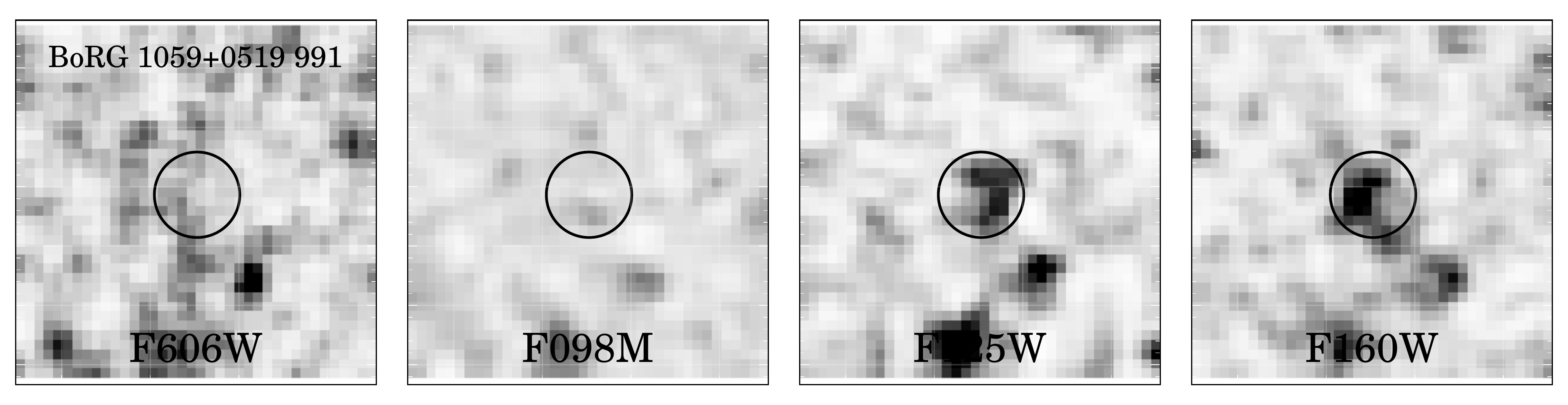}
\includegraphics[width=0.24\textwidth]{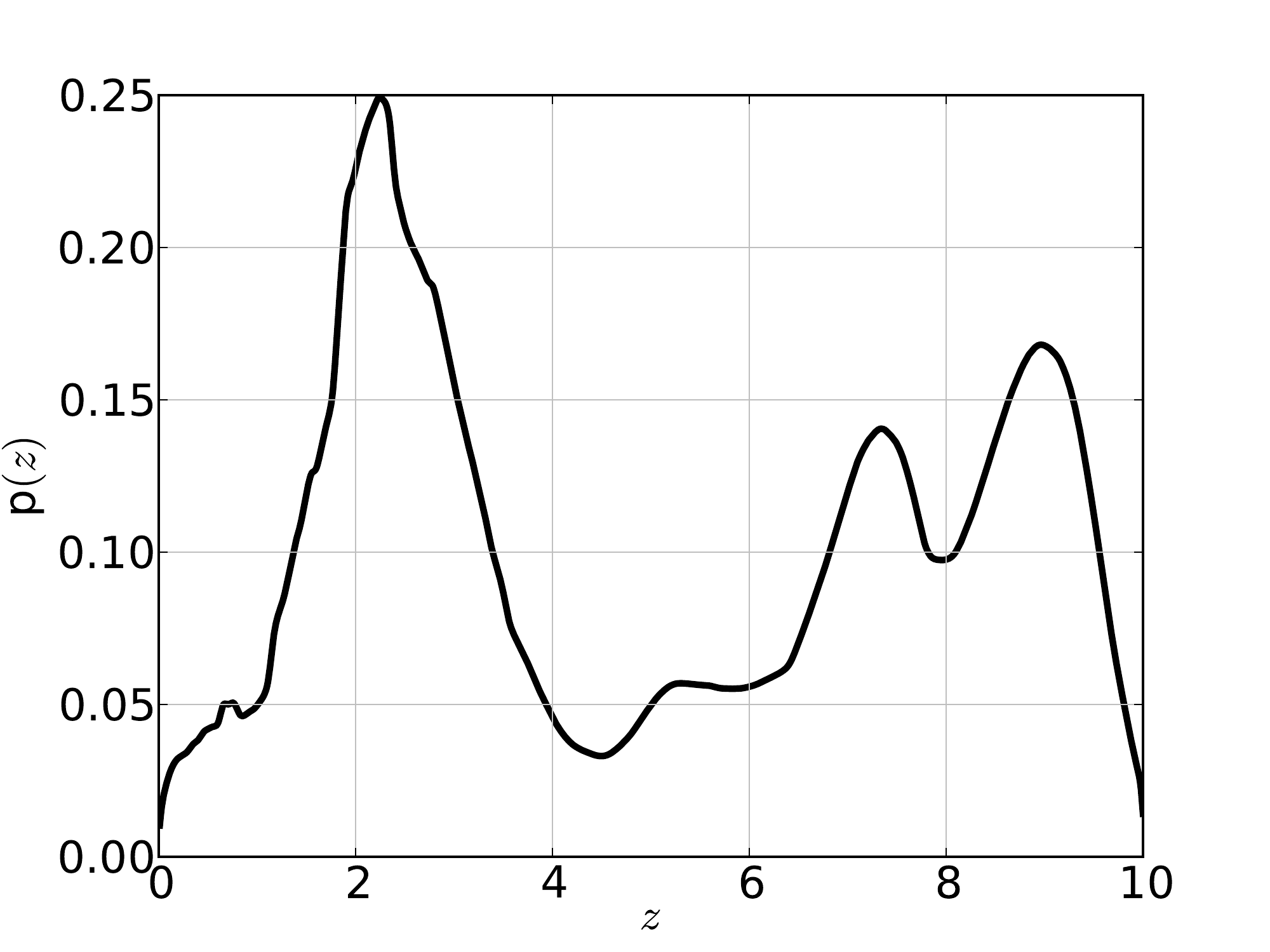}\\
\includegraphics[width=0.75\textwidth]{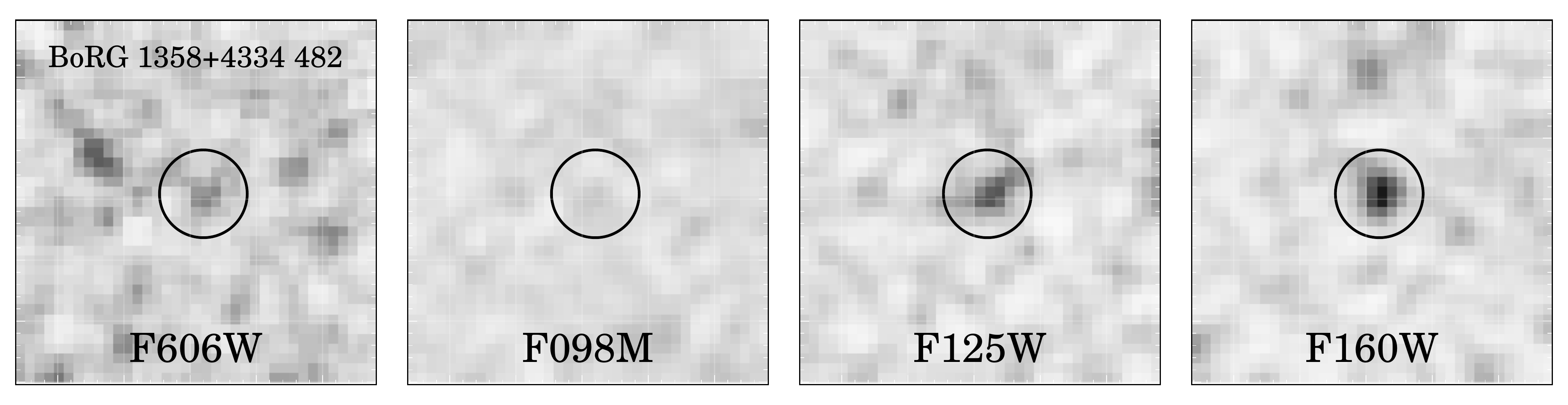}
\includegraphics[width=0.24\textwidth]{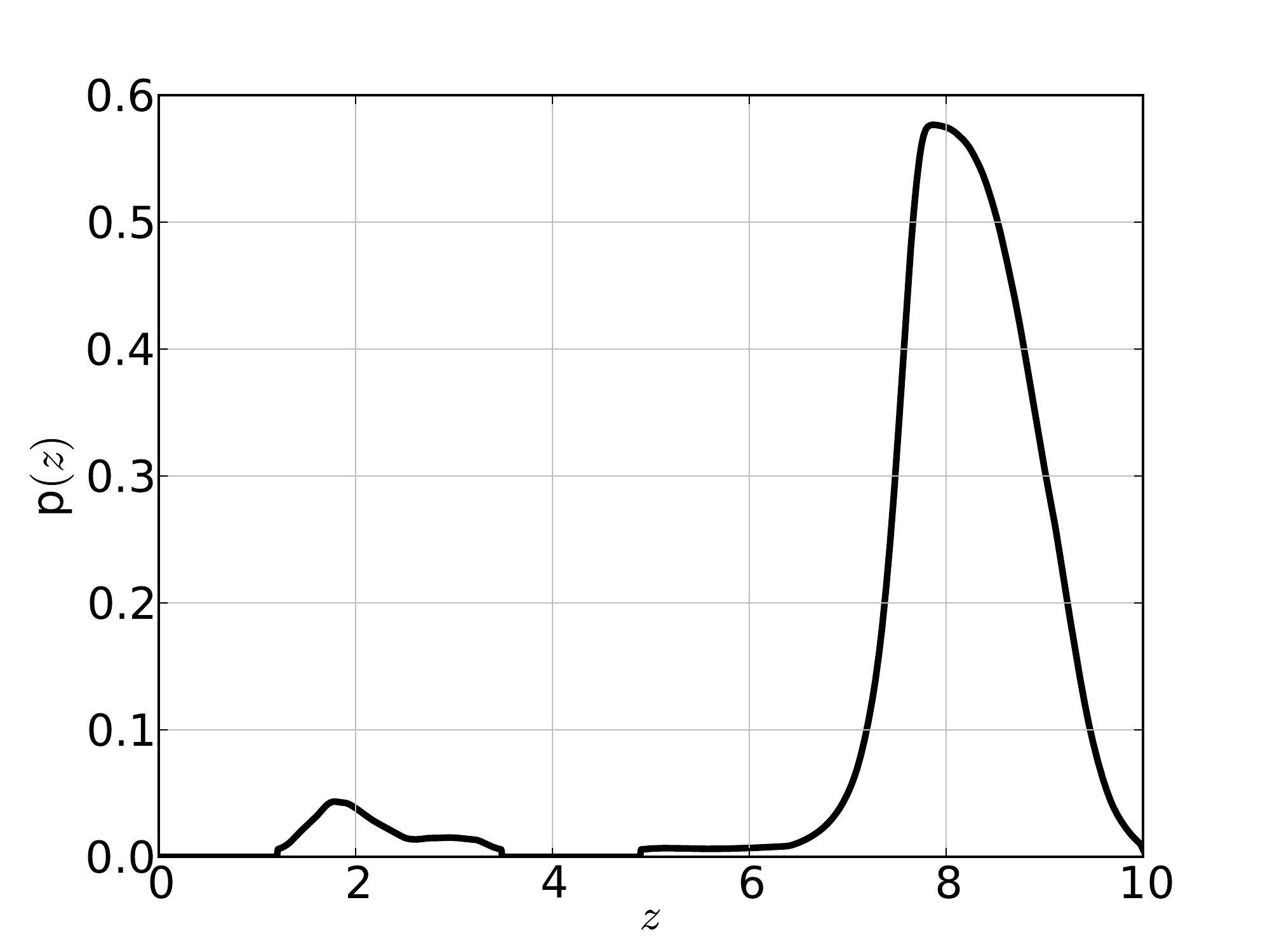}\\
\includegraphics[width=0.75\textwidth]{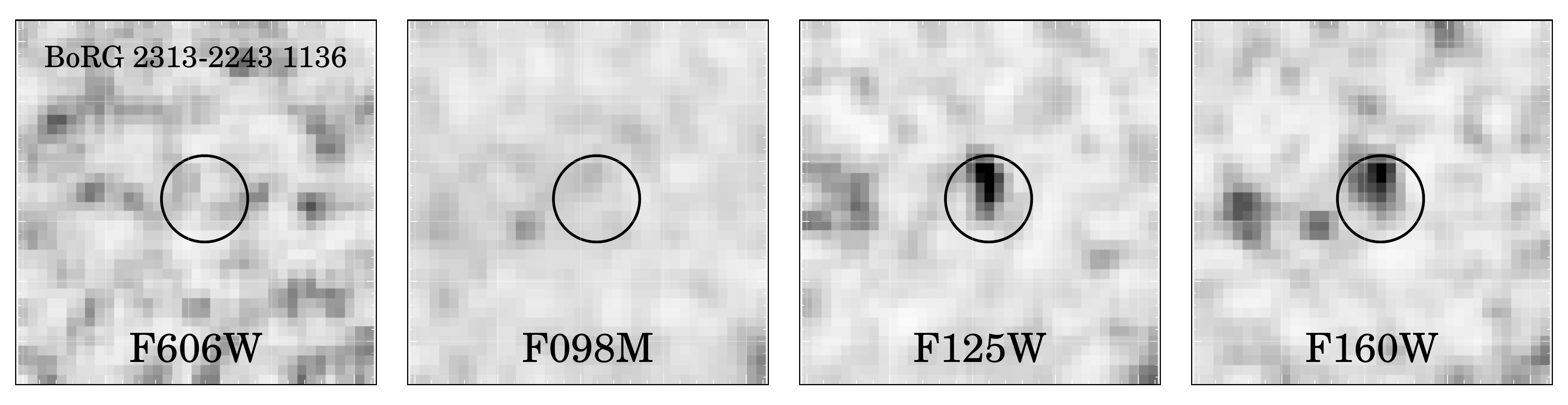}
\includegraphics[width=0.24\textwidth]{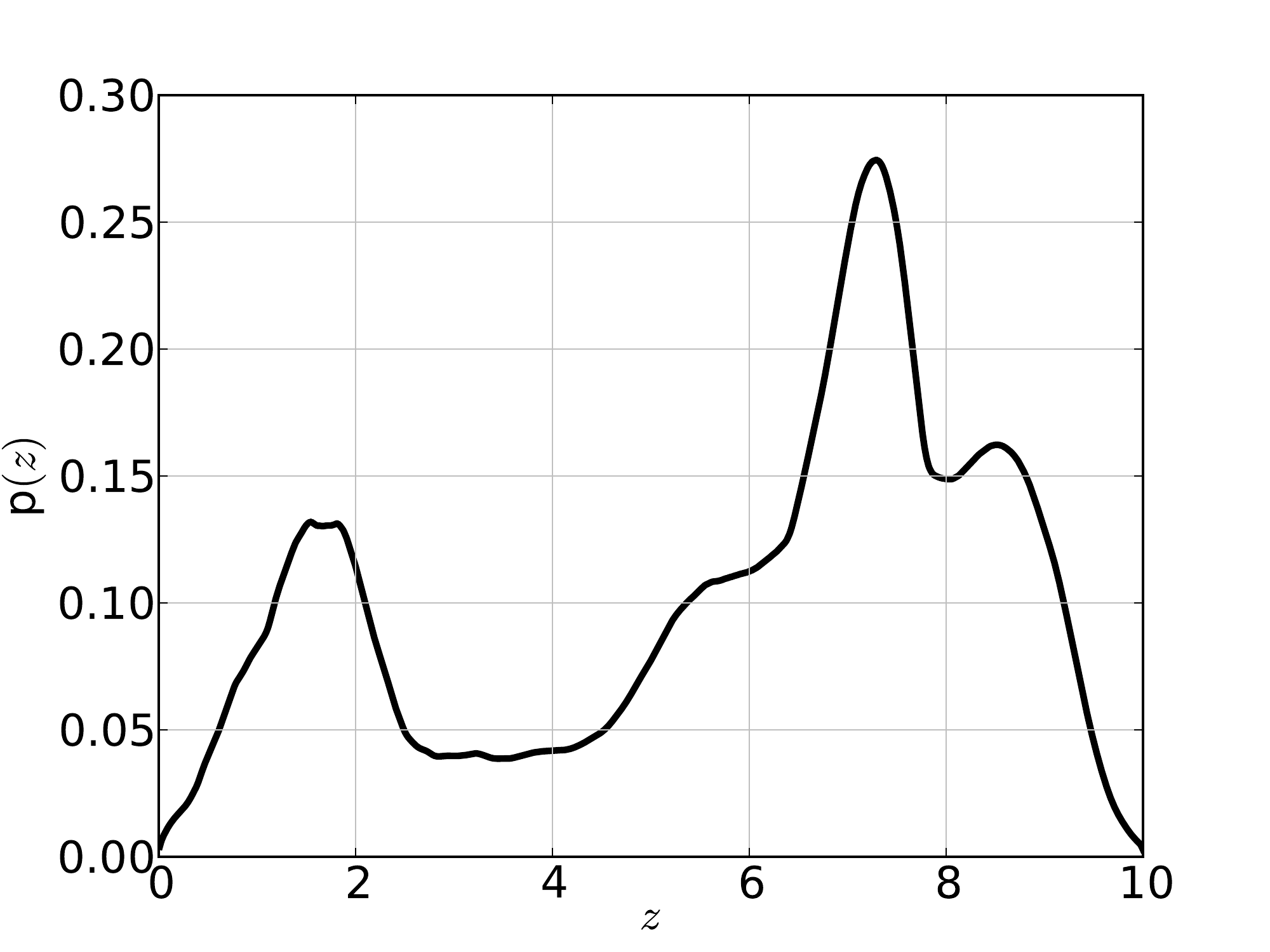}\\
\includegraphics[width=0.76\textwidth]{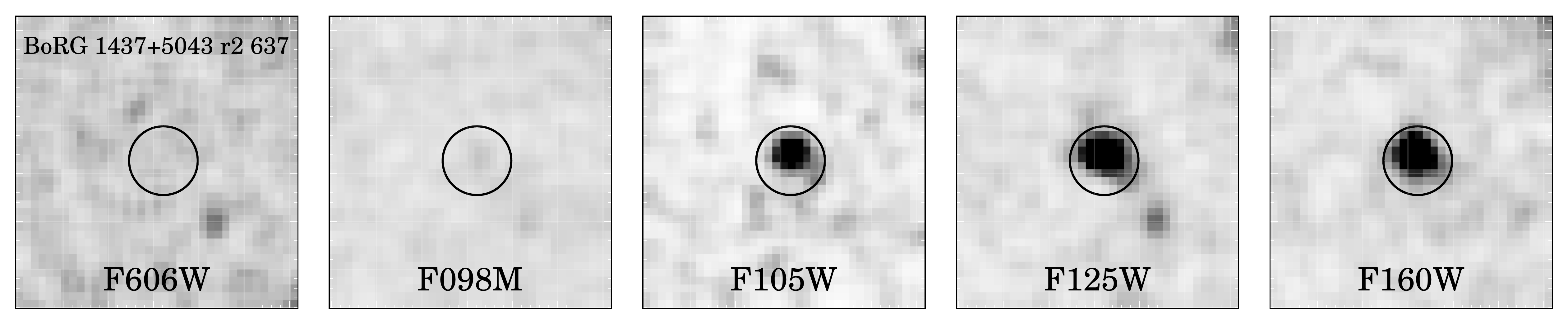}
\includegraphics[width=0.22\textwidth]{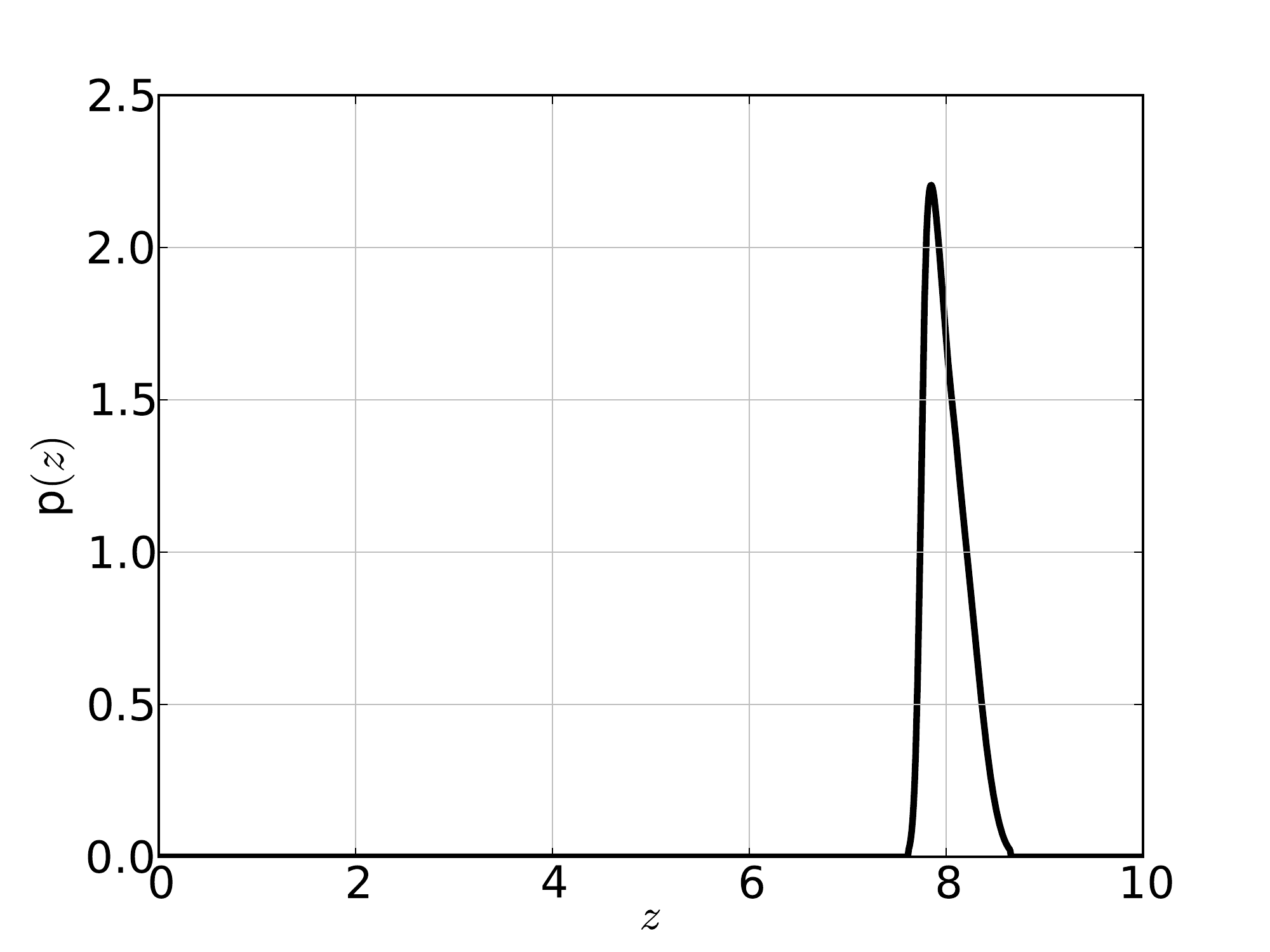}
\caption{6 of the 11 $z\sim8$ Y-band dropouts in the BoRG13 sample presented in this paper (the remaining 5 candidates are shown in Figure~\ref{fig:fitspz2}). The first four columns show V-, Y-, J- and H-band 3$''$x3$''$ HST postage stamps with a power-law stretch. The last column shows the photometric redshift probability distribution $p(z)$ (using a flat prior) obtained with the Bayesian redshift  code BPZ \citep{Benitez:2004p25410,Coe:2006p25409} for each candidate. In the bottom row the F105W (YJ-band) data from our follow-up campaign of BoRG\_1437+5043 (see Appendix~\ref{app:borg58}) are included.}
\label{fig:fitspz}
\end{figure*}

\begin{figure*}
\includegraphics[width=0.75\textwidth]{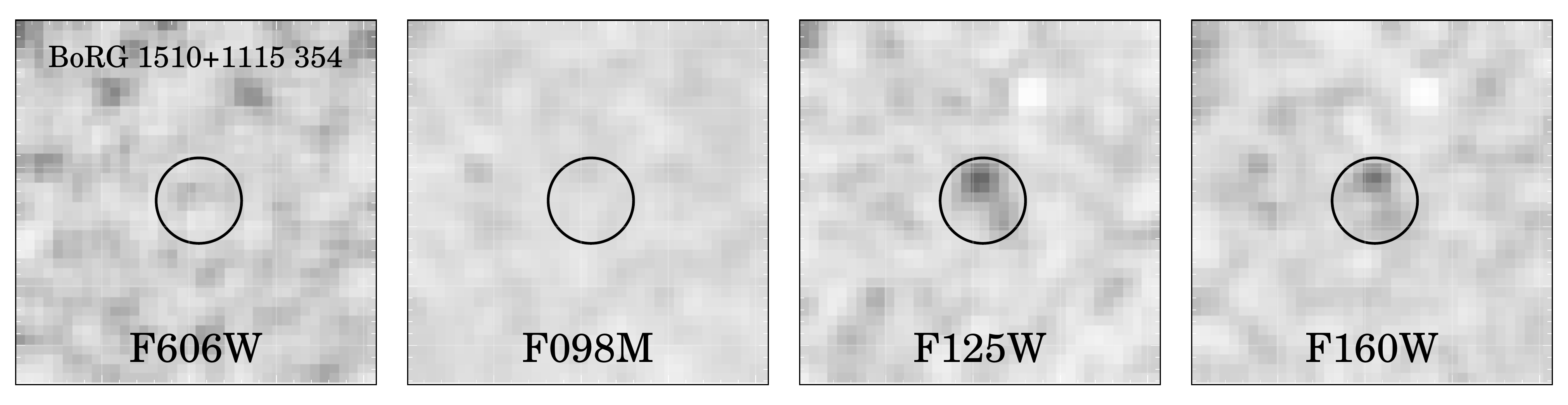}
\includegraphics[width=0.24\textwidth]{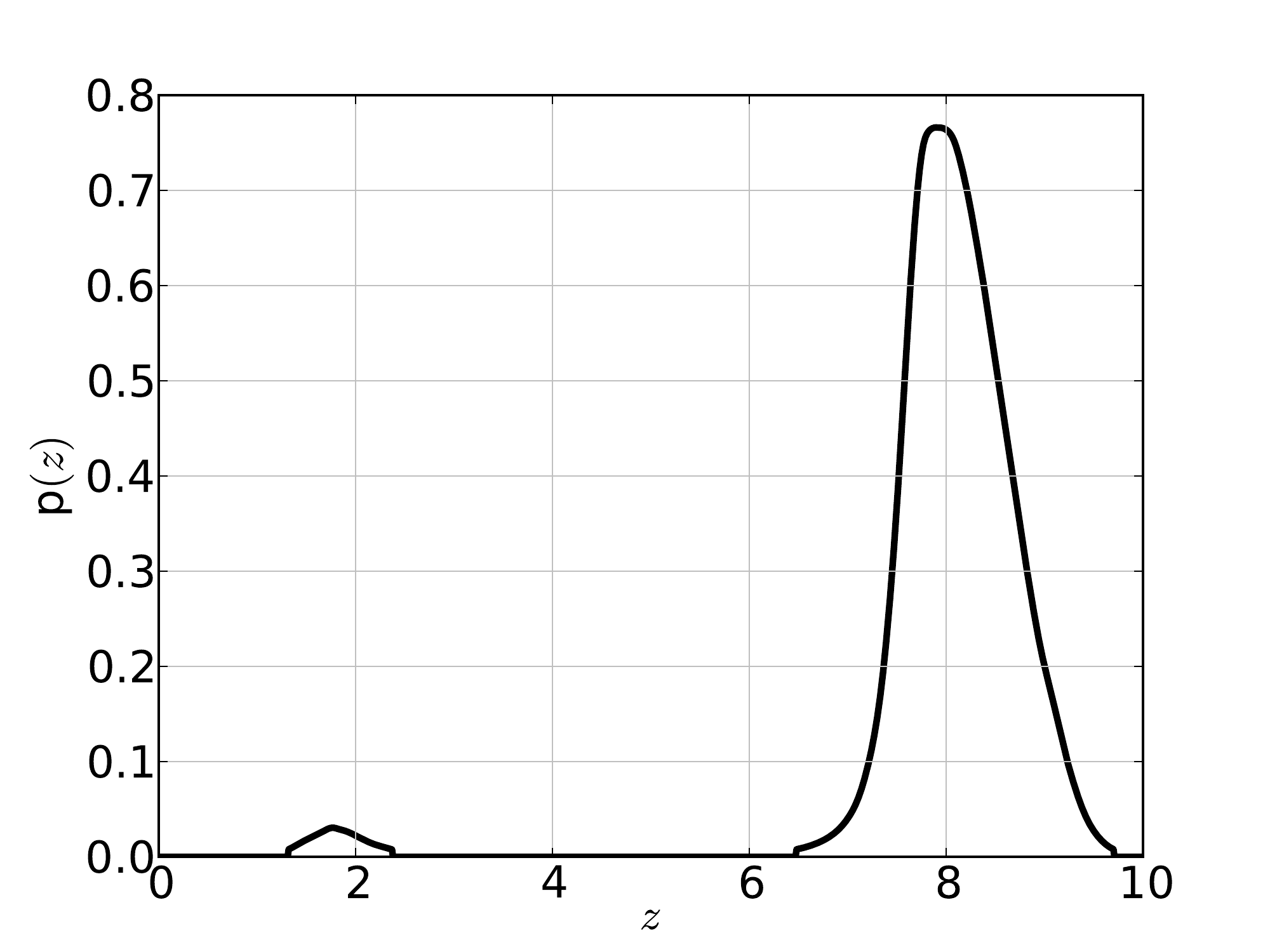}\\
\includegraphics[width=0.75\textwidth]{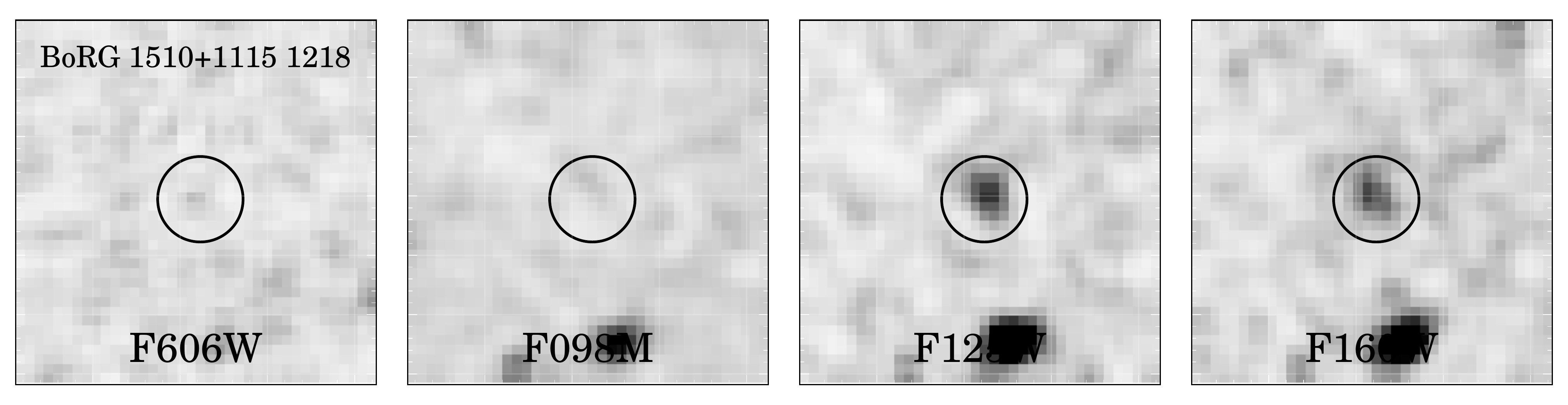}
\includegraphics[width=0.24\textwidth]{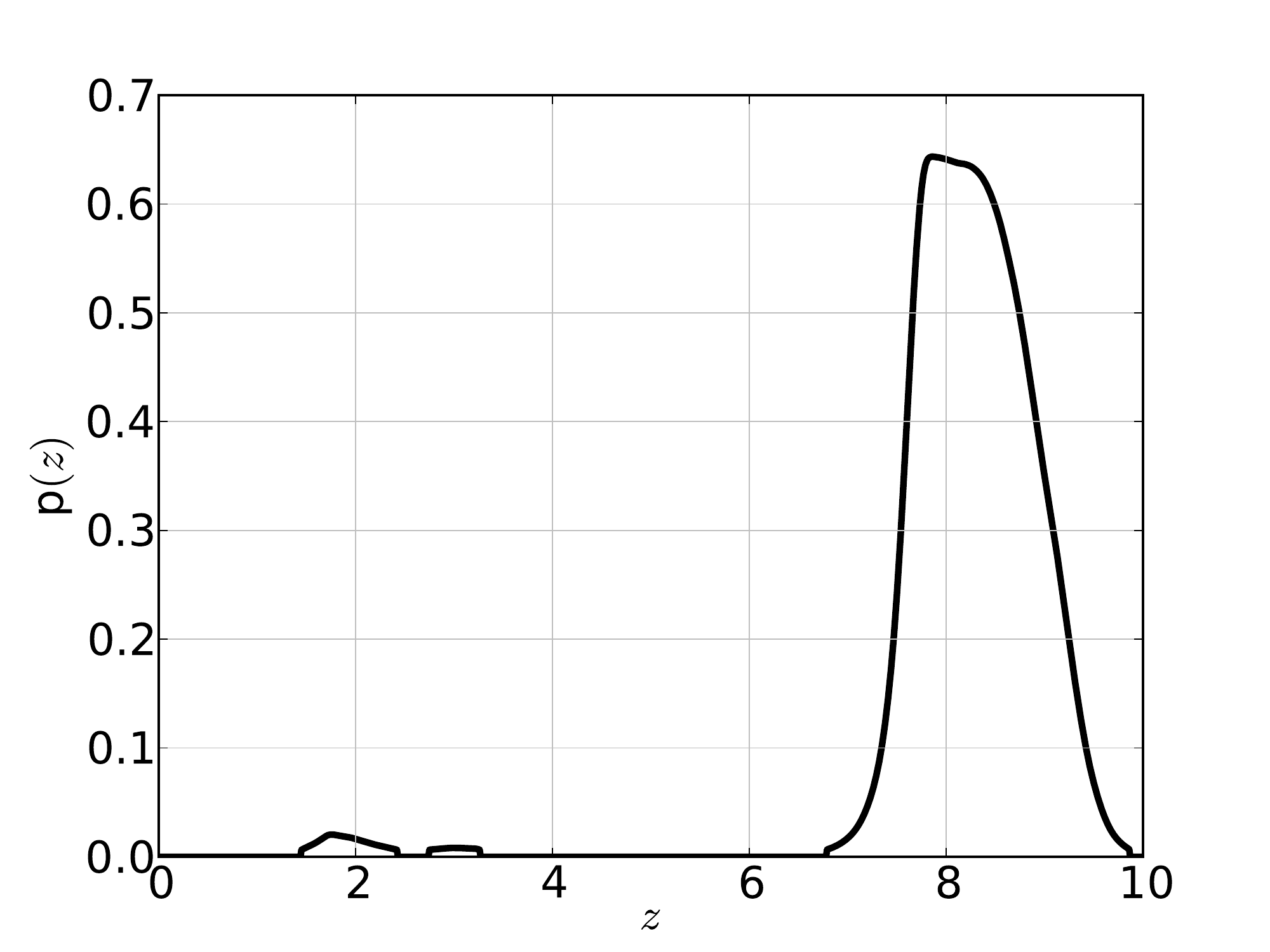}\\
\includegraphics[width=0.75\textwidth]{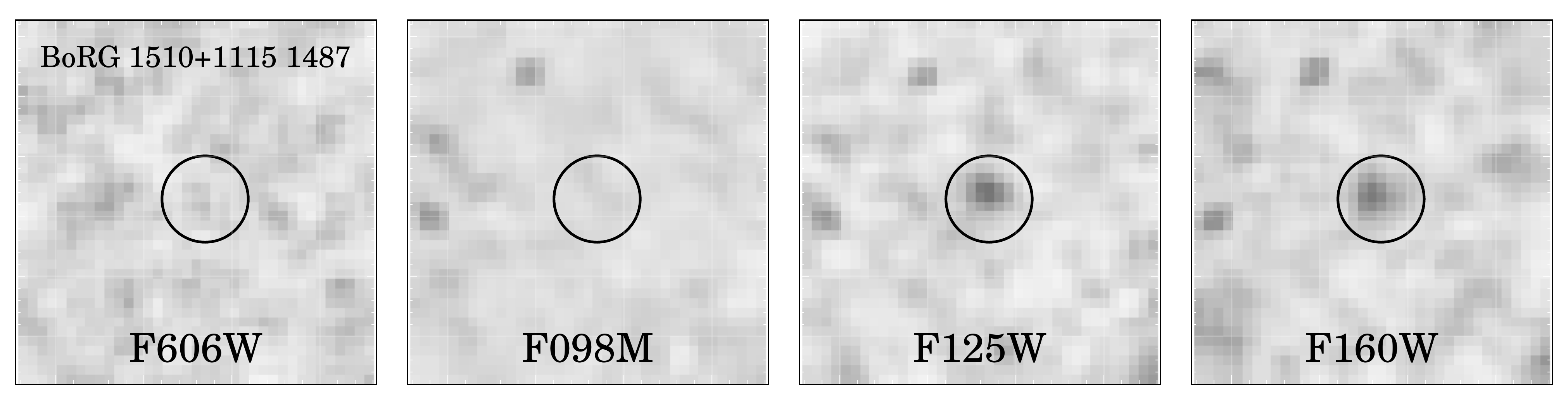}
\includegraphics[width=0.24\textwidth]{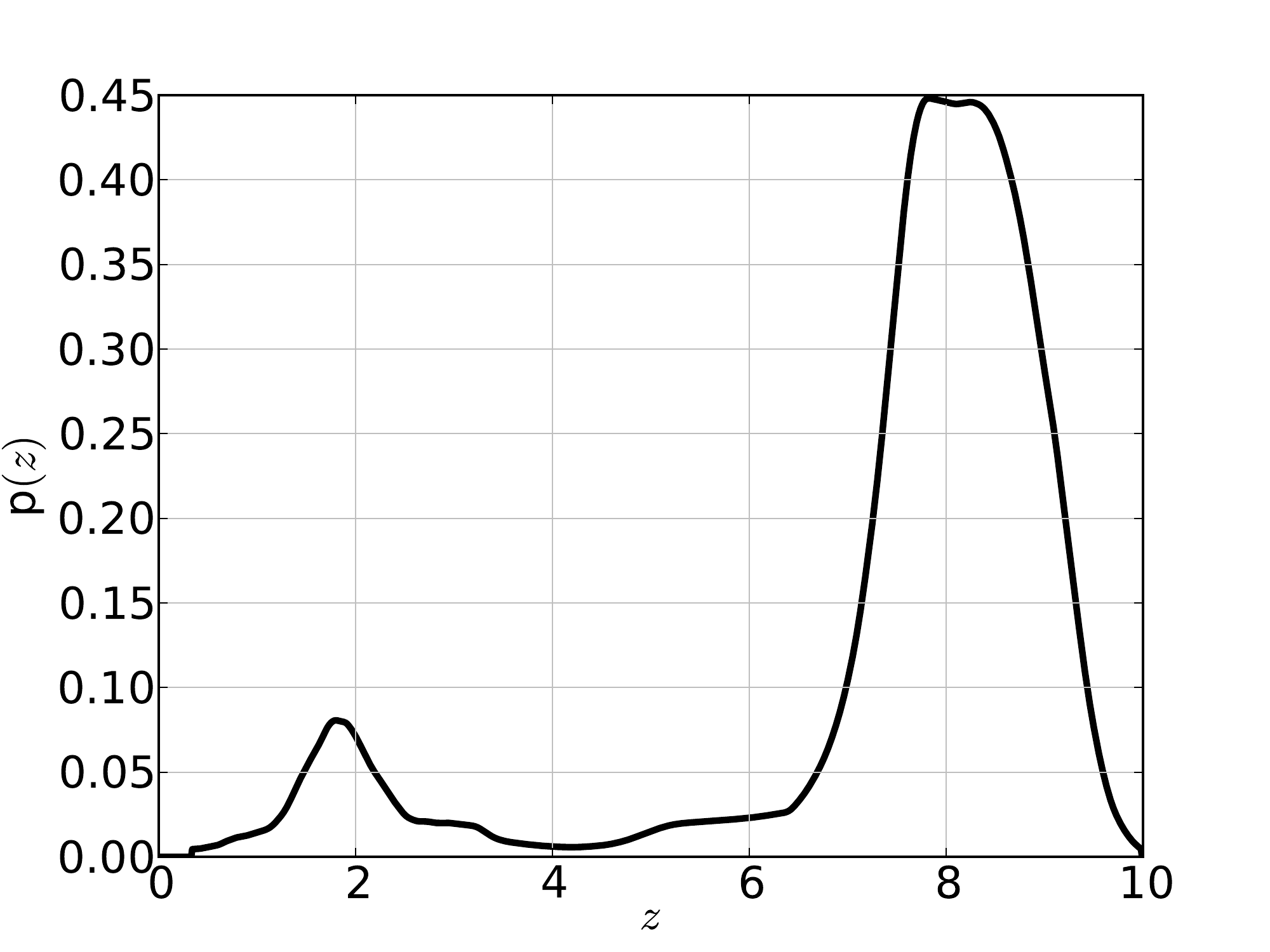}\\
\includegraphics[width=0.75\textwidth]{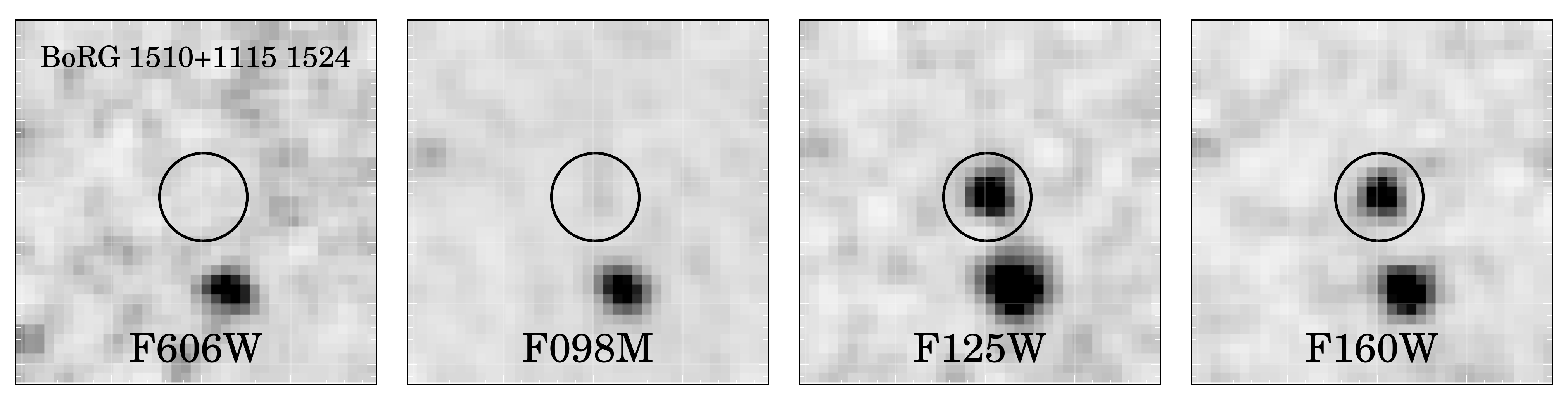}
\includegraphics[width=0.24\textwidth]{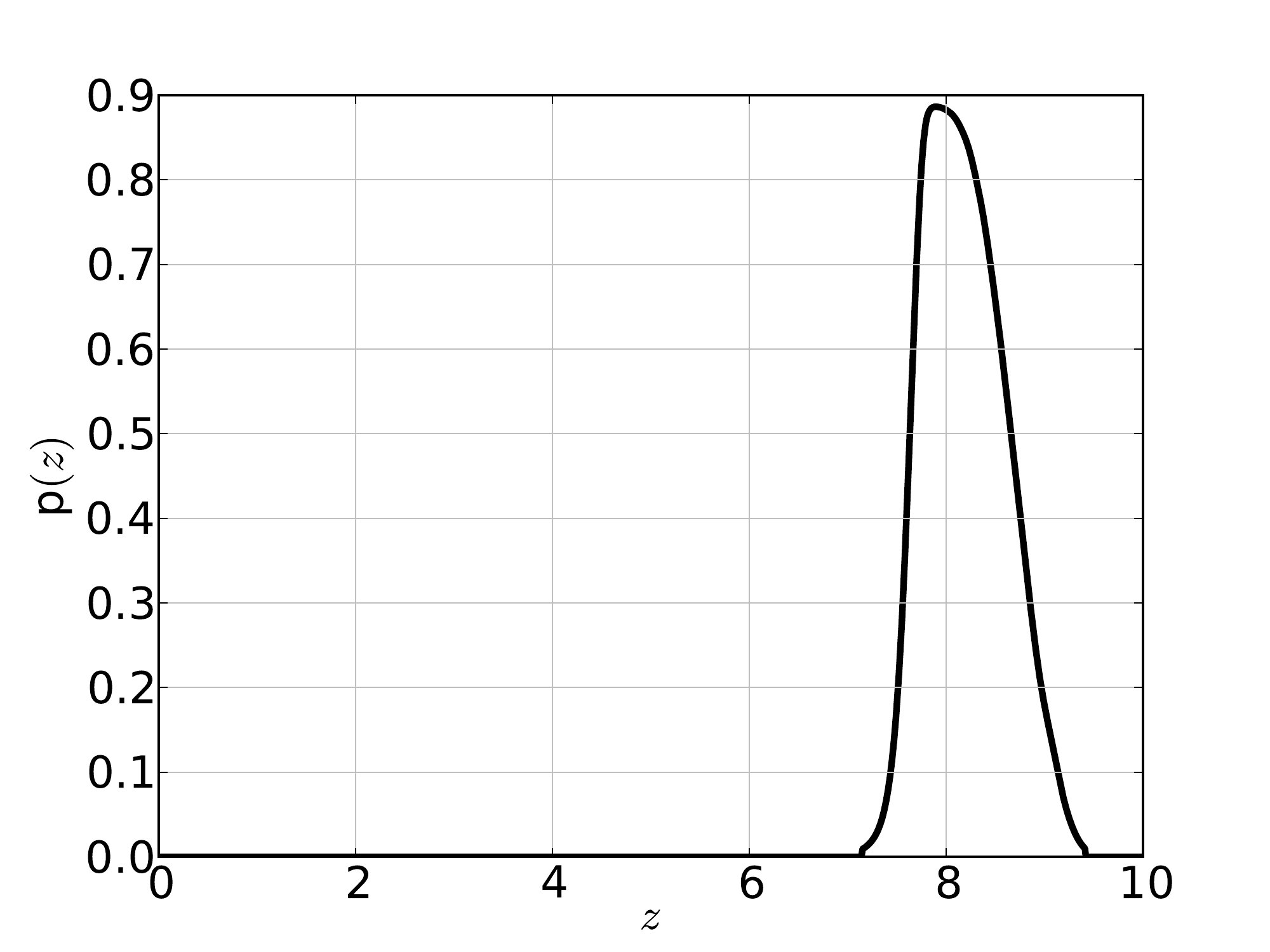}\\
\includegraphics[width=0.75\textwidth]{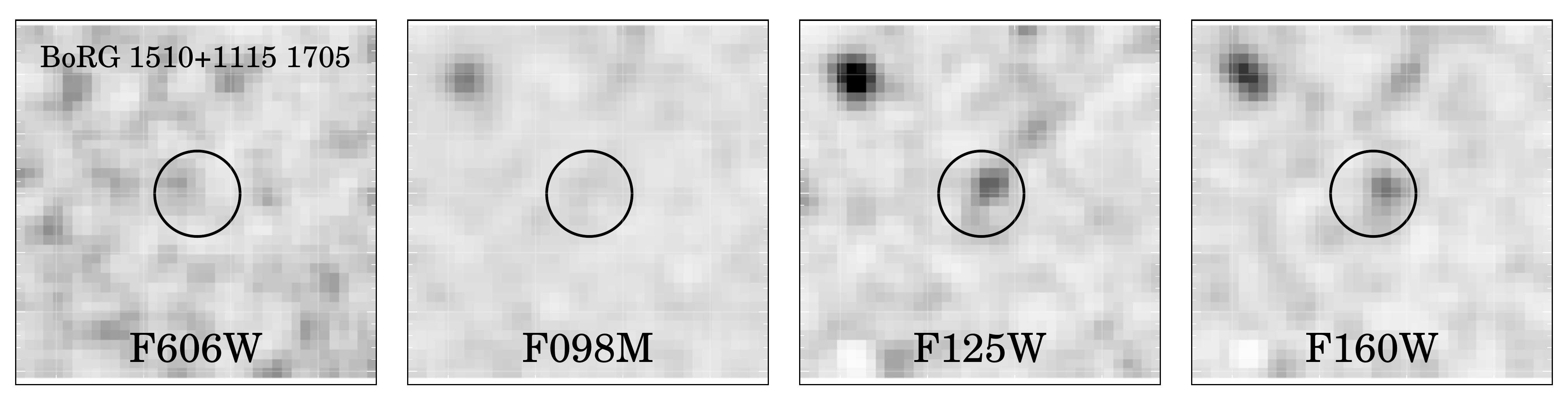}
\includegraphics[width=0.24\textwidth]{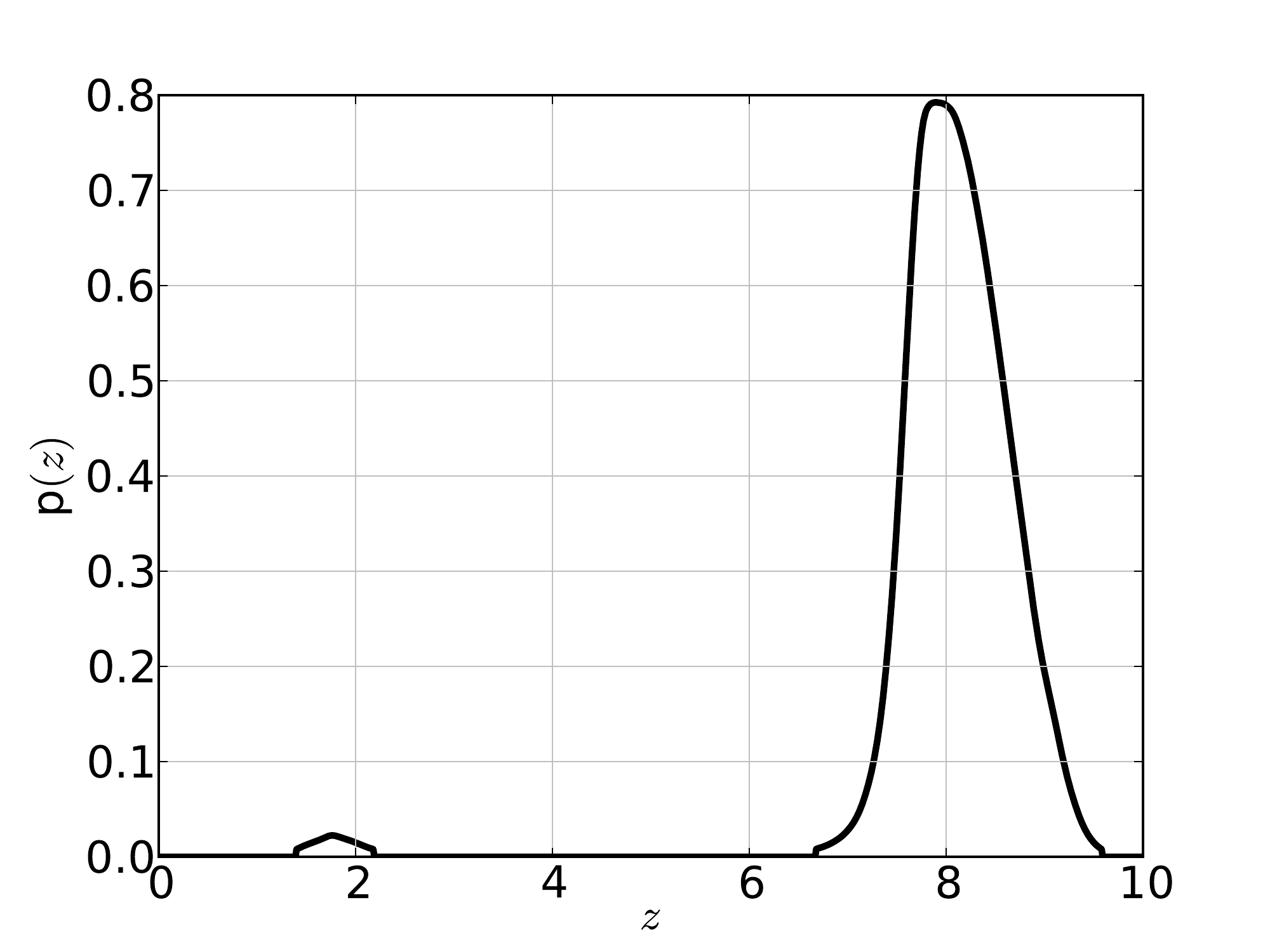}\\
\caption{The 5 $z\sim8$ Y-band dropouts in BoRG\_1510+1115. The first four columns show V-, Y-, J- and H-band 3$''$x3$''$ HST postage stamps with a power-law stretch. The last column shows the photometric redshift probability distribution $p(z)$ (using a flat prior) obtained with the Bayesian redshift  code BPZ \citep{Benitez:2004p25410,Coe:2006p25409} for each candidate.}
\label{fig:fitspz2}
\end{figure*}

The new dataset includes 11 sources passing our selection criteria. 9
of these are new findings, while two (BoRG\_1437+5043\_r2\_637 and
BoRG\_1510+1115\_1218) are candidates initially presented by
\citet{Trenti:2011p12656} and \citet{Bradley:2012p23263} and are now
confirmed at higher S/N by the deeper data. The sample of BoRG13
$z\sim8$ galaxy candidates is summarized in Table~\ref{tab:obj}. In
Figures~\ref{fig:fitspz} and \ref{fig:fitspz2} we show 3$''$x3$''$
postage stamps of all 11 BoRG13 redshift 8 galaxy candidates.

Combining the BoRG13 sample with the updated samples from BoRG09 and
BoRG12 results in a total sample of 38 bright $>5\sigma$ galaxy
candidates at $z\sim8$ from the BoRG survey. Note that a `$5\sigma$'
detection in this context means $\textrm{S/N}>5$. As we discuss in
Appendix~\ref{app:noise}, the noise background distribution is highly
non-Gaussian with significant broader wings. As discussed in the
appendix, our requirement of two-band detections and
non-detections at bluer bands, is essential to avoid contaminants, especially in the 5$\sigma$ sample.
10 of the 38 candidates have $\textrm{S/N}_\textrm{J}>8$.

\begin{table*}
\centering{
\caption[ ]{Photometry of the Y-band dropout ($z\sim8$) Candidates in BoRG13}
\label{tab:obj}
\begin{tabular}[c]{lrrcccrrrr}
\hline
\hline
ID & $\alpha_{J2000}$ & $\delta_{J2000}$ & J & Y $-$ J & J $-$ H & S/N$_\textrm{V}$ & S/N$_\textrm{Y}$ & S/N$_\textrm{J}$ & S/N$_\textrm{H}$ \\
\hline
BoRG\_0456-2203\_904 			&   73.97655 	& -22.04115 	&  26.72 $\pm$ 0.27 & 2.1 $\pm$ 0.9 & -0.2 $\pm$ 0.4 & 0.8 & 0.8 & 5.1 & 3.5 \\
BoRG\_0951+3304\_277$^a$ 		& 147.68443 	&  33.07019 	&  25.87 $\pm$ 0.22 & 2.3 $\pm$ 0.7 & -0.2 $\pm$ 0.3 & -0.6 & 1.2 & 7.9 & 5.3 \\
BoRG\_1059+0519\_991 			& 164.69864 	&    5.32262 	&  26.34 $\pm$ 0.31 & 1.8 $\pm$ 0.6 & -0.2 $\pm$ 0.4 & 0.7 & 1.6 & 5.9 & 3.6 \\
BoRG\_1358+4334\_482 			& 209.44475 	&  43.55779 	&  26.91 $\pm$ 0.26 & 2.0 $\pm$ 0.8 & 0.0 $\pm$ 0.3 & 1.3 & 0.4 & 6.4 & 4.9 \\
BoRG\_1437+5043\_r2\_637$^{a,b}$	& 219.21058 	&  50.72601 	&  25.76 $\pm$ 0.07 & 3.2 $\pm$ 0.8 & 0.1 $\pm$ 0.1 & -0.2 & 1.0 & 20.2 & 16.5 \\
BoRG\_1510+1115\_354$^{a}$	 	& 227.54706 	&  11.23145 	&  27.03 $\pm$ 0.22 & 2.1 $\pm$ 0.7 & -0.1 $\pm$ 0.3 & 0.1 & 1.2 & 7.7 & 4.6 \\
BoRG\_1510+1115\_1218$^{a,c}$ 	& 227.54266 	&  11.26152 	&  26.87 $\pm$ 0.22 & 2.2 $\pm$ 0.8 & 0.0 $\pm$ 0.3 & 0.8 & 0.7 & 7.3 & 5.2 \\
BoRG\_1510+1115\_1487$^{a}$ 	& 227.53173 	&  11.25254 	&  27.60 $\pm$ 0.24 & 2.0 $\pm$ 0.8 & 0.0 $\pm$ 0.4 & 0.4 & 0.5 & 5.8 & 4.0 \\
BoRG\_1510+1115\_1524$^{a}$ 	& 227.53812 	&  11.25552 	&  26.63 $\pm$ 0.15 & 2.3 $\pm$ 0.6 & 0.0 $\pm$ 0.2 & -0.8 & 1.5 & 11.9 & 7.9 \\
BoRG\_1510+1115\_1705$^{a}$ 	& 227.54008 	&  11.25111 	&  27.00 $\pm$ 0.19 & 1.8 $\pm$ 0.6 & -0.8 $\pm$ 0.4 & -2.0 & 1.6 & 8.4 & 2.6 \\
BoRG\_2313-2243\_1136 			& 348.24871 	& -22.71342 	&  27.14 $\pm$ 0.26 & 2.3 $\pm$ 0.8 & -0.1 $\pm$ 0.3 & -0.3 & 1.0 & 6.4 & 4.8 \\
\hline
\multicolumn{10}{l}{\textsc{Note.} -- J-band magnitudes are corrected for galactic extinction using the \cite{Cardelli:1989p26704} extinction law and the}\\
\multicolumn{10}{l}{E(B$-$V) from Table~\ref{tab:C19fields}. $^a$Followed up with MOSFIRE as presented in \cite{Treu:2013p32132}. $^b$Presented in \cite{Trenti:2011p12656}}\\
\multicolumn{10}{l}{and \cite{Bradley:2012p23263} as BoRG58\_1787-1420 and BoRG\_1437+5043\_1137, respectively. For previous photometry}\\
\multicolumn{10}{l}{of this candidate see the `B1437\_r2\_0637\_T12a' rows in Table~\ref{tab:objB1437}. $^c$Presented in \cite{Bradley:2012p23263} as}\\
\multicolumn{10}{l}{BoRG\_1510+1115\_1404. The HST postage stamps and the photometric redshift probability distribution $p(z)$ for each}\\\multicolumn{10}{l}{candidate are shown in Figures~\ref{fig:fitspz} and \ref{fig:fitspz2}.}
\end{tabular}}
\end{table*}

\subsection{Photometric Redshifts}\label{sec:bpz}

The Bayesian photometric redshift code BPZ
\citep{Benitez:2004p25410,Coe:2006p25409} was run on the photometry
for each of the LBG candidates shown in Table~\ref{tab:obj} providing photometric
redshift probability distributions, $p(z)$, for each individual
object.  In Figures~\ref{fig:fitspz} and \ref{fig:fitspz2} we show the
$p(z)$ obtained using a flat prior on the redshift distribution. In
all cases prominent probability peaks are seen at the expected Y-band
dropout redshift of 7.4--8.8.

\section{Estimating the Luminosity Function}\label{sec:LF}

In this section we present the Bayesian framework applied to infer
the luminosity function parameters. We use the empirical Schechter
function \citep{Schechter:1976p29330}, 
\BE
\Phi(L) = \frac{\phi^\star}{L^\star} \left( \frac{L}{L^\star}\right)^{\alpha} \exp\left( -\frac{L}{L^\star}\right)
\EE 
(see also Appendix~\ref{sec:BF}), as our luminosity function
model. Hence, the luminosity function parameters to fit are the
faint-end slope, $\alpha$, the `knee', $L^\star$, characterizing
the transition between the power-law part at the faint end and the
exponential cut-off at the bright end of the distribution, and the
normalization $\phi^\star$. 
As described by, e.g., \cite{Bradley:2012p23263} and \cite{Oesch:2012p30149} $\alpha$ and
$L^\star$ are degenerate.
This implies that both bright objects (the latter, for instance in case of the BoRG sample) and faint objects are
needed to obtain precise estimates of both $\alpha$ and $L^\star$. We
will illustrate this degeneracy in Section~\ref{sec:results} by
applying our framework to a sample of faint LBGs only.
The normalization $\phi^\star$ represents the co-moving number density of objects described by the luminosity function. 
The density is directly related to the total number of high-redshift objects, $N_z$, in the surveyed volume, $V$, as described below 
(\Eq{eqn:phistar} and Appendix~\ref{sec:BF}).

The framework used to fit the Schechter function in the present study
improves on the standard formalism typically adopted to estimate luminosity
functions in the literature. In particular we improve on three main
issues with luminosity function fitting that seems to have become
standard practice in the high redshift community

First of all it is often assumed that the likelihood follows a Poisson
distribution
\citep[e.g.,][]{Bradley:2012p23263,Schenker:2013p26914,McLure:2013p27183,Oesch:2012p30149}. 
As described by \citet{Kelly:2008p29070}, the Poisson distribution is
only an approximation of the formally correct binomial distribution. To the
extent that the detection probability is small and the total number of
objects in the parent sample is large, the binomial distribution
reduces to the Poisson distribution.
Hence, in the case of rare-object
luminosity functions the Poisson distribution is a fair
approximation.
Effects like cosmic variance \citep{Schenker:2013p26914} will further smear and reduce any significant differences and
we therefore do not expect
any strong bias in the shape of the luminosity function from this
approximation. 
Nevertheless it is important to quantify this
statement, by applying the formally correct binomial distribution, and verify
that this is indeed not the case. 

Secondly, it has become generally accepted to use binned samples
of dropouts when estimating the luminosity functions
\citep[e.g.,][]{Bouwens:2006p27665,Bouwens:2007p29848,Bouwens:2011p8082,Bradley:2012p23263,McLure:2013p27183,Schenker:2013p26914} instead of using the
actual data themselves. Such an approach intentionally reduces the
information to an arbitrarily defined set of bins which is sub-optimal
and introduces smoothing on the scale of the bins thus potentially
biasing the inference to flatter distributions \citep{Cara:2008p33969,Yuan:2013p29344}.

Lastly, the photometric errors on the individual sources in the high-redshift samples are often not modeled directly when obtaining the luminosity function parameters. 
\cite{Bouwens:2007p29848,Bouwens:2011p8082} use a set of generalized transfer functions to model the effect of photometric scatter in their samples and
simulations similar to the ones described in Section~\ref{sec:SF} are used to account for photometric scatter in and out of the color selection boxes \citep{Bradley:2012p23263,Oesch:2012p30149}, or between adjacent bins in the binned luminosity function \citep{Schenker:2013p26914}. 
However, due to the binning
of the data when fitting the luminosity function this does not fully
account for the photometric uncertainty of the individual objects in
the sample. Certainly previous approaches provided an approximate
treatment, but direct modeling of the photometric uncertainties
themselves makes full use of all the available information and
provides more rigorous results.

The formalism applied in this study is therefore drawing from a
posterior distribution using a likelihood based on the binomial
distribution described by \cite{Kelly:2008p29070}, avoiding binning of
the data completely and thereby estimating the luminosity function
using the full information directly, and lastly, explicitly modeling
the photometric error distribution of the sample by assuming the
errors are Gaussian distributed.

The posterior distribution for $n$ Y-band dropouts can be summarized as (see Appendix~\ref{sec:BF})
\BEA\label{eqn:postdist}
&p&(\theta \;|\; L_\textrm{J,obs},I_\textrm{V}=0) \propto \; p(\theta) \nonumber \\
&\times& C^{N_z}_{(1-f)n} C^{\frac{f}{1-f} N_z}_{f n}  \; \prod_{l}^\mathcal{C}
\left[1-\frac{ A_l}{A_\textrm{sky}}\; p(I=1|\theta) \right]^{\frac{N_z-(1-f_l)c_{l}}{1-f_l}}  \nonumber \\
&\times& \prod_i^{n} p(L_{\textrm{J,obs},i}|\theta)
\EEA
Here $\theta = (\alpha,L^\star,N_z)$ where $\alpha$ and $L^\star$ are the main Schechter 
luminosity function parameters and $N_z$ is the number of high-$z$ LBGs in the 
surveyed co-moving cosmological volume,
which as mentioned is closely related to $\phi^\star$, the Schechter function normalization.
$L_\textrm{J,obs}$ is the set of observed J-band
luminosities and $I_\textrm{V}=0$ indicates that the object is a V-band
non-detection as required by the dropout selection described in
Section~\ref{sec:zsel}. On the right-hand-side $p(\theta)$ represents the prior assumptions on the problem
and the $C^a_b$ factors are binomial coefficients.
We assume uniform priors on $\alpha$, $\log_{10}L^\star$ and $\log_{10}N_z$.
$p(I=1|\theta)$ is the probability distribution of an object making it into the dropout sample
which is independent of the $n$ individual objects in the sample, and
$p(L_{\textrm{J,obs},i}|\theta)$ is the likelihood function for the observed J-band luminosity of the $i$'th object in the sample.  
$A_l$ is the area of the individual $\mathcal{C}$ fields in the BoRG13 sample, which each contain $c_l$ high redshift candidates ($n = \sum_l^\mathcal{C} c_l$) and have an assumed contamination of $f_l$.
$A_\textrm{sky}$ is the area of the full sky.
In Appendix~\ref{sec:BF} we give the expanded
expression of the posterior distribution from \Eq{eqn:postdist} which was actually used 
when performing the luminosity function parameter inference.
We note that in the current framework the $p(z)$ prior on the redshift of the individual LBG candidates
is not explicitly taken into account.
We refer to Appendix~\ref{sec:BF} for further details.

A posterior distribution function as the one shown in
\Eq{eqn:postdist} is well suited for Markov Chain Monte Carlo (MCMC)
sampling over the luminosity function parameters $\theta$.
Calculating the posterior probability enables a
robust determination of the luminosity function as preferred by the
data given the sample of $n$ sources.  We used the Python \verb+pymc+
package\footnote{Available at \url{http://pymc-devs.github.io/pymc/}} with a
Robust Adaptive Metropolis \citep[RAM;][]{Vihola:2012p32122} algorithm to sample
the luminosity function parameters $\alpha$, $L^\star$, and $N$.  
The RAM algorithm adapts the proposal covariance matrix to obtain a fixed acceptance ratio
(set to 0.4 in this work).

\subsection{Selection and Completeness Functions}\label{sec:SF}

When estimating the luminosity function a crucial part of the
expression for the posterior distribution is the selection function
$\mathcal{S}(L_\textrm{J,obs})$ including an estimate of the
completeness of the source selection.  Here we have used two
distinct selection functions.

For each of the BoRG fields we explicitly split the total selection function into a completeness function $C(L_\textrm{J,obs})$ and a selection function not including completeness $S(L_\textrm{J,obs},z)$. 
$C(L_\textrm{J,obs})$ and $S(L_\textrm{J,obs},z)$ were simulated as described by \citet{Oesch:2007p30657,Oesch:2009p30660,Oesch:2012p30149} and \cite{Bradley:2012p23263}. In summary, the steps in obtaining these functions are:
\begin{itemize}
\item Sources with different spectral energy distribution, luminosity, redshift, and size are added to the original science images. The simulated galaxies are $z\sim 4$ galaxies rescaled to higher redshift using a size relation of $(1+z)^{-1}$ as determined from $z\sim3-7$ LBG samples \citep{Bouwens:2004p31890,Ferguson:2004p31888,Oesch:2010p30155} and  their UV-continuum slopes \citep{Dunlop:2012p26718,Finkelstein:2012p31892,Bouwens:2012p31891}.
\item The detection and selection procedure described in Section~\ref{sec:ydrop} is re-run for each field to determine $C(L_\textrm{J,obs})$ and $S(L_\textrm{J,obs},z)$.
\end{itemize}

\begin{figure}
\includegraphics[width=0.49\textwidth]{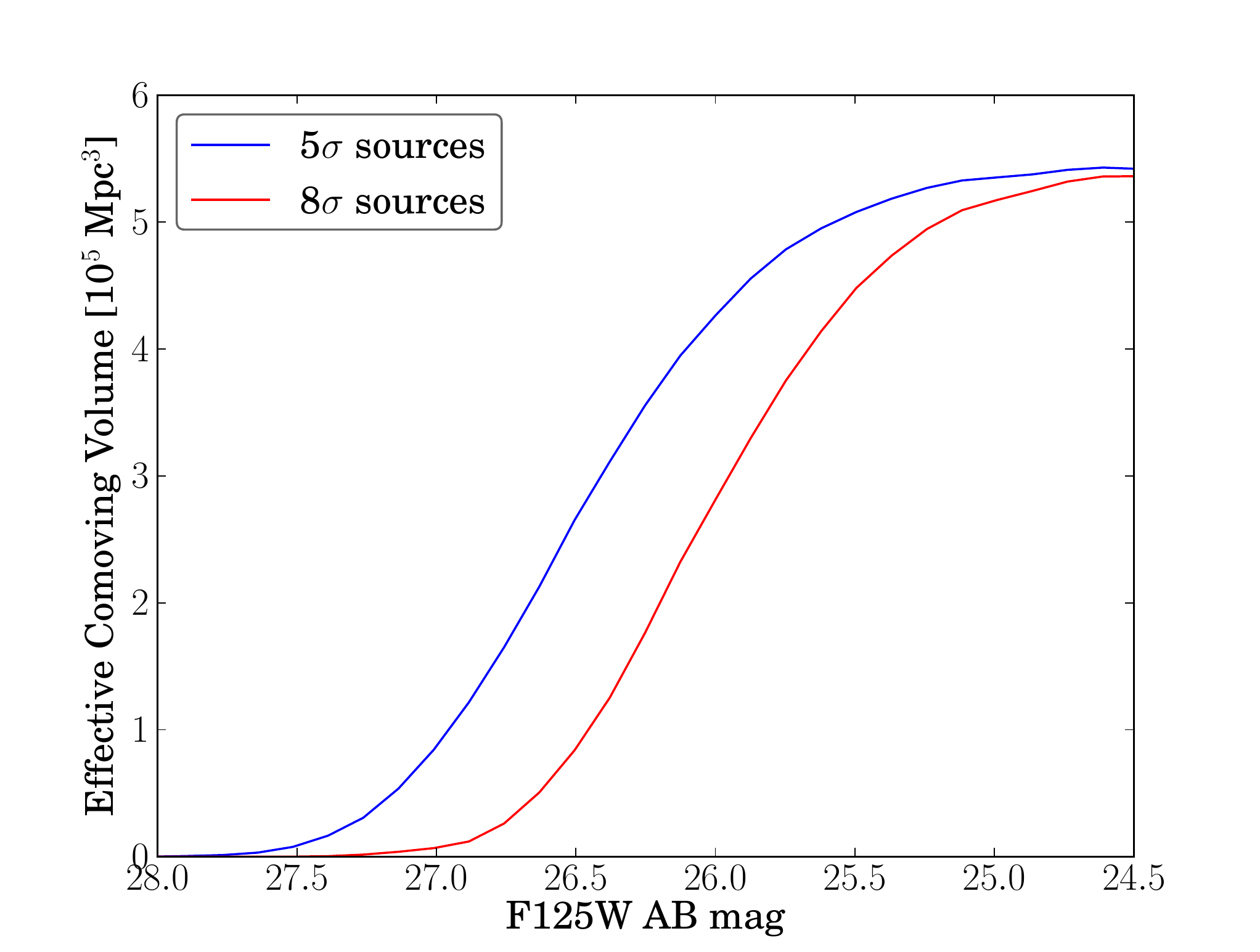}
\caption{The effective co-moving volume of BoRG13 as a function of J-band (F125W) magnitude for the 5$\sigma$ (blue) and 8$\sigma$ (red) sample. This is calculated taking the selection functions $S(L_\textrm{J,obs},z)$ and completeness functions $C(L_\textrm{J,obs})$ described in Section~\ref{sec:SF} of each individual field into account.}
\label{fig:B13vol}
\end{figure}

The selection function $\mathcal{S}(L_\textrm{J,obs})$ (used in Section~\ref{sec:eqs}) is as mentioned corrected for
completeness. 
When estimating the intrinsic luminosity function the BoRG13 LBGs are all assumed to be at $z=8$.
Hence, when sampling the posterior for the BoRG fields
\BE
\mathcal{S}(L_\textrm{J,obs}) = C(L_\textrm{J,obs})\times S(L_\textrm{J,obs},z=8) \; .
\EE

The effective volume of the BoRG survey is derived from the full selection function as a function of J-band magnitude, i.e., $C(L_\textrm{J,obs})\times S(L_\textrm{J,obs},z)$. 
The total effective co-moving volumes for the 5$\sigma$ and 8$\sigma$ BoRG13 samples are shown in Figure~\ref{fig:B13vol}

To break the $\alpha$-$L^\star$ parameter degeneracy we used a sample of 59 faint LBGs
from the Hubble Ultradeep Field (HUDF) and Early Release Science (ERS) programs 
\citep[Tables 14--17 in][]{Bouwens:2011p8082} to
populate the faint end of the distribution when fitting the luminosity function.  
For this sample a piecewise selection function $\mathcal{S}(L_\textrm{J,obs})$ following
the binned data presented in \cite{Bouwens:2011p8082} was used.  Integration over
this selection function was done using linear interpolation between
the bins.
Note that the effect of large-scale structures (cosmic variance) on the $z~8$ luminosity functions in the HUDF/ERS luminosity range is very modest as described by \cite{Bouwens:2011p8082}.

Both the BoRG and HUDF/ERS selection function models $\mathcal{S}(L_\textrm{J,obs})$ are explicitly
 set to 1 and 0 for luminosities brighter than and fainter than the modeled luminosities, respectively. 

\subsection{Contamination}\label{sec:contamination}

When estimating the number of contaminants among the BoRG LBG candidates
we follow the approach taken by \cite{Bradley:2012p23263}. 
We use a fiducial average contamination level of $f_\textrm{BoRG} = 0.42$ for the BoRG sample. 
(In Section~\ref{sec:LFB13} we investigate the effects of modifying this assumption on the shape of the luminosity function)
The contamination of the HUDF/ERS sample is included in the selection function described above and is therefore assumed to be 0.0 for the \cite{Bouwens:2011p8082} sample.
The available information about the contaminants at redshift 8 is limited and distinguishing between the luminosity function of the contaminants and the $z\sim8$ LBGs complicates matters considerably
Hence, we assume that the fraction of contaminants is
independent of luminosity for the BoRG sample. In \Eq{eqn:postdist} The
number of contaminants are approximated by their expectation values as
explained in Appendix~\ref{sec:BF}.

Having obtained $N_z$ and assuming that the luminosity function does not evolve over the redshift interval of interest, the number density is obtained by calculating (see \Eq{eqn:Nuni} in Appendix~\ref{sec:BF})
\BEA\label{eqn:phistar}
\phi^\star &=& \frac{N_z}{V \times \int_{L_\textrm{min}}^\infty \;\frac{1}{L^\star} \left( \frac{L}{L^\star}\right)^{k-1} \exp\left( -\frac{L}{L^\star}\right) \;dL} \; .
\EEA
Here $V$ is the co-moving cosmological volume of the ideal full-sky survey. To avoid divergence of the integral we set our lower integration limit of the luminosity function to $L_\textrm{min}=10^{40}$erg/s which corresponds to an absolute magnitude of $M\sim -10$. 
Hence, $L_\textrm{min}$ sets the lower bound for what we include as ``a galaxy'' in our analysis.
With an estimated normalization of the sample luminosity function, the luminosity density, $\epsilon$, is obtained by integrating the luminosity function multiplied by the luminosity itself (see Appendix~\ref{sec:BF}).

\subsection{Sanity Check of Bayesian Framework}\label{sec:sanity}

Prior to applying the newly developed Bayesian inference framework to the BoRG13 data it was tested extensively on a set of simulated data samples.
We confirmed that in all cases the MCMC sampling recovered the input luminosity functions from which the emulated samples were drawn.
We also tested the code on the $z\sim8$ LBG sample analyzed in \cite{Bradley:2012p23263} (BoRG12).
The obtained luminosity function had a faint-end slope of $\alpha = -2.06^{+0.27}_{-0.25}$ and $M^\star=-20.40^{+0.36}_{-0.45}$ in agreement with the $\alpha=-1.98^{+0.23}_{-0.22}$ and $M^\star=-20.26^{+0.29}_{-0.34}$ presented by \citet{Bradley:2012p23263}. 
Within the 1$\sigma$ confidence intervals these measurements of both $\alpha$ and $M^{\star}$ are fully consistent. 
We note that we do not expect the results to be identical, since our framework avoids some of the approximations of previous studies.

\section{Results}\label{sec:results}

From the framework described in Section~\ref{sec:LF} the assumed
intrinsic luminosity function for a sample of objects can be
determined.  
In the following we will describe and summarize our findings from applying our framework 
to the 97 $z\sim8$ LBG candidates from the combined samples of BoRG13 and
HUDF/ERS from \cite{Bouwens:2011p8082}.
We note that to ease comparison with the literature we
convert $L^\star$ into $M^\star$ in the remainder of this work using
that $M^\star = M_{\textrm{UV}\odot} - 2.5\log_{10}(L^\star/L_\odot)$ with
$M_{\textrm{UV}\odot} = 5.48$.

\begin{table*}
\centering{
\caption[ ]{Comparison of Luminosity Function Parameters}
\label{tab:results}
\begin{tabular}[c]{lllll}
\hline
\hline
Reference &  $M^\star$ & $\alpha$ & $\log_{10} \phi^\star$ [Mpc$^{-3}$] & $\log_{10}\epsilon$ [erg/s/Hz/Mpc$^{3}$]\\
\hline
BoRG13 $5\sigma$ sample (This work)	& $-20.15^{+0.29}_{-0.38}$	& $-1.87^{+0.26}_{-0.26}$	& $-3.24^{+0.25}_{-0.34}$	& $25.52^\textrm{a}\; ^{+0.05}_{-0.05}$ \\
BoRG13 $8\sigma$ sample (This work)	& $-20.40^{+0.39}_{-0.55}$	& $-2.08^{+0.30}_{-0.29}$	& $-3.51^{+0.36}_{-0.52}$	& $25.50^\textrm{a}\; ^{+0.05}_{-0.06}$ \\
\cite{Bradley:2012p23263}         		& $-20.26^{+0.29}_{-0.34}$	& $-1.98^{+0.23}_{-0.22}$ 	& $-3.37^{+0.26}_{-0.29}$ 	& $25.50^\textrm{a}$ \\ 
\citet{Oesch:2012p30149}       			& $-19.80 ^{+0.46}_{-0.57}$	& $-2.06^{+0.45}_{-0.37}$ 	& $-3.17^{+0.40}_{-0.55}$ 	& $25.58^\textrm{b} \pm 0.12$ \\
\citet{Bouwens:2011p8082}    			& $-20.10\pm0.52$            		& $-1.91\pm0.32$           		& $-3.23^{+0.74}_{-0.27}$	& $25.65^\textrm{b} \pm 0.11$ \\
\cite{Lorenzoni:2011p12992}   			& $-19.5$                   			& $-1.7$ (fixed)           		& $-3.0$                   			& $25.23^\textrm{c} $ \\ 
\cite{Trenti:2011p12656}      			& $ -20.2\pm0.3$             		& $-2.0$ (fixed)           		& $-3.4$ (fixed)          		& $25.45^\textrm{a}$ \\
\cite{McLure:2010p31124}      			& $-20.04$ (fixed)          		& $-1.71$ (fixed)          		& $-3.46$                  			& $25.17^\textrm{a}$ \\
\cite{Bouwens:2010p30142}    			& $-19.5\pm0.3$            		& $-1.74$ (fixed)			& $-2.96$ (fixed)          		& $25.18^\textrm{d} \pm 0.24$ \\
\cite{Schenker:2013p26914}			& $-20.44^{+0.47}_{-0.33}$	& $-1.94^{+0.21}_{-0.24}$	& $-3.50^{+0.35}_{-0.32}$	&  $25.46^\textrm{a}$ \\
\cite{McLure:2013p27183}			& $-20.12^{+0.37}_{-0.48}$	& $-2.02^{+0.22}_{-0.23}$  	& $-3.35^{+0.28}_{-0.47}$	&  $25.46^\textrm{a}$ (but see their Figure~7) \\
\hline
\multicolumn{5}{l}{\textsc{Note.} -- $^\textrm{a}$Estimated from the luminosity function parameters via the method described in Appendix~\ref{sec:BF} integrated}\\
\multicolumn{5}{l}{down to $M=-17.7$ at $\lambda=1600$\AA{} using $M_{\textrm{UV}\odot} = 5.48$. A small $\sim$0.13dex offset between these estimates and the}\\
\multicolumn{5}{l}{literature was found. This difference can be recovered using $M_{\textrm{UV}\odot} \sim 4.7$ suggesting that a value similar to this}\\
\multicolumn{5}{l}{was used in, e.g., \citet{Oesch:2012p30149,Bouwens:2011p8082} and \cite{McLure:2013p27183}. $^\textrm{b}$From \citet{Oesch:2012p30149} and}\\
\multicolumn{5}{l}{\citet{Bouwens:2011p8082}, respectively, where the luminosity function was integrated down to $M=-17.7$ for $\lambda=1600$\AA. }\\
\multicolumn{5}{l}{$^\textrm{c}$From \cite{Lorenzoni:2011p12992} where the luminosity function was integrated down to $M=-18.5$ for $\lambda=1600$\AA. }\\
\multicolumn{5}{l}{$^\textrm{d}$From \cite{Bouwens:2010p30142} where the luminosity function was integrated down to $M=-18.2$ for $\lambda=1700$\AA. }\\
\end{tabular}}
\end{table*}

\subsection{The Luminosity Function at $z\sim8$ from BoRG13}\label{sec:LFB13}

We apply the inference framework to the full BoRG13 sample of 38 objects augmented by the fainter 59 candidates presented in \cite{Bouwens:2011p8082}. For a consistency check on a more robust determination of the luminosity function bright end, we also consider a restricted BoRG13 sample consisting of the 10 objects in BoRG13 detected at $\textrm{S/N}_\textrm{J-band}>8$, similarly to the approach followed by \cite{Bradley:2012p23263}. This additional step allows us to validate the luminosity function derived from the full sample, which might be affected by photometric scatter as discussed in Appendices~\ref{app:borg58}~and~\ref{app:noise}.

When sampling the posterior distribution (given in Appendix~\ref{sec:BF} \Eq{eqn:margpost}) we use the selection functions described in Section~\ref{sec:SF}, i.e., interpolation over a piecewise selection function for the HUDF/ERS data from \cite{Bouwens:2011p8082} and the simulated selection and completeness functions for the 5$\sigma$ and 8$\sigma$ BoRG samples at $z=8$.

\begin{figure}
\includegraphics[width=0.49\textwidth]{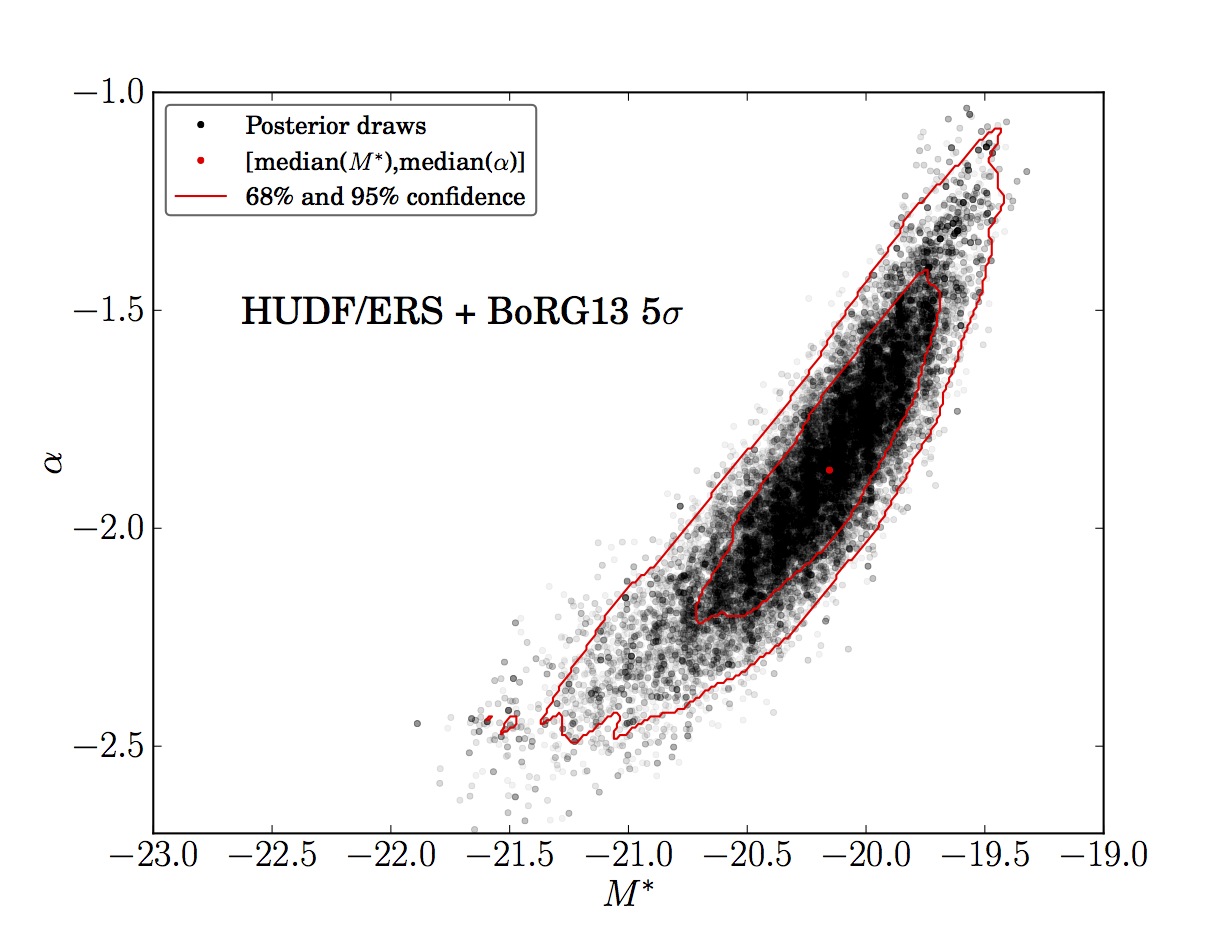}\\
\includegraphics[width=0.49\textwidth]{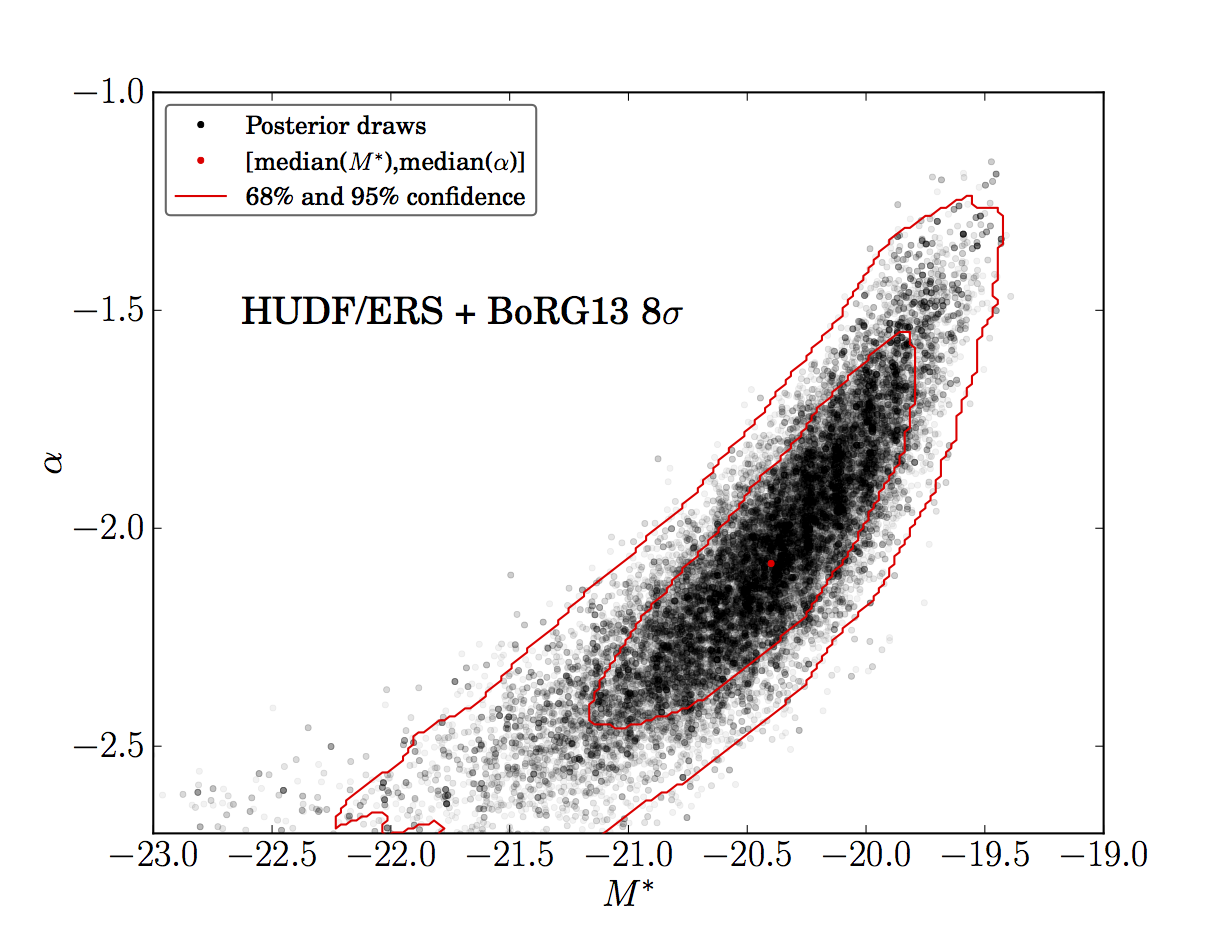}\\
\includegraphics[width=0.49\textwidth]{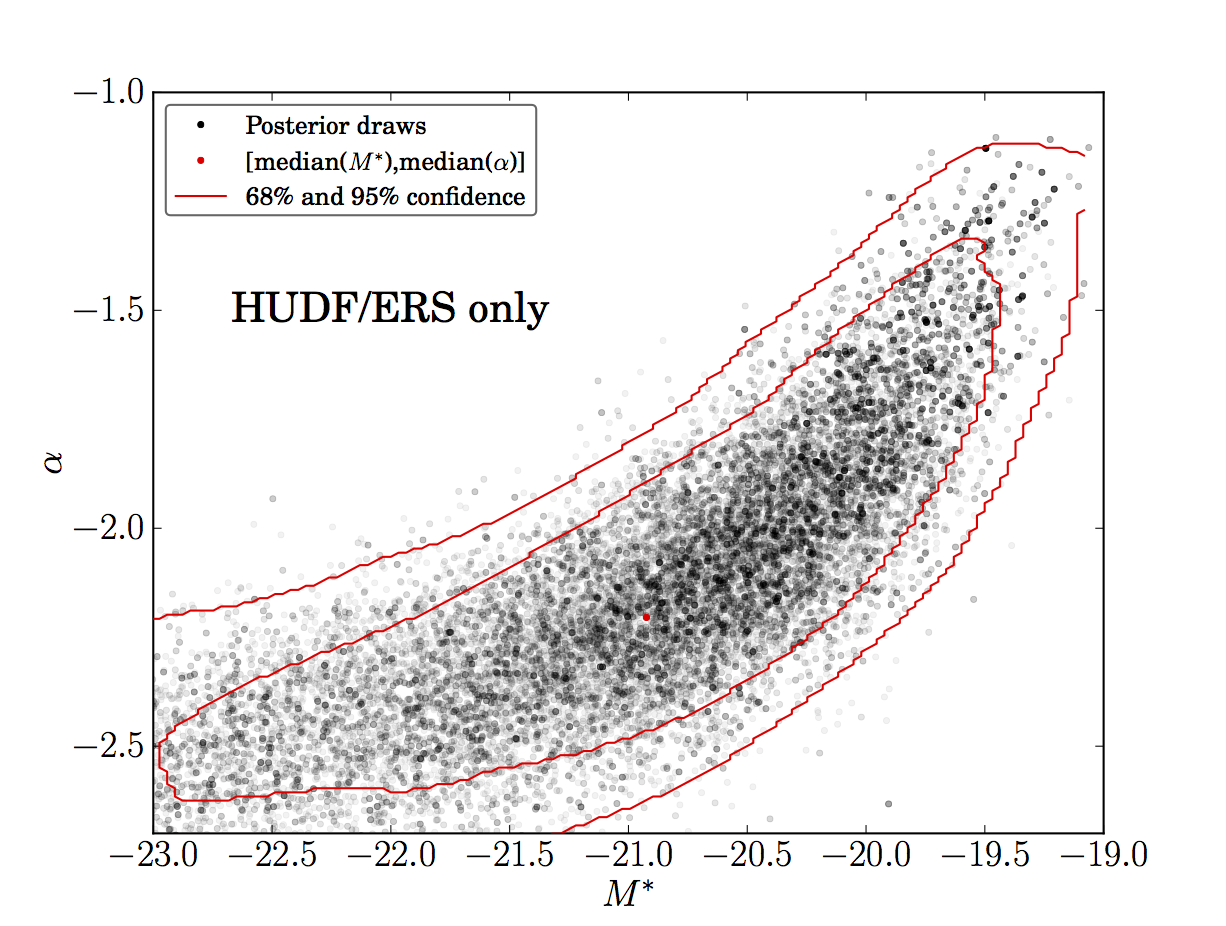}
\caption{The MCMC draws (black dots) from inferring the luminosity function parameters $\alpha$ and $M^\star$ based on candidate $z\sim8$ LBG samples. Here $L^\star$ has been converted to absolute magnitude $M^\star$ to ease comparison with the literature.
From top to bottom results from the BoRG13 5$\sigma$ sample, the BoRG13 8$\sigma$ sample,
both including the \cite{Bouwens:2011p8082} faint HUDF/ERS $z\sim8$ LBG sample,
and the \cite{Bouwens:2011p8082} sample only is shown. The latter clearly illustrates the degeneracy between $\alpha$ and $M^\star$.
Adding the bright BoRG Y-band dropouts greatly improve the $\alpha$ and $M^\star$ estimates.
The best-fit BoRG13 values (small red dots) agree well with the sample of literature luminosity functions presented in Table~\ref{tab:results}.
Contours show 68.2\% and 95.4\% (1$\sigma$ and 2$\sigma$) confidence intervals.}
\label{fig:mcmc}
\end{figure}

The results from sampling $\theta$ after an MCMC burn-in phase are shown in Figure~\ref{fig:mcmc}. From top to bottom we show the results from fitting the Schechter luminosity function to the BoRG13 5$\sigma$, BoRG13 8$\sigma$ (both including the \cite{Bouwens:2011p8082} sample), and the \cite{Bouwens:2011p8082} samples. The latter is presented to illustrate the prominent degeneracy between the $\alpha$ and $M^\star$ luminosity function parameters without the bright BoRG LBGs to constrain the bright end of the distribution and break the degeneracy. 
The 1$\sigma$ and 2$\sigma$ confidence intervals are indicated by the red contours.

In Table~\ref{tab:results} the results from the full BoRG13 luminosity
function and the BoRG13 luminosity function only using the 8$\sigma$
candidates are summarized together with a selection of recent
luminosity function fits from the literature.  The BoRG13 luminosity
function parameters $\alpha$ and $M^\star$ agree well with the
literature values, confirming that the Poisson approximation is indeed
a fair approximation when fitting luminosity functions at $z\sim8$, in
terms of best-fit value.  We note that the BoRG13 5$\sigma$ best-fit
faint-end slope is fully consistent within the 1-$\sigma$ error bars
with the literature values.
We also note that the BoRG13 5$\sigma$ faint-end slope of -1.87 is
fully consistent with the faint-end slope obtained from the BoRG12
data using the framework presented here ($\alpha=-2.06$) within the
1$\sigma$ error bars and fall comfortably within the 68\% confidence
interval contour when also considering the degeneracy with $M^\star$
in the top panel of Figure~\ref{fig:mcmc}.

\begin{figure}
\includegraphics[width=0.49\textwidth]{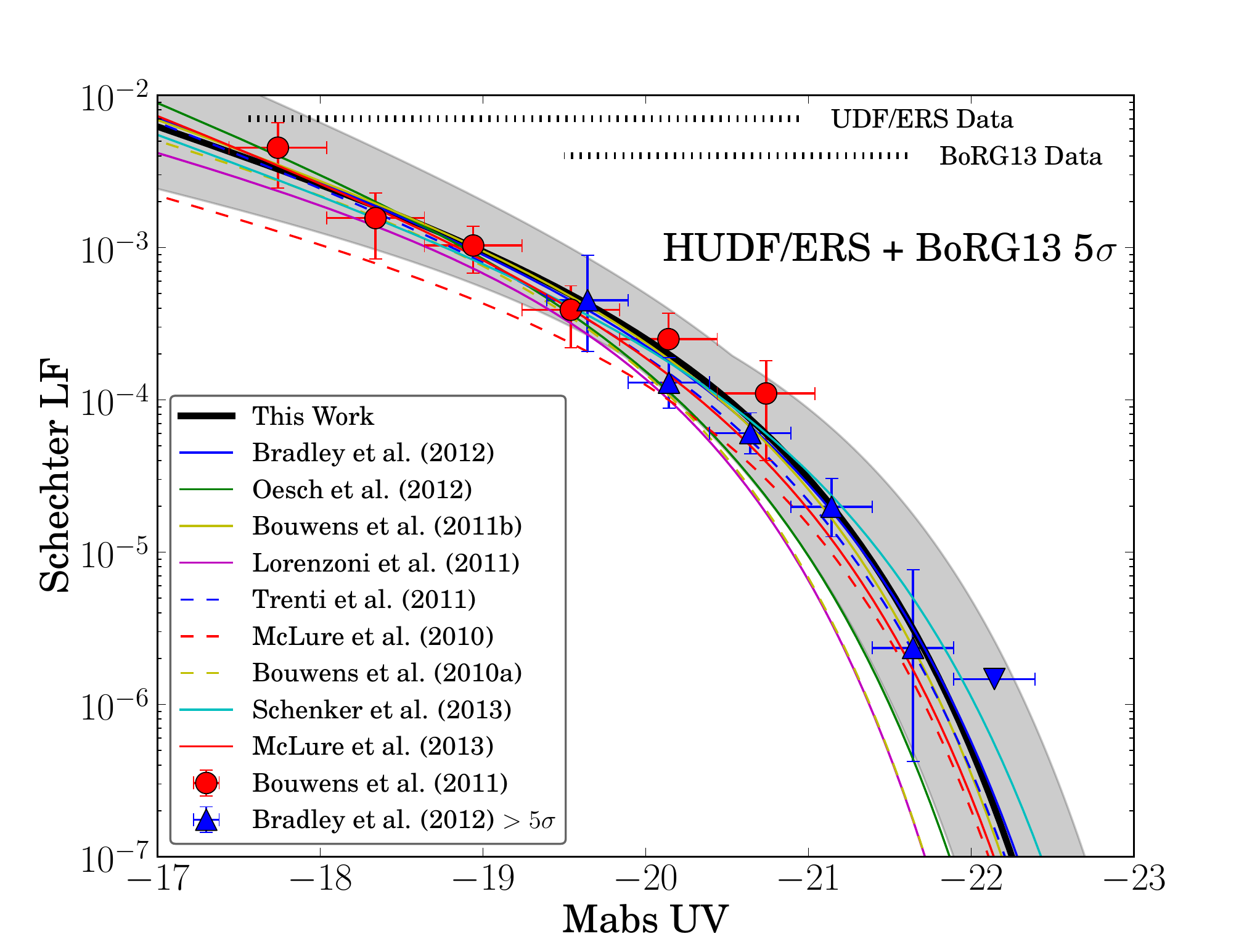}\\
\includegraphics[width=0.49\textwidth]{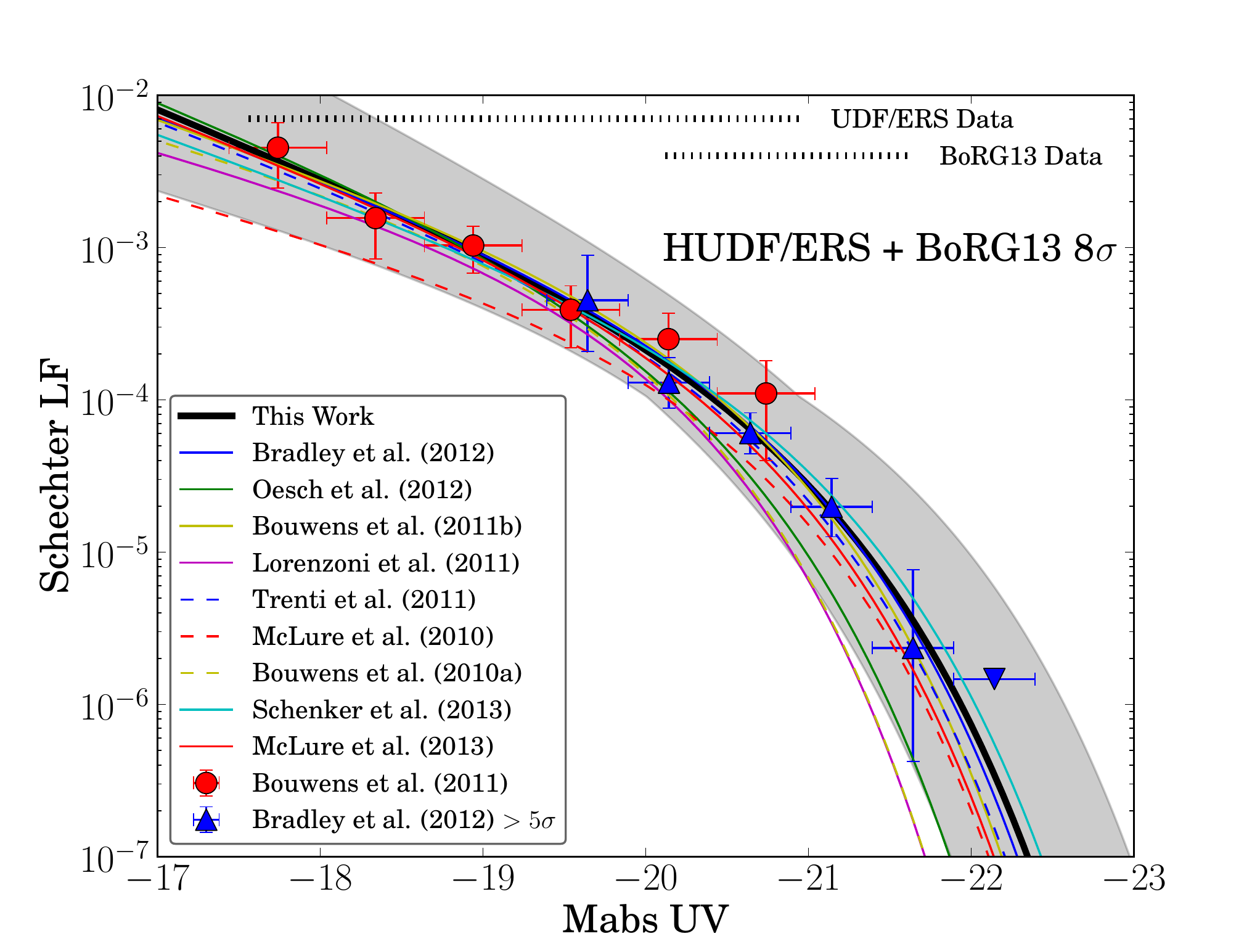}\\
\includegraphics[width=0.49\textwidth]{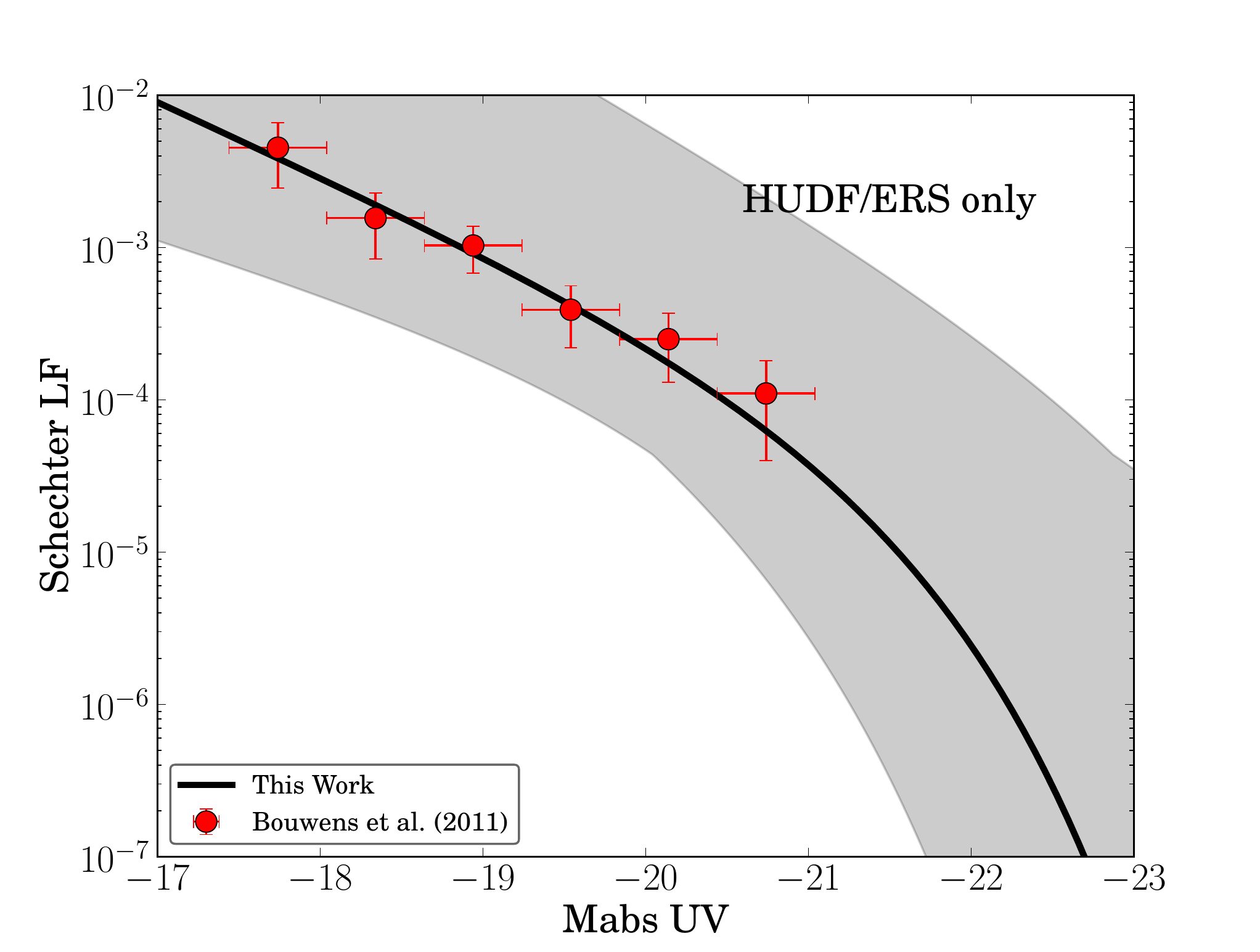}
\caption{The BoRG13 luminosity function (black solid line) corresponding to the median values of the MCMC samples shown in Figure~\ref{fig:mcmc} for the BoRG13 5$\sigma$ sample and the BoRG13 8$\sigma$ sample both including the \cite{Bouwens:2011p8082} faint HUDF/ERS $z\sim8$ LBG sample are shown in the two top panels. 
The dotted lines indicate the ranges the BoRG13 5$\sigma$ sample and the HUDF/ERS \cite{Bouwens:2011p8082} sample span.
Each luminosity function is compared to a sample of luminosity functions from the literature (see Table~\ref{tab:results}). 
The binned data from BoRG12 \citep{Bradley:2012p23263} and HUDF/ERS 
\citep{Bouwens:2011p8082} are shown for reference as the blue and red symbols, respectively.
We emphasize that we are \emph{not} fitting to these binned data. 
We take advantage of the full information of the data sets by using the full unbind BoRG13 data.
In the bottom panel the resulting luminosity function using only the faint HUDF/ERS \cite{Bouwens:2011p8082} sample is shown. 
The gray shaded regions in all three panels show the 68\% confidence intervals of the MCMC draws.
The BoRG13 Schechter luminosity functions are seen to agree well with the literature luminosity functions.}
\label{fig:LF}
\end{figure}

The luminosity functions corresponding to each of the MCMC samples
shown in Figure~\ref{fig:mcmc} are compared to the literature
luminosity functions from Table~\ref{tab:results} in
Figure~\ref{fig:LF}.

\begin{figure}
\includegraphics[width=0.49\textwidth]{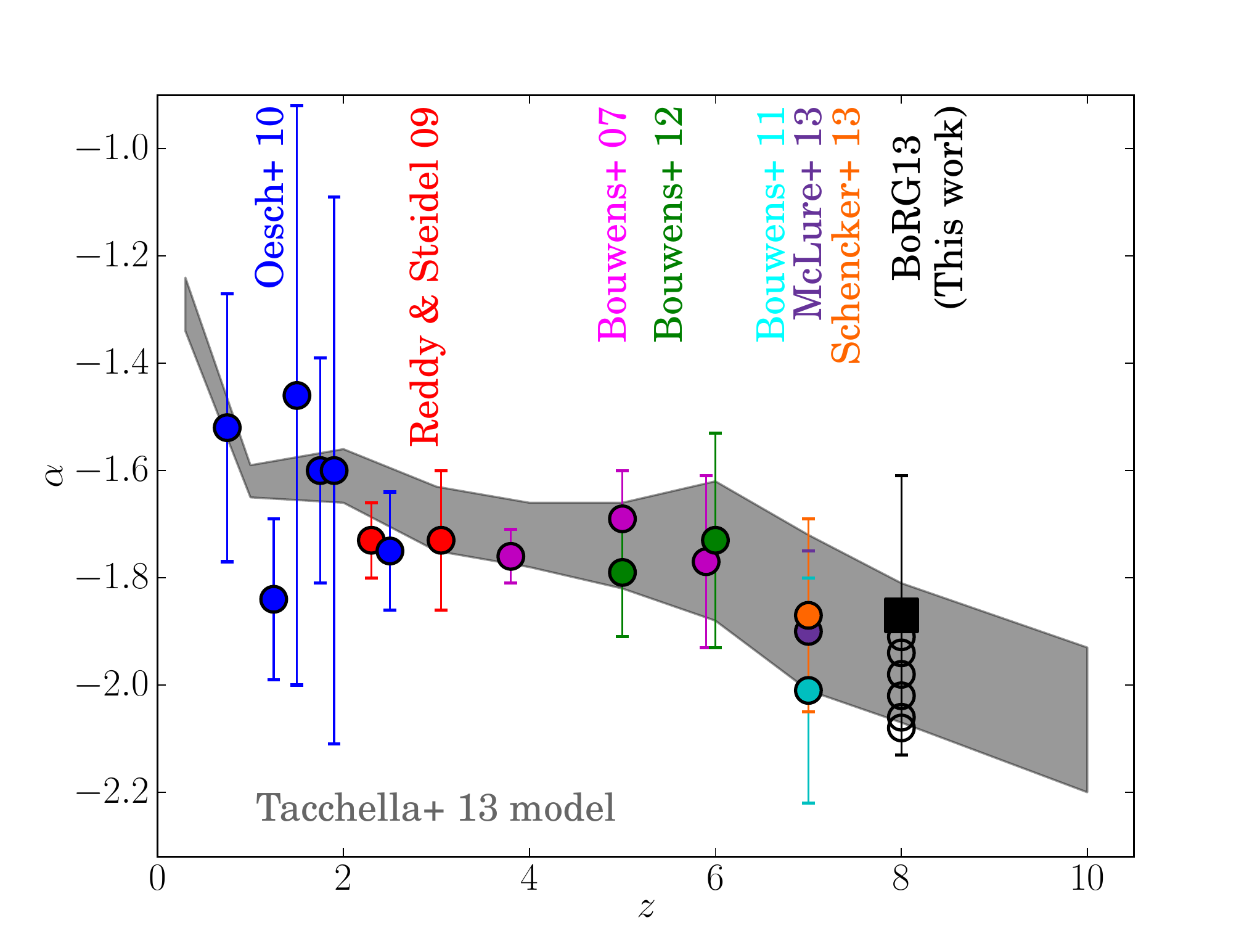}
\caption{The luminosity function faint-end slope $\alpha$ as a function of redshift. The references of the individual points are indicated above the points. The shaded region shows the physical model of the faint-end-slope evolution from \cite{Tacchella:2013p32241}. The square at $z\sim8$ shows the result presented here for the 5$\sigma$ sample. 
The open circles at redshift 8 show the literature faint-end slopes (without error-bars not to clutter the plot) from Table~\ref{tab:results} for comparison.
The uncertainties are too large to confirm or reject the suggested steepening of the luminosity function at $z\gtrsim7$.}
\label{fig:alphaz}
\end{figure}

The relatively shallow best-fit faint-end slope of -1.87 for the BoRG13 5$\sigma$ sample might
question the speculated steepening of the luminosity function at $z>6$. 
Such a steepening is expected from galaxy formation models and cosmological simulations 
\citep{Trenti:2010p29335,Jaacks:2013p27519,Tacchella:2013p32241}, and would be related 
to the evolution of the slope of the dark matter halo mass function at the scale of 
$M_h\sim 10^{10}~\mathrm{M_{\odot}}$ which is characteristic for hosting the faint galaxies observed in the HUDF. 
To place our latest determination of $\alpha(z=8)$ in the broader context of other studies, 
Figure~\ref{fig:alphaz} shows the redshift evolution predicted by a recent theoretical models \citep[e.g.,][]{Tacchella:2013p32241,Cai:2014p34530}, 
together with estimated UV luminosity function faint-end slopes at 
$z\sim0.7-2.5$ \citep{Oesch:2010p30155}, 
$z\sim2-3$ \citep{Reddy:2009p31927},  
$z\sim4$ \citep{Bouwens:2007p29848}, 
$z\sim4-7$ \citep{Bouwens:2012p31891}, and 
$z\sim7$ \citep{Bouwens:2011p8082}. 
As the error bars on our $z\sim8$ $\alpha$-estimate are still fairly large with $\Delta \alpha =0.26$
a potential decline is still very possible within 1$\sigma$.
Even though cosmic variance is
insignificant over the full BoRG13 sample, it is important to check
that it does not have a significant effect on the estimate of the
faint-end slope, which is based on only three quasi-independent fields
(HUDF). Based on the calculations by \citet{Trenti:2008p32309}, the
additional uncertainty due to cosmic variance is $\Delta \alpha \sim
0.08$, which is small compared to the statistical uncertainty of our
study, and therefore can be neglected.

This all highlights that significantly larger samples of faint as well as bright redshift 8 galaxies are needed to
robustly determine the luminosity function shape and to confirm the suggested steepening of $\alpha(z)$ with redshift.
We note that the more robust BoRG13 8$\sigma$ sample has a steeper faint-end slope and therefore favors
a steepening of $\alpha$ despite the marginally larger error bars compared to the BoRG13 5$\sigma$ sample.

\begin{figure*}
\includegraphics[width=0.49\textwidth]{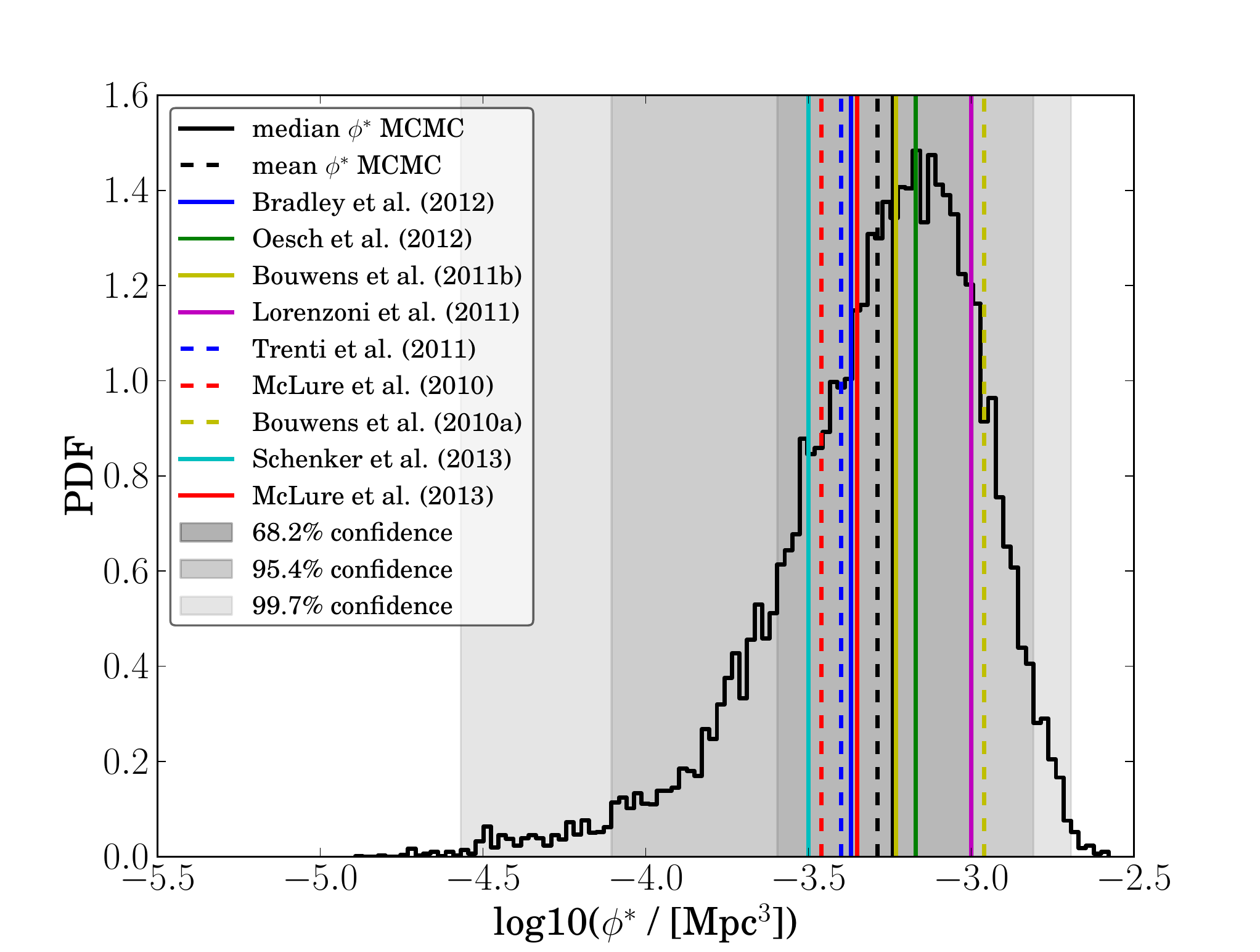}
\includegraphics[width=0.49\textwidth]{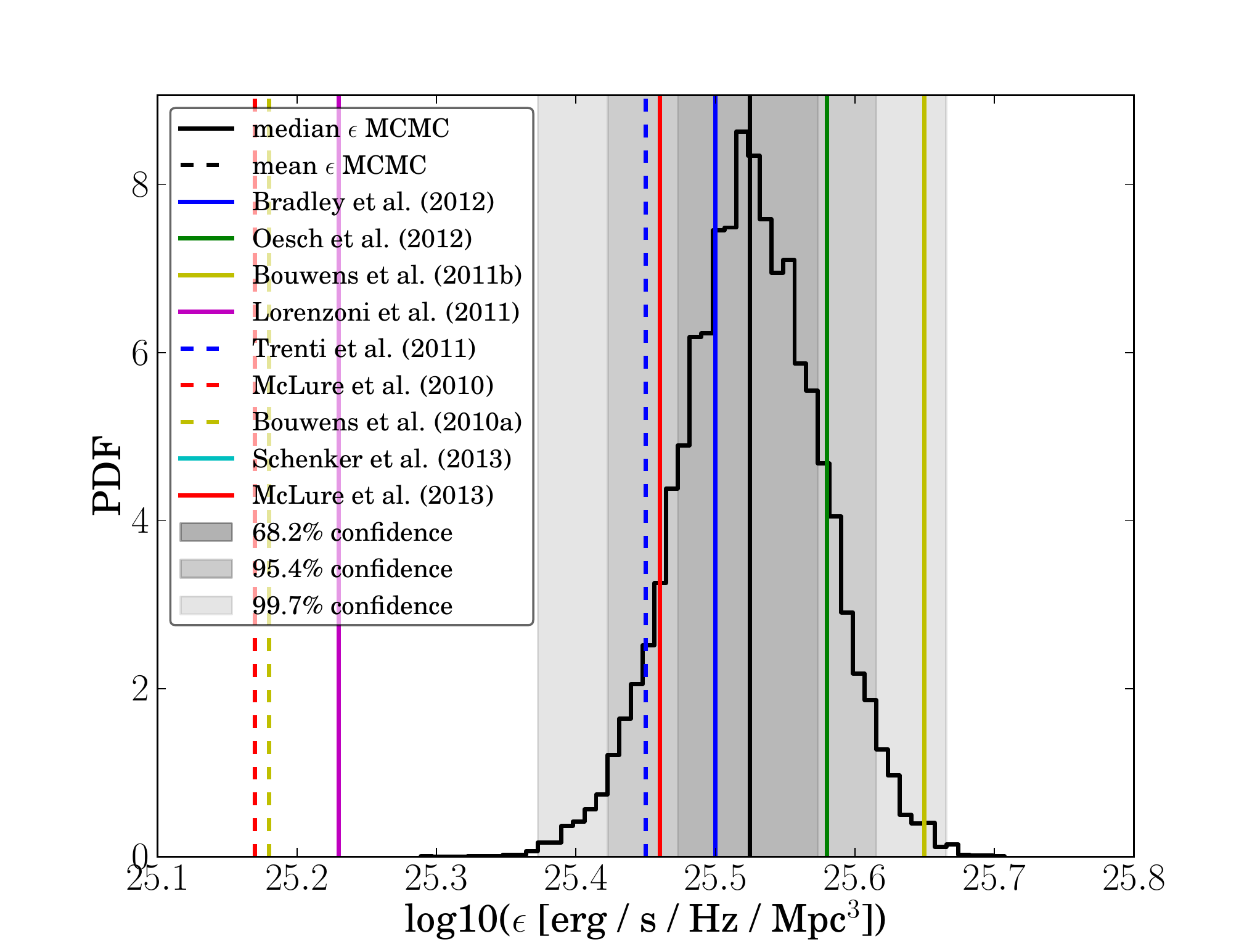}\\
\includegraphics[width=0.49\textwidth]{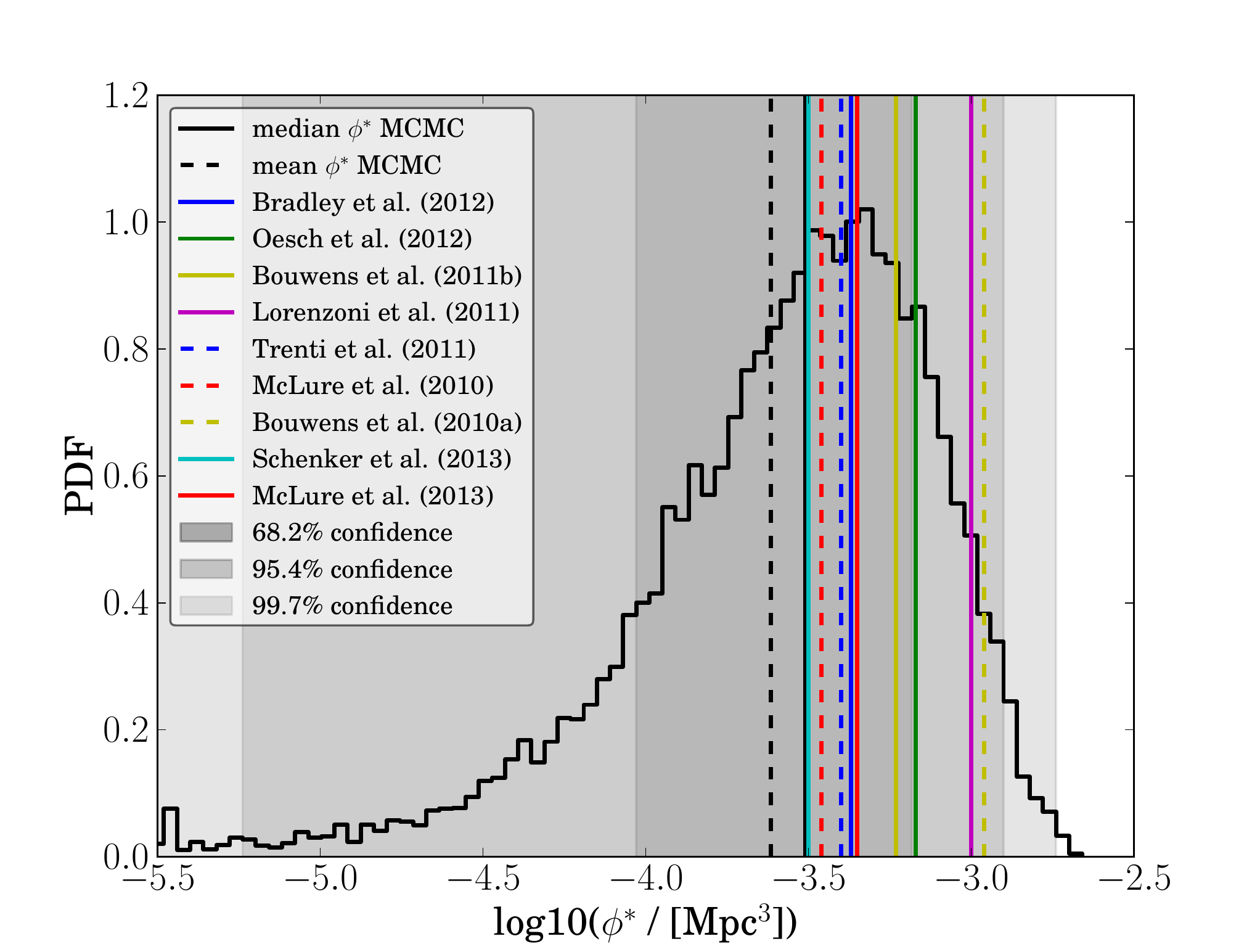}
\includegraphics[width=0.49\textwidth]{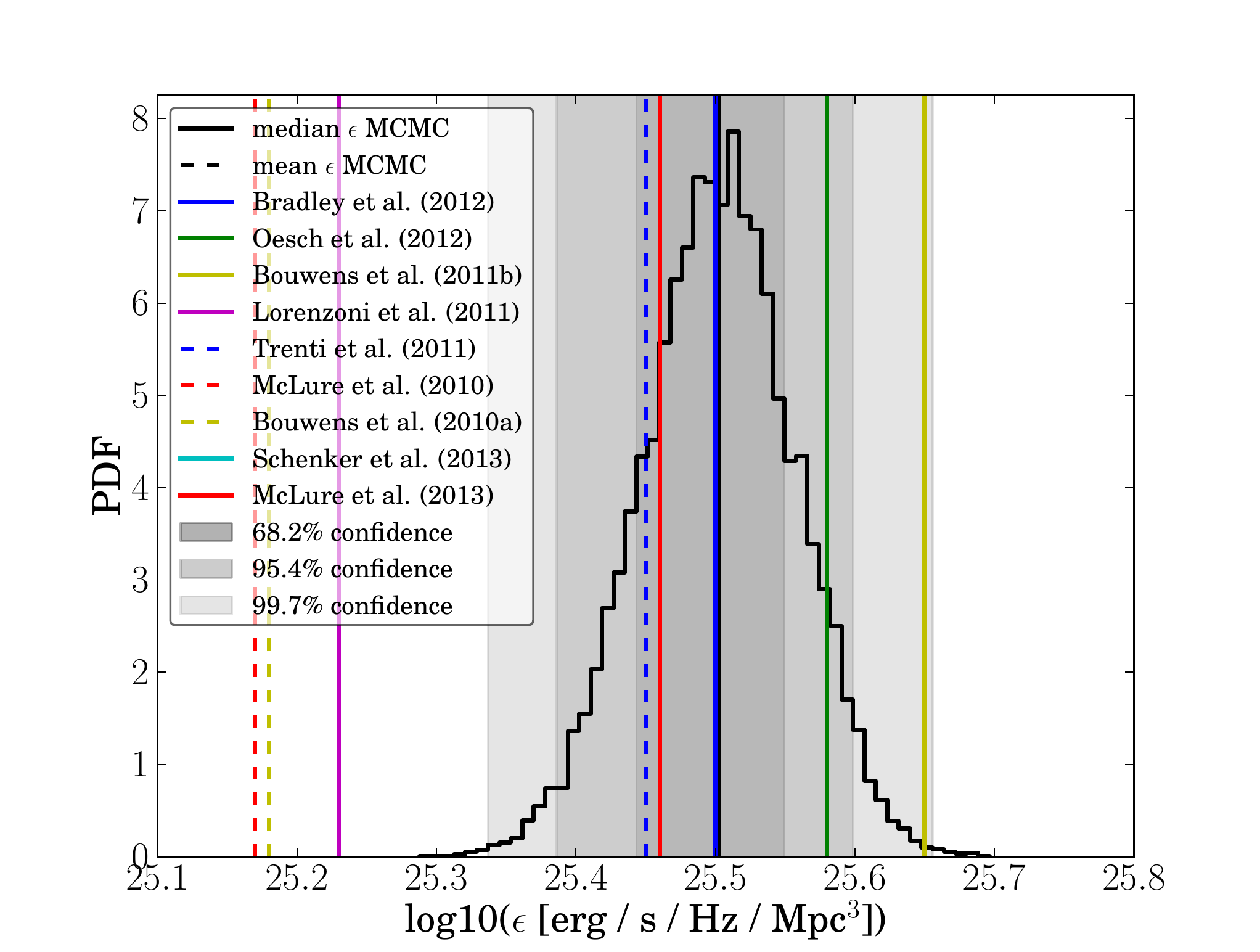}
\caption{The probability distribution functions (PDF) of the number density $\phi^\star$ (left column) and luminosity density $\epsilon$ (right column) of $z\sim8$ LBGs estimated as described in Section~\ref{sec:contamination} (and Appendix~\ref{sec:BF}) from the BoRG13 5$\sigma$ (top) and BoRG13 8$\sigma$ (bottom) samples. The shaded regions show the confidence intervals of the distributions. The mean and median values are indicated by the dashed and solid black vertical lines. $\phi^\star$ and $\epsilon$ values from the literature (Table~\ref{tab:results}) are shown by the remaining vertical lines according to the legends.
Again the agreement between the results obtained here and in the literature is prominent.}
\label{fig:phi}
\end{figure*}

From the MCMC samples of $\theta$ we estimate the number density of high redshift objects, $\phi^\star$, and the luminosity density, $\epsilon$, using the expressions given in Section~\ref{sec:contamination} and Appendix~\ref{sec:BF}. 
The distributions of the obtained $\phi^\star$ and $\epsilon$ values for the BoRG13 5$\sigma$ and 8$\sigma$ samples are shown in Figure~\ref{fig:phi}.
The median estimates and their uncertainties are quoted in Table~\ref{tab:results}.
The $\epsilon$ values were obtained by integrating down to the HUDF magnitude limit of $M=-17.7$ \citep{Bouwens:2011p8082}. Integrating to fainter magnitudes, i.e., extrapolating outside the data range, increases the luminosity density itself as well as the uncertainties.

\begin{figure*}
\includegraphics[width=0.99\textwidth]{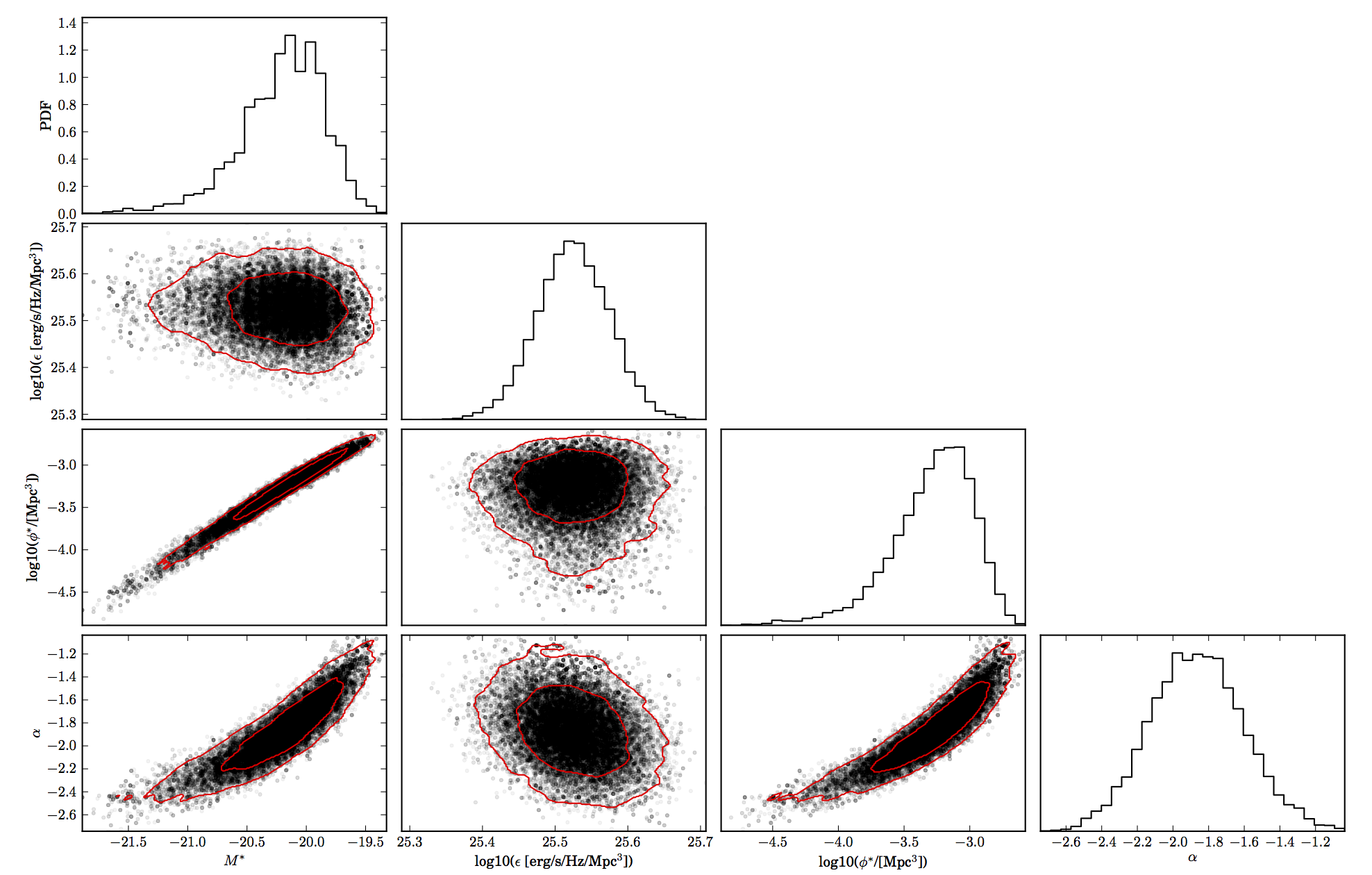}\\
\caption{The correlations between the faint-end slope, $\alpha$, the characteristic magnitude, $M^\star$, the number density of redshift 8 LBGs, $\phi^\star$, and the luminosity density, $\epsilon$, integrated down to the HUDF limit of $M=-17.7$ \citep{Bouwens:2011p8082} estimated from the BoRG13 5$\sigma$ sample. The marginalized one-dimensional probability distribution function (PDFs) of each parameter is shown in the upper right panels. 
The red contours show the $1\sigma$ and $2\sigma$ confidence contours.
Prominent correlations are seen between the three luminosity function parameters $\alpha$, $M^\star$, and $\phi^\star$, whereas no correlation is seen with $\epsilon$ as it is integrated down to $M=-17.7$, i.e., over the actual range of the data. Integrating down to fainter magnitudes will introduce a correlation and increase the mean luminosity density value as well as the uncertainties on the estimate. The lower left panel reproduces the top panel in Figure~\ref{fig:mcmc} and the PDFs of $\phi^\star$ and $\epsilon$ are shown in the top row of Figure~\ref{fig:phi}.}
\label{fig:LFcorr}
\end{figure*}

In Figure~\ref{fig:LFcorr} we present the full multi-dimensional parameter space $\alpha$, $M^\star$, $\phi^\star$, and $\epsilon$ for the BoRG13 $5\sigma$ MCMC inference (the lower left plot reproduces the top panel in Figure~\ref{fig:mcmc} and the $\phi^\star$ and $\epsilon$ probability distribution functions are also shown in the top row of Figure~\ref{fig:phi}). 
This illustrates the strong correlations between the three luminosity function parameters. 
As can be seen there is no significant correlation between the luminosity density, $\epsilon$, and the luminosity function parameters when integrating down to $M=-17.7$, i.e., it is well-determined. Extrapolating $\epsilon$ (or rather the luminosity function) to fainter magnitudes introduces strong correlations.

As noted in Section~\ref{sec:contamination} we have used a fiducial contamination fraction of 42\% for the BoRG sample following \cite{Bradley:2012p23263}. In the presented Bayesian framework the assumed contamination affects both the shape and the normalization of the luminosity function. To quantify this effect we estimated the luminosity function for the BoRG13 5$\sigma$ sample using a contamination fraction of $f_\textrm{BoRG} = 0.0, 0.2, 0.33, 0.5, 0.55,$ and $0.6$ \citep[the same values used in Table~8 of][]{Bradley:2012p23263}. The resulting luminosity functions arameters are summarized in Table~\ref{tab:LFcontam}. The BoRG13 5$\sigma$ luminosity function parameters are presented in Table~\ref{tab:results}.
The effect on the characteristic magnitude and the number density of high-redshift objects is smaller than the estimated 1$\sigma$ uncertainty with only 0.09~dex and 0.15~dex difference between the two extrema. The faint-end slope  varies by 0.27 from no contamination to a contamination of 60\% which is comparable to the average 1$\sigma$ uncertainty on the slope.  
The variations in the luminosity function parameters found here are similar to the range in parameters found by \cite{Bradley:2012p23263}. 
Hence, our more detailed calculations confirm
that validity of previous approaches.

\begin{table*}
\centering{
\caption[ ]{Effect of Contamination Fraction on Luminosity Function Estimate}
\label{tab:LFcontam}
\begin{tabular}[c]{lllll}
\hline
\hline
$f_\textrm{BoRG}$ &  $M^\star$ & $\alpha$ & $\log_{10} \phi^\star$ [Mpc$^{-3}$] & $\log_{10}\epsilon$ [erg/s/Hz/Mpc$^{3}$]\\
\hline
0.00 & $-20.23^{+0.30}_{-0.39}$	& $-1.73^{+0.27}_{-0.26}$	& $-3.17^{+0.23}_{-0.32}$	& $25.57^{+0.05}_{-0.05}$ \\
0.20 & $-20.18^{+0.30}_{-0.37}$	& $-1.76^{+0.26}_{-0.26}$	& $-3.19^{+0.24}_{-0.32}$	& $25.55^{+0.05}_{-0.05}$ \\
0.33 & $-20.17^{+0.29}_{-0.35}$	& $-1.82^{+0.26}_{-0.25}$	& $-3.22^{+0.24}_{-0.31}$	& $25.54^{+0.05}_{-0.05}$ \\
0.42$^\dagger$ & $-20.15^{+0.29}_{-0.38}$	& $-1.87^{+0.26}_{-0.26}$	& $-3.24^{+0.25}_{-0.34}$	& $25.52^{+0.05}_{-0.05}$ \\
0.50 & $-20.15^{+0.29}_{-0.35}$	& $-1.92^{+0.27}_{-0.25}$	& $-3.27^{+0.26}_{-0.32}$	& $25.51^{+0.05}_{-0.05}$ \\
0.55 & $-20.15^{+0.29}_{-0.38}$	& $-1.96^{+0.26}_{-0.25}$	& $-3.30^{+0.26}_{-0.35}$	& $25.51^{+0.05}_{-0.05}$ \\
0.60 & $-20.14^{+0.29}_{-0.38}$	& $-2.00^{+0.27}_{-0.25}$	& $-3.32^{+0.27}_{-0.36}$	& $25.49^{+0.05}_{-0.05}$ \\
\hline
\multicolumn{5}{l}{\textsc{Note.} -- $^\dagger$Fiducial contamination used in this study as presented in Table~\ref{tab:results}.}
\end{tabular}}
\end{table*}

To summarize, in general the luminosity function parameters estimated using the BoRG13 sample are in good agreement with previous results within the 1$\sigma$ uncertainties (see Table~\ref{tab:results}). 

\subsection{Inferences About Reionization}

Having derived rigorous posterior distribution functions for the
parameters describing the luminosity function at $z\sim8$ we are in a position to
infer the implications of our measurement for reionization.  Before
proceeding with this inference we briefly summarize the basic equations
describing the problem.
We refer to \cite{Shull:2012p33057} and \cite{Robertson:2013p27340} and references therein for more details. 
The fraction of ionized hydrogen $\QHII$ depends on the balance between
the density of ionizing photons $\nion$ and the recombination time $\trec$ as
\begin{equation}\label{eq:balance}
{\dot \QHII}=\frac{\dot \nion}{\nh}-\frac{\QHII}{\trec} \; .
\end{equation}

The density of ionizing photons can be related to the
measured UV luminosity density $\epsilon$ given a conversion factor
$\xi$ based on models for the spectral energy distribution of the
sources and the ionizing photons escape fraction $f_\textrm{esc}$:
\be\label{eq:nion}
\nion=\xi \epsilon f_\textrm{esc} \;.
\ee

At $z=8$, for case-B\footnote{Case-B recombination is for an opaque IGM as opposed to
case-A recombinations which relates to an IGM transparent to Ly$\alpha$ 
radiation as described by \cite{Baker:1938p32983}.}
recombination of hydrogen at 20,000K and taking
the baryon physical density from WMAP9 (Planck differs by only
$\sim1\%$), the condition for ionization equilibrium ${\dot \QHII}=0$
can be expressed as:
\be
\log_{10}\left(\frac{\QHII C}{f_\textrm{esc}}\right)=\log_{10} \epsilon + \log_{10} \xi - 50.31
\label{eq:equilibrium}
\ee
where $C$ is the so-called clumping factor ($\langle n_{\rm
H}^2\rangle/ \nh^2$), $\epsilon$ is the luminosity density in units of erg s$^{-1}$
Hz$^{-1}$ Mpc$^{-3}$, and $\xi$ is given in Hz/erg.  Typically,
some fiducial value is assumed for the unknowns in order to
determine whether the observed galaxies are sufficient to keep the Universe ionized
and, in case they are not, to explore if extrapolation of the observed luminosity function to fainter magnitudes is sufficient
to increase the ionized fraction of hydrogen to that level.

We carry out our inference by assigning prior
probabilities to the unknowns based on theoretical considerations.
$\xi$ depends on metallicity, the initial mass function age and the dust content of the stellar
populations. Based on the measured UV slopes by
\cite{Dunlop:2012p26717} and a range of models,
\cite{Robertson:2013p27340} estimate it to be in the range $\log_{10}
\xi$[Hz/erg] = 24.75--25.35 and take 25.2 as their fiducial value. We
describe the range of theoretically accepted values as a lognormal
distribution with mean 25.2 and standard deviation 0.15 dex.

\begin{figure}
\centering
\includegraphics[width=0.49\textwidth]{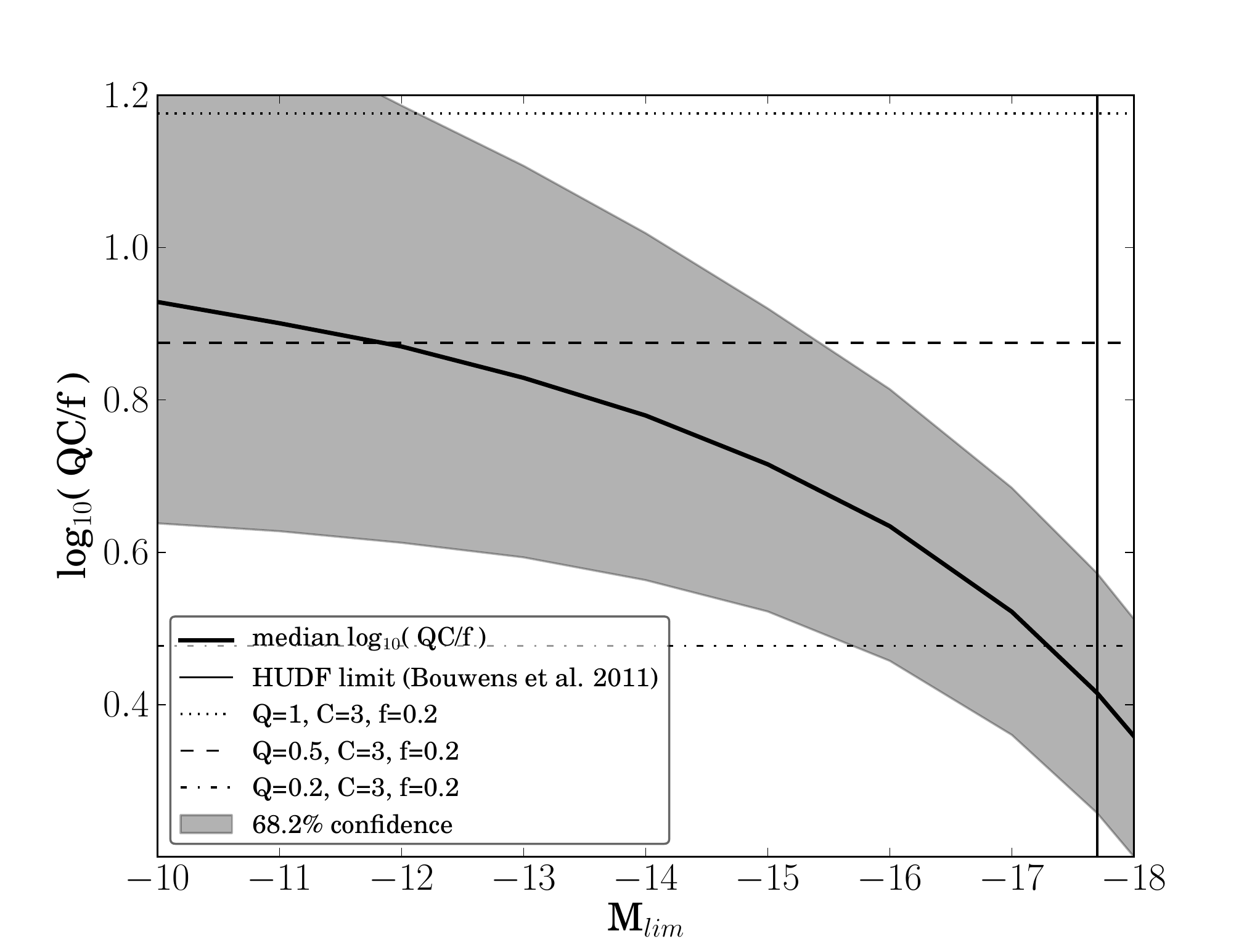}
\caption{The inferred value of $QC/f_\textrm{esc}$ as a function of the limiting magnitude $M_{\rm lim}$ which the BoRG13 $5\sigma$ luminosity density is integrated down to. The horizontal lines show the fraction of ionized hydrogen for fixed clumping factor, $C$, and photon escape fraction, $f_\textrm{esc}$, according to the legend. The vertical line shows the HUDF limit of -17.7 from \cite{Bouwens:2011p8082}.}
\label{fig:QCf}
\end{figure}

Figure~\ref{fig:QCf} shows the inferred value of $QC/f_\textrm{esc}$ for the BoRG13 $5\sigma$ luminosity function as a
function of the integration limit when estimating $\epsilon$, $M_{\rm lim}$. 
As the average luminosity density grows as the integration limit is pushed
to ever fainter magnitudes the overall value of  $QC/f_\textrm{esc}$ increases.
The horizontal lines indicate the ionized fraction of hydrogen for fixed $C$ and $f_\textrm{esc}$.
We use $C=3$ \citep[e.g.,][]{McQuinn:2011p32987,Shull:2012p33057,Finlator:2012p32957,Kaurov:2013p32929}
and $f_\textrm{esc}=0.2$ \citep[e.g.,][]{Ouchi:2009p27671,Shull:2012p33057,Robertson:2013p27340}
as our fiducial values.
The vertical line shows the HUDF limit from \cite{Bouwens:2011p8082} corresponding to the 
distribution of $\epsilon$ shown in the top right panel of Figure~\ref{fig:phi}. 
It is clear from Figure~\ref{fig:QCf} that only when integrating all the way down to $M=-12$
$Q$ is well below 1 for fixed $C$ and $f_\textrm{esc}$. 
Hence, for $C=3$ and $f_\textrm{esc}=0.2$ it seems unlikely that the $z\sim8$ population
of Y-band dropouts are capable of keeping the Universe fully ionized.

\begin{figure}
\centering
\includegraphics[width=0.49\textwidth]{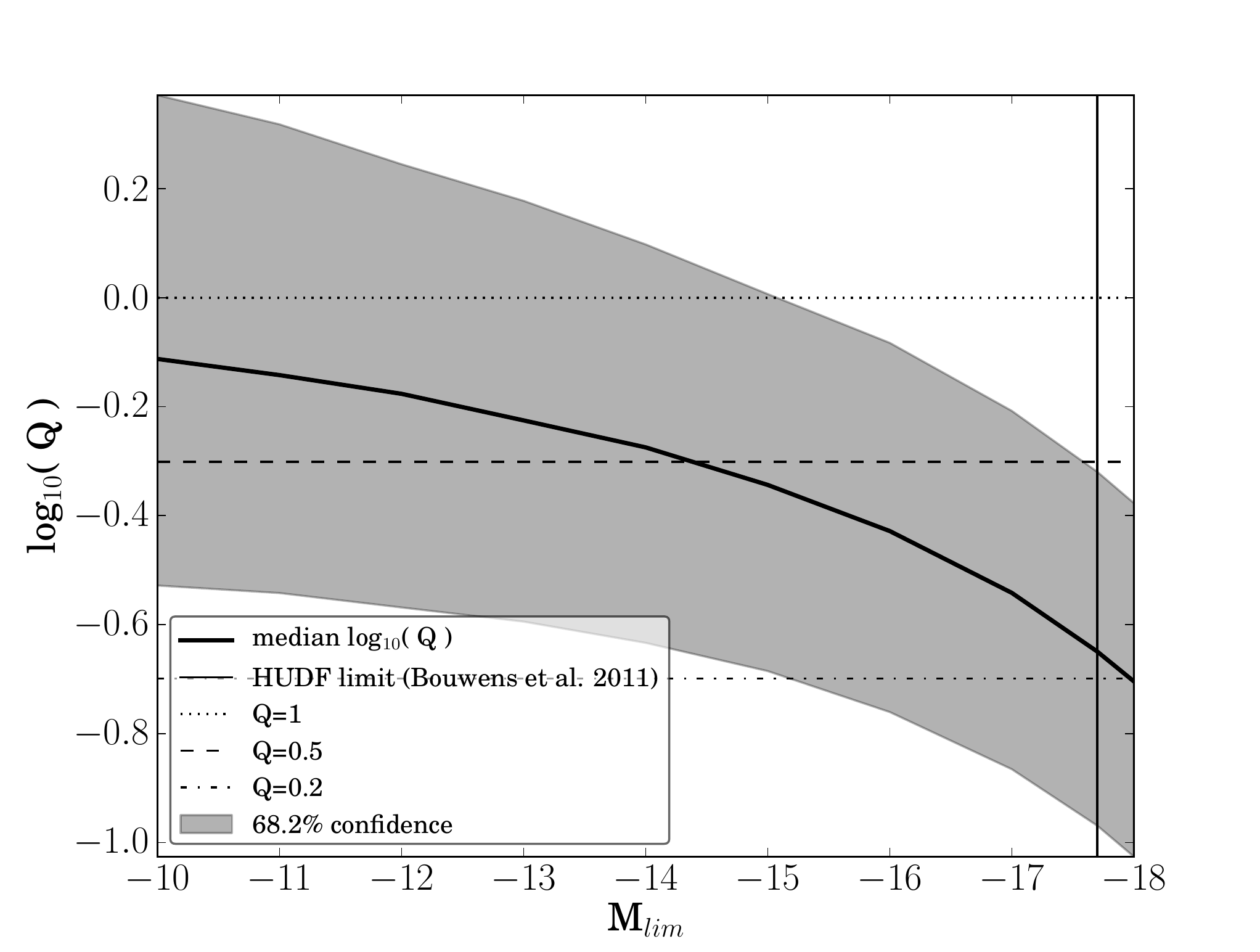}
\caption{The inferred value of $Q$ as a function of the limiting magnitude $M_{\rm lim}$ assuming a flat prior on the clumping factor $C\in[1-6]$ and the photon escape fraction $f_\textrm{esc}\in[0.1-0.5]$. The horizontal lines show an ionization fraction of 20\%, 50\% and 100\%. The vertical line shows the HUDF limit of -17.7 from \cite{Bouwens:2011p8082}. The prior on $C$ and $f_\textrm{esc}$ allows for a fully ionized Universe, but only if objects all the way down to $M\sim-15$ are included. There is very little change to keep the Universe fully ionized if only galaxies down to the HUDF limit are considered.}
\label{fig:Qlim}	
\end{figure}

We can make further progress by adopting a prior on $C/f_\textrm{esc}$. Recent
theoretical calculations suggest that $C$ is of order 1-6, and we therefore
assume a uniform prior within this range. Very little empirical or
theoretical information is available on the escape fraction. We
formalize the uncertainty on this parameter by assuming a uniform
prior in the range $f_\textrm{esc}\in[0.1-0.5]$ \citep[e.g.,][]{Fernandez:2011p33059}. 
With these priors we derive the
joint constraints on $Q$ and M$_{\rm lim}$ shown in
Figure~\ref{fig:Qlim}.
Figure~\ref{fig:Qlim} shows that allowing for a larger range in $C$ and $f_\textrm{esc}$
increases the available ranges of $Q$ for the BoRG13 luminosity function. However,
even here only for the most extreme combination of $C$ and $f_\textrm{esc}$ at $M\sim-15$
there seems to be enough radiation available to sustain a fully ionized Universe at $z\sim8$.
Overall it is clear from Figures~\ref{fig:QCf} and \ref{fig:Qlim} that reionization is far from
being completed at $z\sim8$ by an LBG population following the BoRG13 5$\sigma$ luminosity function.
Invoking a luminosity function with a steeper faint-end slope (e.g., the BoRG13 8$\sigma$ luminosity function)
changes this picture somewhat. However, $Q\sim1$ still requires radiation from objects down to $M\sim-16$,
as supported by non-detection of gamma-ray burst host galaxies at $z>5$
\citep{Trenti:2012p32219,Tanvir:2012p33137} and therefore still supports a late
reionization, i.e., that reionization completes at $z<8$.

\begin{figure}
\centering
\includegraphics[width=0.49\textwidth]{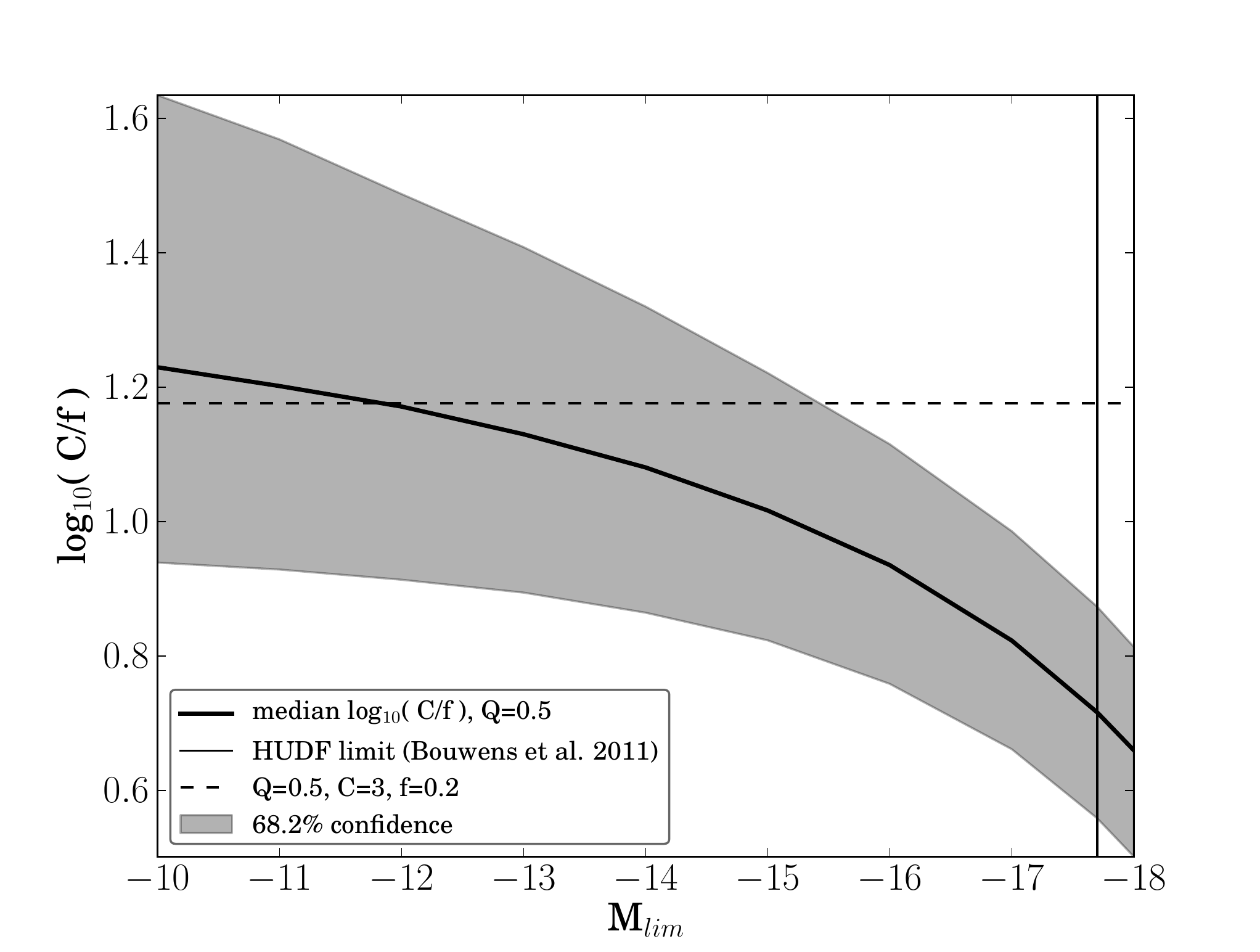}
\caption{The inferred value of $C/f_\textrm{esc}$ as a function of the limiting magnitude $M_{\rm lim}$ assuming $Q=0.5$ for the BoRG13 $5\sigma$ luminosity function. The horizontal dashed line shows where $C=3$ and $f_\textrm{esc}=0.2$. As in Figures~\ref{fig:QCf} and \ref{fig:Qlim} the vertical line marks the HUDF limit.}
\label{fig:Cflim}
\end{figure}

We can compare our inference on $Q$ with the values
inferred from the measured \lya\ optical depth as inferred from
spectroscopic follow-up of Y-band dropouts \citep[e.g.,][]{Treu:2013p32132}. 
Naturally this comparison is extremely delicate and needs to rely on a number of
assumptions, as the \lya\ optical depth depends not only on the
average fraction of neutral hydrogen in the IGM but also on the
details of the \lya\ emission itself and on the detailed radiative
transfer at or in the vicinity of the source
 \citep[e.g.,][]{Dijkstra:2011p33040,Bolton:2013p33051}. 
Several authors have suggested that ionized fractions of $\sim0.5$ are needed to explain the
observed decline in \lya\ emission from LBGs between $z\sim6$ and
$z\sim7$, if all the optical depth arises from neutral hydrogen in the
IGM and if the universe is completely reionized by $z\sim6$. 
Our recent follow-up of a subsample of the BoRG LBGs with MOSFIRE \citep{Treu:2013p32132} indicates that the
optical depth increases even more out to $z\sim8$, indicating perhaps
an even lower fraction of ionized hydrogen, which seems to be in good 
agreement with the results summarized in Figures~\ref{fig:QCf} and \ref{fig:Qlim}.
However, given the extreme assumptions, these numbers should be considered lower limits to the
cosmic average fraction of neutral hydrogen. In order to gain some
insight into $C/f_\textrm{esc}$ we can adopt $Q=0.5$ as our upper limit and
infer our exclusion plots in $C/f_\textrm{esc}$ versus M$_{\rm lim}$ shown in
Figure~\ref{fig:Cflim}.

In summary the inferred luminosity function from the BoRG13 $5\sigma$ data is not capable
of fully ionizing the Universe at redshift $z\sim8$ as a significant fraction of neutral
hydrogen seems to still be present. This is in good agreement with recent results
from spectroscopic follow-up campaigns which all seem to suggest that a significant 
fraction of neutral hydrogen is present at $z>7$. Hence, the results presented here
support a late reionization scheme where the reionization is still ongoing at $z\sim8$.

\section{Summary and conclusion}\label{sec:conc}

The BoRG survey has carried out the largest-area search to date for
Y-band dropouts (HST F098M-dropouts). We present new observations from 12
parallel fields not included in our previous studies and additional deeper
datasets for two fields. We combine our BoRG sample with a sample of
fainter dropouts taken from the literature and use the combined
sample of $97$ dropouts to carry out a rigorous study of the luminosity function at
$z\sim8$ and its implications for reionization. Our main results can
be summarized as follows:
\begin{enumerate}
\item We present 9 new Y-band dropouts from the $\sim$50 arcmin$^2$ of new data.
Furthermore, we re-confirm two dropouts previously presented by 
\citet{Trenti:2011p12656} and \citet{Bradley:2012p23263}.
Combining these 11 dropouts with out previously published $z\sim8$ LBGs from
BoRG gives a sample of 38 bright 
($25.5\leqslant m_{J} \leqslant 27.6$)
redshift 8 LBGs from the BoRG survey.

\item We develop and implement an improved method for estimating the luminosity
function parameters for a sample of $n$ galaxies. Using a Bayesian
framework the posterior distribution of the population
based on a binomial distribution is given. 
Combining the BoRG Y-band dropouts with the faint $z\sim8$ LBGs from HUDF/ERS presented
by \cite{Bouwens:2011p8082} and sampling over this posterior distribution enables
a robust recovery of the faint-end slope, $\alpha$, the
characteristic magnitude of the Schechter function, $M^\star$, the normalizing number
density of $z\sim8$ LBGs, $\phi^\star$, and the luminosity density, $\epsilon$, of the
redshift 8 luminosity function.

\item The inferred luminosity function at $z\sim8$ is described by the parameters: 
   \begin{itemize}   
   \item[] $M^\star= -20.15^{+0.29}_{-0.38}$, 
   \item[] $\alpha= -1.87^{+0.26}_{-0.26}$, 
   \item[] $\log_{10} \phi^\star [\textrm{Mpc}^{-3}]=  -3.24^{+0.25}_{-0.34}$,
   \item[] $\log_{10} \epsilon \left[\frac{\textrm{erg}}{\textrm{s Hz Mpc}^{-3}}\right]= 25.52^{+0.05}_{-0.05}$
   \end{itemize}
Here $\epsilon$ is the inferred UV luminosity density integrated down to the HUDF limit M$_{\rm lim}=-17.7$. 
Our inferred credible intervals include recent estimates of the same parameters.

\item We show that for the BoRG13 $z\sim8$ luminosity function 
the average fraction of ionized hydrogen $Q$ is only of the order 10-50\% for samples
down to the HUDF limit of $M=-17.7$ assuming standard values of the clumping factor
and the photon escape fraction. 
To sustain a fully ionized Universe at redshift 8 with the presented luminosity function
it is necessary to account for the radiation of objects as faint as $M=-15$.

\item The inferred ionization fractions suggest a relatively late reionization scenario where a considerable 
fraction of neutral hydrogen is still present at $z\sim8$ in good agreement with the results
of our recent spectroscopic MOSFIRE campaign (and others from the literature) where we
followed up a subsample of the BoRG redshift 8 LBGs presented here.
\end{enumerate}

The inference on the implications of the BoRG13 luminosity function for reionization presented here are still limited by the sizable error bars at redshift 8. To reduce the uncertainties on the inferred quantities bright and faint galaxy samples from e.g., the Frontier Fields and future parallel HST campaigns are crucial.

\acknowledgments

This work was supported by the HST BoRG grants GO-11700, 12572, and
12905.  This paper is based on observations made with the NASA/ESA
Hubble Space Telescope, obtained at the Space Telescope Science
Institute. 
BK acknowledges support from the Southern California Center for Galaxy Evolution, a multi-campus research program funded by the University of California Office of Research.
Support for this work was provided by the European Commission through the Marie Curie Career Integration Fellowship PCIG12-GA-2012-333749 (MT).
\begin{appendix}

\section{HST-GO Follow-Up of BoRG\_1437+5043}\label{app:borg58}

The BoRG\_1437+5043 field is a special case, and thus its analysis
deserves a detailed description.  The field was originally identified by
\citet{Trenti:2012p13020} as an overdensity of 4 Y-band dropouts detected 
above the $5\sigma$ limit and 1 detected above $8\sigma$.  As explained in that work, overdensitites
of high redshift galaxies are expected to occur from theoretical and
numerical dark matter modeling.  It was thus argued that the
overdensity could potentially be a high redshift
protocluster. \citet{Bradley:2012p23263} subsequently carried out a
re-analysis of the field. With improved data reduction and photometry
algorithms, 2 out of the 4 initial 5$\sigma$ candidates presented by
\citet{Trenti:2012p13020} fell below the formal 5$\sigma$
threshold. The 8$\sigma$ candidate was confirmed.

\begin{figure}
\centering
\includegraphics[width=0.6\textwidth]{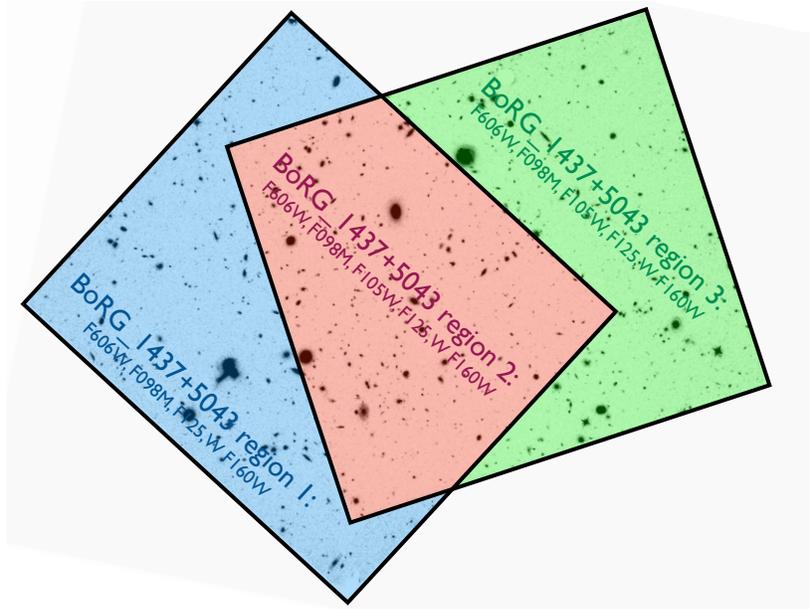}
\caption{The combined J-band imaging of the original and follow-up observations of BoRG\_1437+5043. The coloring shows the 3 regions used when analyzing the photometry of the drizzled image and mark regions with comparable depth and band coverage. Region 1 and 2 outline the observations presented and analyzed in \cite{Trenti:2012p13020} and \cite{Bradley:2012p23263}. The data obtained in November 2012 is outlined by region 2 and 3 which was centered on the potential protocluster presented in \cite{Trenti:2012p13020}.}
\label{fig:obs}
\end{figure}

In order to further investigate the nature of the sources in this
field, follow-up observations were proposed for and obtained in
November 2012 . The main goal of the new HST campaign (GO 12905, PI:
Trenti) was to look for fainter members of the overdensity as well as
expand the area surveyed by centering the field directly on the
overdensity. The GO data was taken in the 4 original photometric bands
of the BoRG survey, F606W, F098M, F125W, and F160W, plus F105W in
order to sharpen the redshift estimate of the dropout candidates.  In
Figure~\ref{fig:obs} we show the combined J-band image of
BoRG\_1437+5043.  The three differently colored regions mark regions
with similar depth and band coverage. The filters used in each region
are listed.  Regions 1 and 2 outline the original area of the
BoRG\_1437+5043 field analyzed by
\citet{Trenti:2012p13020}. Regions 2 and 3 show the position of the new
data. Hence, region 2 centered on the potential protocluster contains
the deepest data.

The new GO observations were reduced and analyzed together with the
rest of the BoRG13 data as presented in this paper. Each of the
three regions were analyzed independently in order to accommodate the
different depth and band coverage. 
The limiting magnitudes of each of the three regions are listed in Table~\ref{tab:C19fields}
together with the rest of the BoRG13 fields.
Photometry at the location of the
five sources identified by \cite{Trenti:2012p13020} is given in
Table~\ref{tab:objB1437}. We note that the two dropouts that were not
confirmed in the reanalysis carried out by \citet{Bradley:2012p23263}
are indeed not high-z candidates according to the deeper data, thus
confirming the improvements introduced in the BoRG pipeline. 

\begin{figure*}
\includegraphics[width=0.76\textwidth]{fig/BoRG_1437+5043_r2_637_mosaic.pdf}
\includegraphics[width=0.22\textwidth]{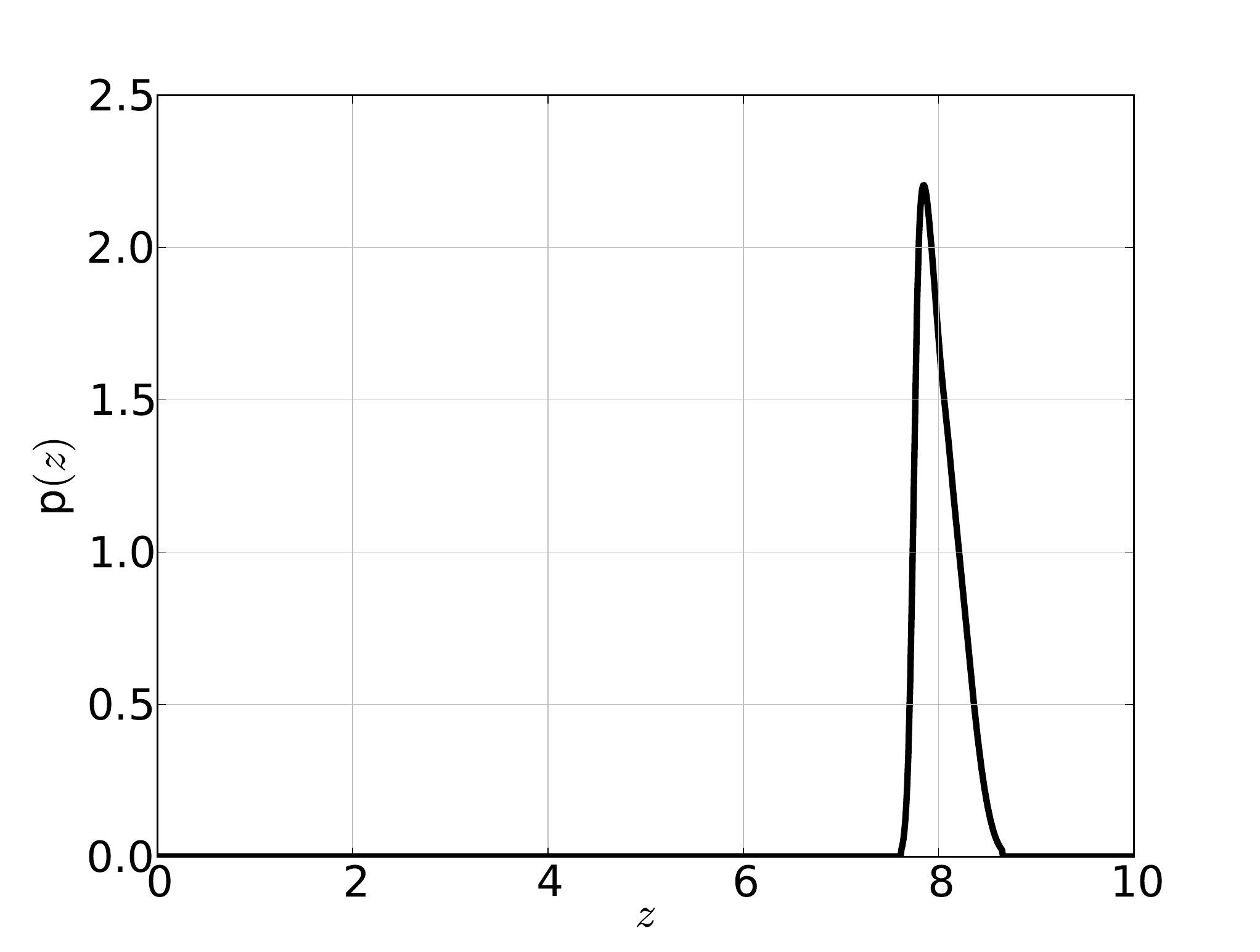}\\
\includegraphics[width=0.76\textwidth]{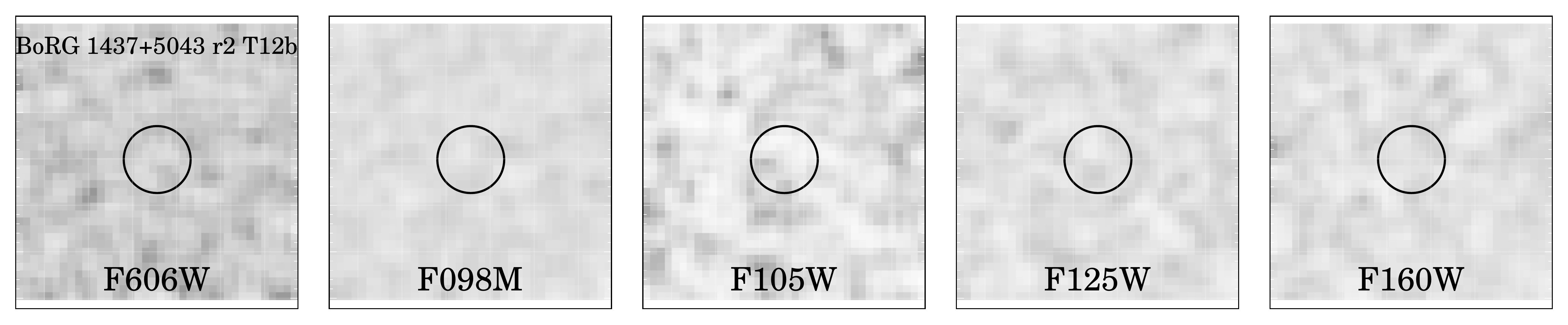}\\
\includegraphics[width=0.76\textwidth]{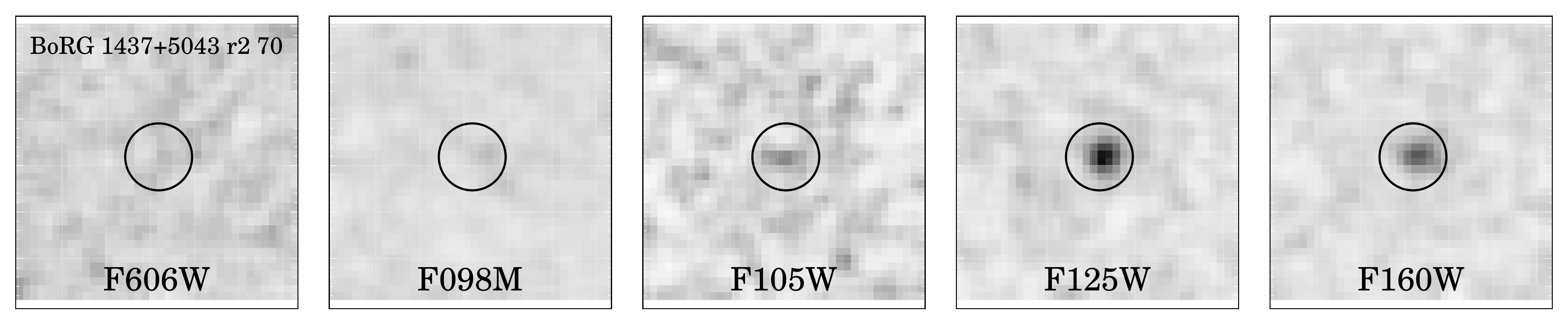}
\includegraphics[width=0.22\textwidth]{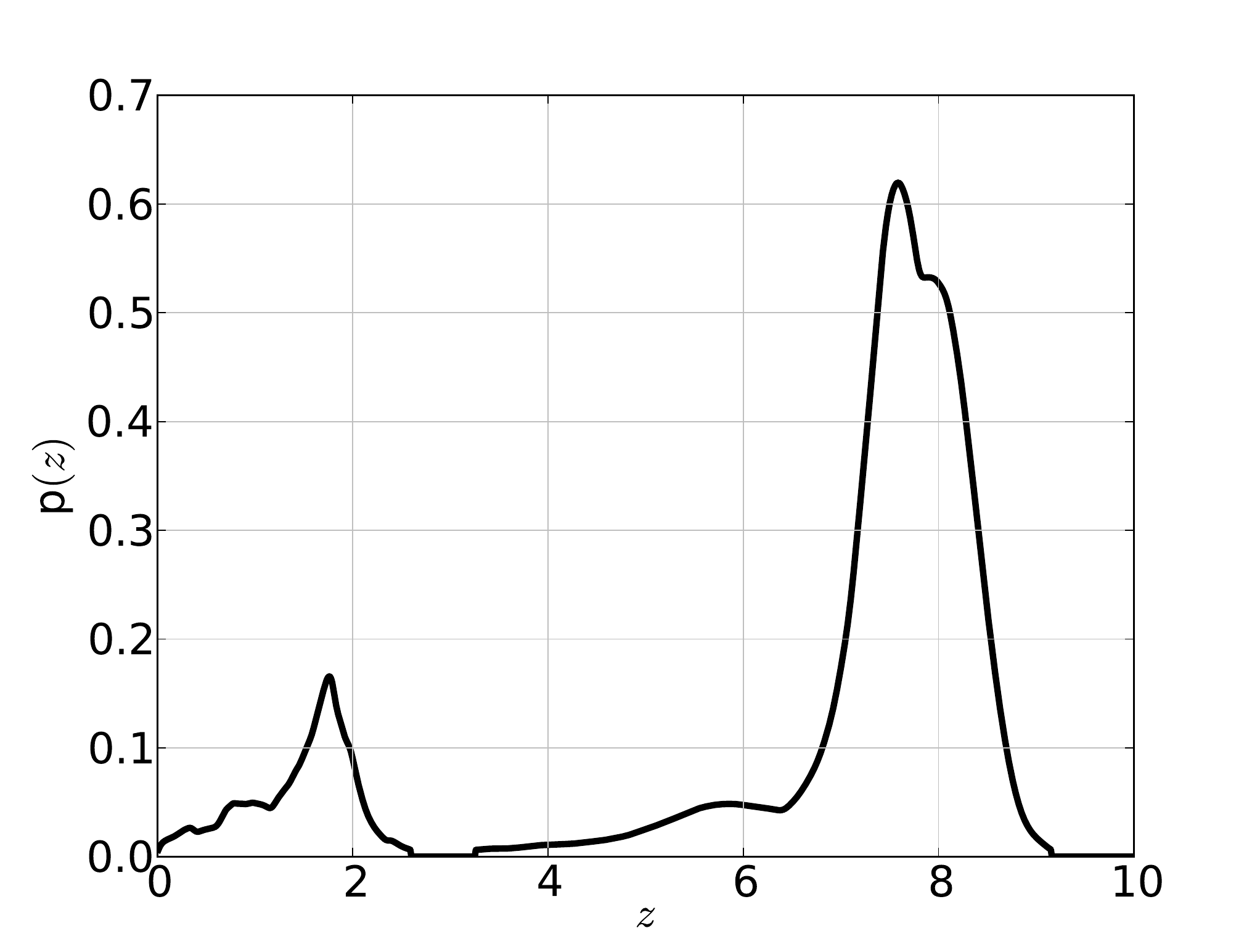}\\
\caption{Follow-up GO observations of the three high-redshift galaxy candidates in BoRG\_1437+5043 \citep[][top to bottom shows candidate a, b and e]{Trenti:2012p13020}. The first five columns show F606W, F098M, F105W, F125W and F160W 3$''$x3$''$ postage stamps with a power-law stretch. The last column shows the redshift probability distribution $p(z)$ (using a flat prior) obtained with the Bayesian redshift code BPZ \citep{Benitez:2004p25410,Coe:2006p25409} for the two candidates detected in the BoRG13 data. The first row duplicates the bottom row of Figure~\ref{fig:fitspz} and is displayed here for completeness.}
\label{fig:fitspzB58}
\end{figure*}

Postage stamps at the location of the three candidates retained by
\citet{Bradley:2012p23263} are shown in Figure~\ref{fig:fitspzB58}
along with their photometric redshift estimates. As we can see from
the photo-$z$ estimates, 2 of the 3 candidates are confirmed to be at $z\sim8$,
even though one of them (T12e) falls just outside of the formal Y-H
color selection criterion.  T12a is now detected at S/N$>20 (16)$ in the
J(H) band and T12e is detected at S/N$>9 (7)$. They are thus both
excellent $z\sim8$ redshift candidates. The third candidate (T12b) is
undetected in the deeper data and thus we conclude it was a spurious
noise peak.  The beautifully sharp photo-$z$ of the initial 8$\sigma$
candidate T12a is a nice validation of the reliability of the
high-significance sample where $\textrm{S/N}_\textrm{J} > 8$. 
As we show in the next section the
distribution of noise and background is highly non-Gaussian, so it is
not surprising that a fraction of sources just above the 5$\sigma$
threshold are spurious. The 2/3 confirmed fraction is consistent with
the fiducial contamination fraction of 42\% that we adopt in our
inference based on numerical simulations.

Albeit, with admittedly small number statistics this follow-up campaign
validates our approach of carrying out separately the inference for
the more robust 8$\sigma$ detections and comparing the results with
those based on the 5$\sigma$ sample for which contamination is
expected to be higher.  No fainter candidates are detected in the
field. These results imply that the field is still likely to be an
overdensity of $z\sim8$ but not as pronounced as previously thought.

\begin{table*}
\centering{
\caption[ ]{BoRG13 Photometry of Y-band dropouts in BoRG\_1437+5043 from \cite{Trenti:2012p13020}}
\label{tab:objB1437}
\begin{tabular}[c]{lllcccrrrrc}
\hline
\hline
ID & $\alpha_{J2000}$ & $\delta_{J2000}$ & J & Y $-$ J & J $-$ H & S/N$_\textrm{V}$ & S/N$_\textrm{Y}$ & S/N$_\textrm{J}$ & S/N$_\textrm{H}$ & Sample \\
\hline
B1437\_r2\_0637\_T12a		& 219.21058 	& 50.72601 	&  25.76 $\pm$ 0.07 & 3.19 $\pm$ 0.78 & 0.07 $\pm$ 0.11 & $-0.2$ & 1.0 & 20.2 & 16.5 & BoRG13 \\
						& $219.210672$ & $50.7260085$ & $26.1 \pm  0.1$ & $> 2.7$        & $ 0.0 \pm  0.2$ & $-1.5$ & $-1.0$ & $10.9$ & $ 7.9$ & BoRG12\\
						& 219.2107 &+50.7260  &  $25.8\pm 0.1$ & $>2.90$ & $-0.10$ & - & - &  13.0 & 8.0 & BoRG09\\
\hline
B1437\_r2\_T12b			& 219.22405*	& 50.72597* 	& - & - & - & - & - & - & - & BoRG13\\
						& $219.2240496$ & $50.7259683$ & $27.3 \pm  0.3$ & $> 1.8$        & $-0.3 \pm  0.5$ & $ 0.3$ & $ 0.5$ & $ 5.0$ & $ 2.7$ & BoRG12\\
						& 219.2241 &+50.7260 &    $27.2\pm 0.3$& $>1.90$ & $-0.30$ & - & - &  5.1   & 2.6 & BoRG09\\
\hline
B1437\_r2\_0560\_T12c		& 219.23092	& 50.72405 	& 27.75 $\pm$ 0.23 & 0.83 $\pm$ 0.47 & 0.06 $\pm$ 0.33 & 3.5 & 2.6 & 6.1 & 5.0  & BoRG13 \\
						& $219.2310489$ & $50.7240585$ & $27.1 \pm  0.2$ & $> 1.8$        & $-0.7 \pm  0.6$ & $ 1.1$ & $-0.5$ & $ 5.0$ & $ 1.7$ & BoRG12\\
						 & 219.2311& +50.7241  & $26.9\pm 0.2$  & $>2.00$ & $-0.30$ & - & - & 5.5   & 2.7 & BoRG09\\
\hline
B1437\_r2\_T12d			& 219.22027*	& 50.71563* 	& - & - & - & - & - & - & -  & BoRG13\\
						& $219.2202746$ & $50.7156344$ & $27.4 \pm  0.3$ & $> 1.8$        & $ 0.0 \pm  0.4$ & $ 0.6$ & $ 0.6$ & $ 4.9$ & $ 3.3$ & BoRG12\\
						&219.2203 &+50.7156   &  $27.2\pm 0.2$ & $>1.80$ & $-0.30$ & - & - &  5.4   & 2.7 & BoRG09\\
\hline
B1437\_r2\_0070\_T12e		& 219.22225	& 50.70808 	& 26.91 $\pm$ 0.14 & 1.53 $\pm$ 0.49 & $-0.03$ $\pm$ 0.24 & 0.8 & 2.1 & 9.4 & 7.0  & BoRG13\\
						& $219.2223469$ & $50.7080907$ & $27.1 \pm  0.2$ & $ 2.0 \pm 0.8$ & $-0.4 \pm  0.5$ & $ 1.0$ & $ 1.0$ & $ 5.8$ & $ 2.8$ & BoRG12\\
						&219.2224 &+50.7081 &    $27.0\pm 0.2$ & $>2.10$ & $-0.40$ & - & - &  6.0   & 2.9 & BoRG09\\
\hline
\multicolumn{11}{l}{\textsc{Note.} -- Columns are the same as in Table~\ref{tab:obj} except for the last column which shows where the data were taken from. The `T12x' subscript }\\
\multicolumn{11}{l}{in the IDs refers to the candidate's designation in \cite{Trenti:2012p13020} Figure~3. *Coordinates taken from \cite{Bradley:2012p23263} as objects}\\
\multicolumn{11}{l}{are not in the BoRG13 SExtractor segmentation maps.}
\end{tabular}}
\end{table*}

\section{Non-gaussian noise distribution and the importance of multi-band selection}\label{app:noise}

When quoting `3$\sigma$' or `5$\sigma$' detections in dropout surveys,
as well as other areas of astrophysics, it may be implicitly assumed
that the noise of the science images is Gaussian distributed such
that a 3$\sigma$ and 5$\sigma$ detection corresponds to a probability
of the detection being real of 99.730020\% and 99.999943\%. For a
million independent apertures, like those in the BoRG survey, this
would correspond to 2700 and 0.57 objects, respectively, and thus
effectively no contaminants in a 5$\sigma$ sample, even for an area
as large as the one surveyed by BoRG.

In practice, however, we do not expect the noise distribution to be
Gaussian. Furthermore, we do not expect it to be symmetric.  This is
in part due to unresolved background objects but also due to the positive
nature of defects such as faint cosmic ray hits and hot pixels.  Thus,
when the data are pushed to the limit as in dropout searches for Lyman
break galaxies, the non-Gaussian tails of the noise distributions can
result in false positive detections \citep[e.g.,][]{Dunlop:2013p23759}
even if they are formally 5$\sigma$ detections.

In this appendix we take advantage of the very large area covered by
the BoRG Survey to characterize and demonstrate the non-Gaussianity of
the background. We show that the level of contaminants is much higher
than for a Gaussian background, and that only through the use of
multiple filters for detections and non-detection requirements it can
be kept at the relatively manageable level of 42\%.

\begin{figure}
\includegraphics[width=0.49\textwidth]{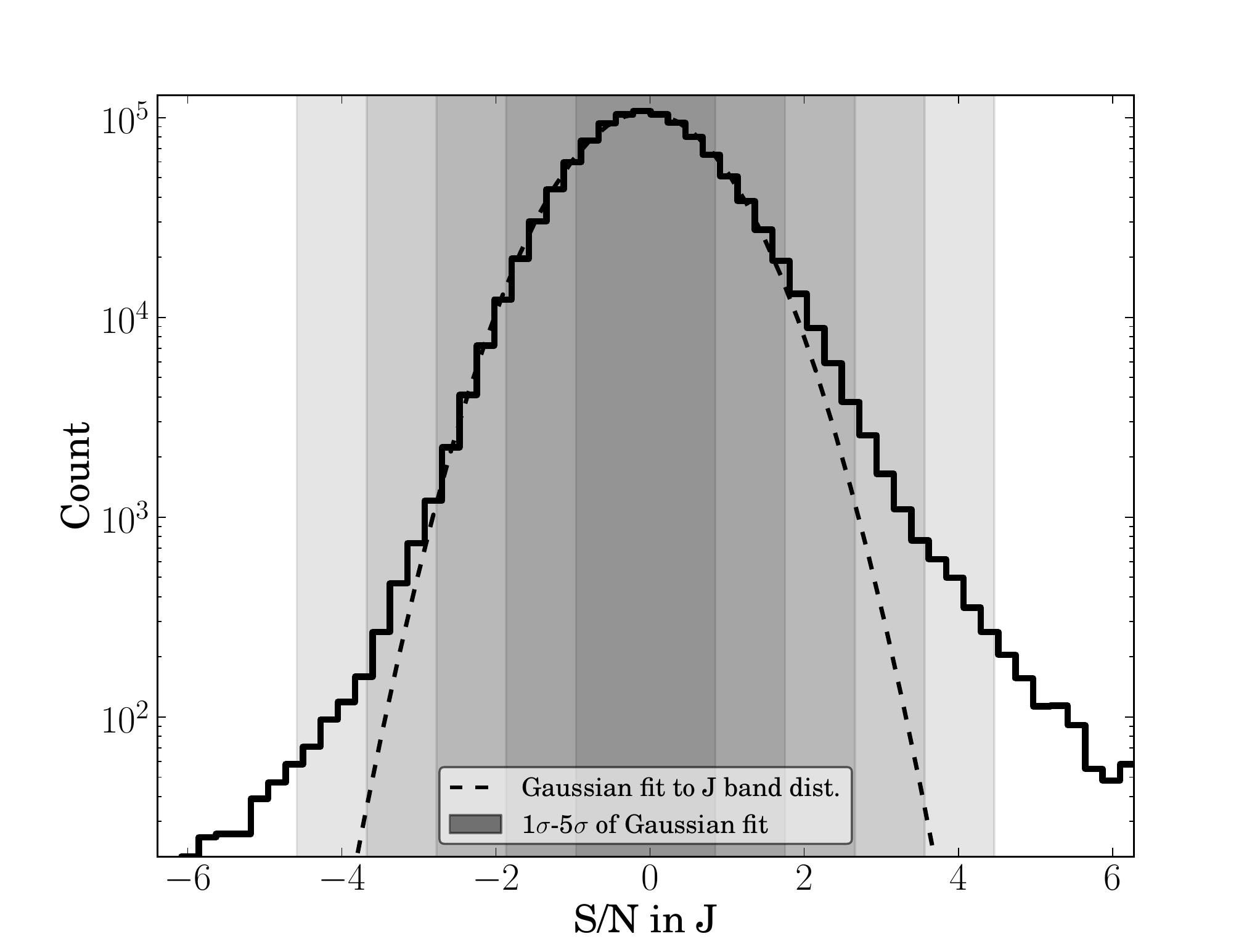}
\includegraphics[width=0.49\textwidth]{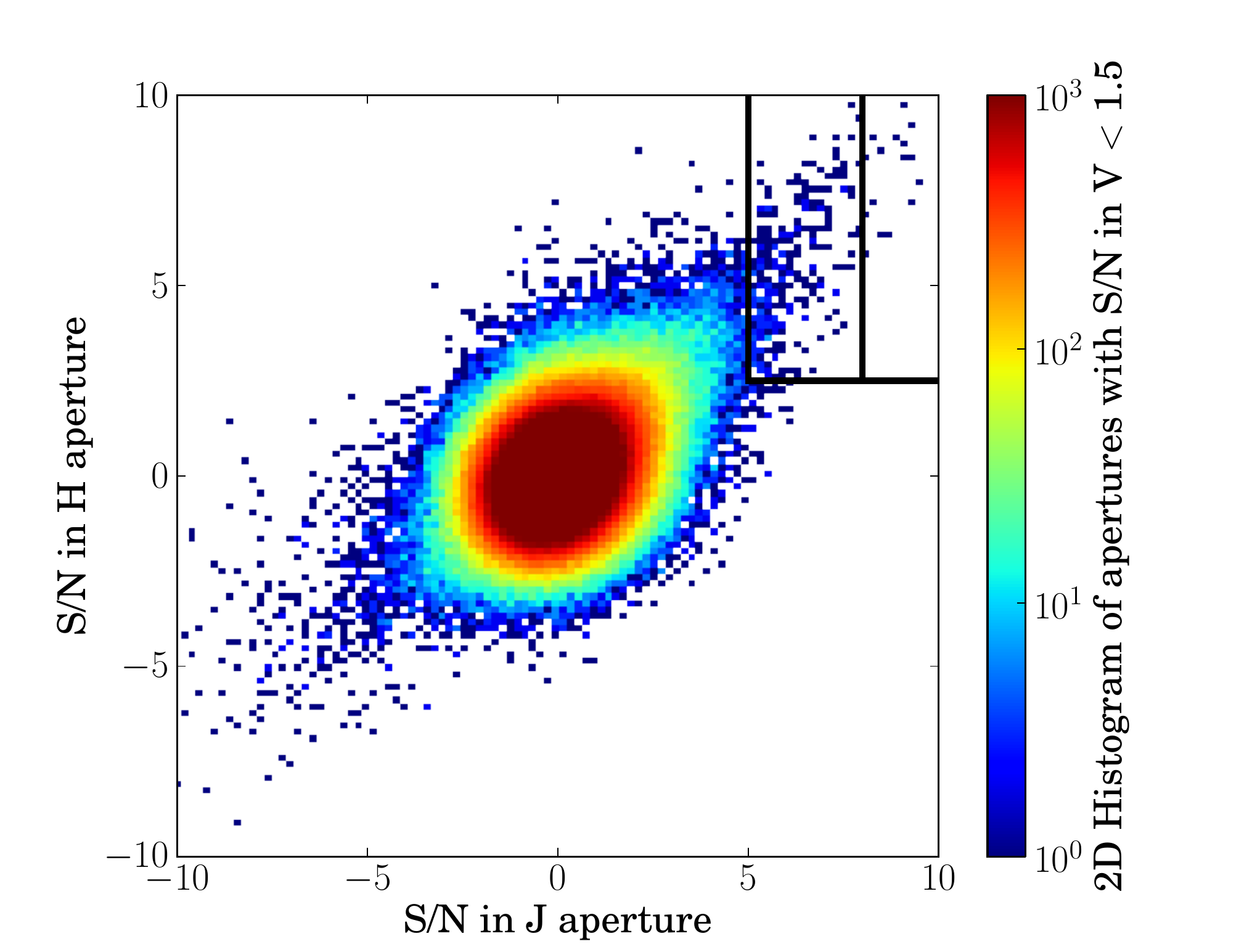}
\caption{The left panel shows the distribution of the measured J-band S/N in $10^6$ apertures of ``empty sky'' from the BoRG fields. The dashed line shows a Gaussian fit to the distribution. It is clear that the wings of the distribution are highly non-Gaussian. In the right panel the 2D distribution in J and H for apertures with S/N$_\textrm{V}<1.5$ is shown. Again the wings are non-Gaussian giving rise to a relatively high occurrence of spurious high-$\sigma$ detections. 
The two boxes in the upper right corner show the region of potentially false positive $5\sigma$ and $8\sigma$ candidates. Of the $10^6$ apertures 479 and 65 fall in these two regions, respectively.
Applying the Y$-$J color cut of the dropout selection described in Section~\ref{sec:ydrop} removes all these potential false positives 
which illustrates the importance of stringent color criteria in dropout searches digging into the noise, like the one presented in this paper.}
\label{fig:noisetails}
\end{figure}

We construct the actual probability distribution function of
background noise by conducting aperture photometry on all $10^6$ possible
apertures of radius $r=0\farcs32$ from the empty regions in 71 of
the BoRG fields. Empty regions are identified according to the
\verb+SExtractor+ segmentation maps (grown by 5 pixels) produced when
creating our source catalogs as described in Section~\ref{sec:ydrop}.
We note that the correlated noise introduced by the drizzling of the science images
is automatically accounted for by this approach.
In Figure~\ref{fig:noisetails} we show the distribution of S/N values
for all apertures in the J-band (left) and the J-H-band 2D S/N
histogram for objects with S/N$_\textrm{V}<1.5$ (right).  The dashed
line in the left panel shows a Gaussian fit to the J-band distribution
of the empty apertures. Is is clear that the one-dimensional
distribution of the noise apertures has broader tails than would be
expected for a Gaussian distribution.

The majority of the potential false positives are rejected by our
multi-band requirements: a `5$\sigma$' detection in J as well as a
`2.5$\sigma$' detection in the H-band and a non-detection in
the V band  (S/N$_\textrm{V}<1.5\sigma$). This selection is illustrated by the box 
in the upper right corner of the right
panel in Figure~\ref{fig:noisetails}. In spite of the three band selection, still 479
empty apertures survive the cuts, i.e. $\sim$7 per BoRG field. The
8$\sigma$ requirement on J is more stringent but still leaves 65
apertures, i.e. approximately one per BoRG field.

However, as mentioned in Section~\ref{sec:ydrop} a crucial part of the
dropout selection is the color cut on $\textrm{Y}-\textrm{J}>1.75$. Applying this
stringent cut removes {\it all} of the noise peaks passing the S/N
cuts. This can be understood as the results of two effects in the BoRG
data. First, the vast majority of sources in the sky are not as red,
and therefore the population with fluxes just below our detection
threshold which happens to be upscattered in the sample by noise
fluctuations is unlikely to satisfy this requirement.  
Second, spurious positive signals like hot pixels will be present in all near-infrared 
bands in undithered data and therefore requiring much fainter flux in Y
than J helps eliminate those as well.
We note that the BoRG survey by design takes all three near-infrared exposures in the same orbit to maximize the chances that hot pixels and detector persistence create images in both the Y-band (taken first), J-band and H-band image. Hence, such artifacts should not be selected as Y-band dropouts in the BoRG survey.
 
From this study of the noise distribution we draw the following
conclusions. Potential contaminants abound in 5$\sigma$ samples, and
multiple band detections are essential to keep them under
control. Thus, it is to be expected that repeated observations of the
same field will yield different sets of marginal candidates, like in
the case of the BoRG\_1437+5043 field. Higher significance 8$\sigma$
samples are much more reliable, but still not immune to noise
fluctuation. In addition to carrying out and comparing analyses for
both $5\sigma$ and 8$\sigma$ samples it is therefore essential to impose strict
color cuts, or impose equivalently tight photo-$z$ requirements. This
naturally reduces the completeness of the samples, but it is a
reasonable price to pay as long as the completeness can be properly
accounted for by means of detailed simulations.

\section{Bayesian Luminosity Function Inference Framework}\label{sec:BF}

In this appendix we outline the Bayesian framework used to infer the parameters of the intrinsic luminosity function in this study. 
Before describing the posterior distribution used for the MCMC sampling we describe how the individual parameters are related under the assumption of a Schechter luminosity function.

\subsection{The Schechter Luminosity Function and its Parameters}

The luminosity function of a sample of galaxies represents the density of objects in a given co-moving volume as a function of the luminosity.
Hence, it gives information about the population of galaxies at a certain redshift and can be integrated to reveal the total number of galaxies.
For luminosity functions at higher redshift this becomes interesting as this gives a direct measure of the reionization power of the galaxies at the given epoch and therefore aid the understanding of which sources reionized the Universe and by how much and when the majority of reionization happened. 
The Schechter function \citep{Schechter:1976p29330} is one of the most widely used luminosity function models.
The Schechter function is given by
\BE \label{eqn:schechter}
\Phi(L) = \frac{\phi^\star}{L^\star} \left( \frac{L}{L^\star}\right)^{k-1} \exp\left( -\frac{L}{L^\star}\right) 
\EE
where $k$ is the so-called shape parameter. 
$k-1=\alpha$ is the so-called `faint-end slope' of the luminosity function that is analyzed in the present study. 
$L^\star$ is the scale parameter that determines the transition between the power-law behavior of \Eq{eqn:schechter}
that dominates at low luminosities and the exponential cut-off at the bright end. 
$\phi^\star$ is a normalizing co-moving number density of objects.
The Schechter function is closely related to the gamma distribution as
\BE\label{eqn:gamma}
\textrm{gamma}(L | k,L^\star) = \frac{\Phi(L)}{\phi^\star\Gamma(k)} = \frac{\left( \frac{L}{L^\star}\right)^{k-1} \exp\left( -\frac{L}{L^\star}\right) }{L^\star\; \Gamma(k)}
\EE
with $\Gamma(k)$ being the gamma function which is defined as
\BE
\Gamma(k)	= \int_{0}^\infty \left( \frac{L}{L^\star}\right)^{k-1} \exp\left( -\frac{L}{L^\star}\right)  \;d \frac{L}{L^\star} \;. 
\EE

Formally the integral of the gamma (Schechter) function diverges for $k<0$. This problem is circumvented by introducing a minimum luminosity, $L_\textrm{min}$, instead of integrating from 0. As the concept of galaxies is only valid above a certain luminosity this approximation is physically motivated. For instance a galaxy of $L_\textrm{gal}=L_\odot$ makes no physical sense. 
An often used ``definition'' of a galaxy is an object with an absolute magnitude $M_\textrm{abs}>-10$ which corresponds to a luminosity of $L\sim10^{40}$erg/s which is  roughly $10^6$ times the energy output of the Sun.
Limiting the integration of the gamma functions makes it the \emph{incomplete} gamma function.

The normalizing galaxy density, $\phi^\star$, is closely related to the total number of galaxies present in the surveyed volume of the Universe following the given luminosity function. 
As noted above, the luminosity function can be integrated to reveal the total number of galaxies, $N$, within the effective co-moving volume $V$ given the number density $\phi^\star$ such that
\BEA\label{eqn:Nuni}
N &=& V \times \int_{L_\textrm{min}}^\infty \;\Phi(L)\;dL \nonumber \\
   &=& V \times \int_{L_\textrm{min}}^\infty \;\frac{\phi^\star}{L^\star} \left( \frac{L}{L^\star}\right)^{k-1} \exp\left( -\frac{L}{L^\star}\right) \;dL
\EEA
which assumes that the luminosity function does not evolve over the considered redshift interval.
The effective co-moving volume can be determined as
\BE
V = \int \frac{dV}{dz} p(z) \; dz \label{eqn:vol}
\EE
where $\frac{dV}{dz}$ is the cosmological volume element and p(z) is the probability of selecting the objects for a given redshift in the surveyed redshift range; essentially a selection function.
For an evolving luminosity function the integral in \Eq{eqn:Nuni} is done over $\frac{dV}{dz}$ separately.

In a similar manner it is possible to obtain the luminosity density, $\epsilon$, i.e., the available radiation per volume, radiated by the galaxy sample, by integrating the product of the luminosity function and the luminosity such that
\BE\label{eqn:epsilon}
\epsilon = \int_{L_\textrm{min}}^\infty\Phi(L)\times L\;dL
\EE

Hence, the three key parameters to determine when characterizing the luminosity function of a sample of galaxies are $k$, $L^\star$ and $N$ (or similarly $\alpha$, $M^\star$ and  $\phi^\star$). In the following we will described the marginal posterior distribution we used to obtain these parameters.

\subsection{The Marginal Posterior Distribution}\label{sec:eqs}

From Bayesian statistics it is known that the posterior probability distribution is proportional to the product of the prior distribution and the likelihood. When determining the luminosity function of samples of high redshift objects, what is usually done is essentially to assume that everything is detected above a certain luminosity threshold, which in the case of the $z\sim8$ objects we deal with here, is the detection threshold in the J-band, under the assumption that each object is not detected in V.
Thus the relationship between the posterior and prior probability distribution for the high-redshift galaxy candidates can be expressed as
\BEA
p(\theta \;|\; L_\textrm{J,obs},I_\textrm{V}=0) \propto p(\theta) \times p(L_\textrm{J,obs},I_\textrm{V} \;|\; \theta) \label{eqn:posterior1}
\EEA
where the last term is the likelihood. $\theta = (k,L^\star,N_z)$ contains the parameters describing the luminosity function $\Phi(L)$, $L_\textrm{J,obs}$ is the observed luminosity in the J-band with $L_\textrm{J,true}$ being the true luminosity, and $I_\textrm{V}$ indicates whether the observed luminosity in the V-band $L_\textrm{V,obs}$ represents a formal detection ($I_\textrm{V}=1$) or not ($I_\textrm{V}=0$). For an object to be included in the sample it cannot be detected in the V-band so $p(I_\textrm{V}=0 \;|\; L_\textrm{J,obs}, \theta) = 1$ is always the case.
$p(\theta)$ contains any prior information on the luminosity function parameters that might be available. 
In the present study we assume uniform priors on $\alpha$, $\log_{10}L^\star$ and $\log_{10}N_z$.

Expanding the expression for the posterior in \Eq{eqn:posterior1} marginalizing over the nuisance parameter $L_\textrm{J,true}$ (the true luminosity of the object in the J-band) leads to a \emph{marginal} posterior probability distribution for a sample of $n$ galaxies which is given by \citep[see Section~3.1 and Appendix~B of][]{Kelly:2008p29070}:
\BEA \label{eqn:postcontam}
p(\theta \;|\; L_\textrm{J,obs},I_\textrm{V}=0) &\propto& \; p(\theta)\; C^{N_z}_{n_z} \;  \prod_{l}^\mathcal{C}  \left[1- A_l/A_\textrm{sky}\; p(I=1|\theta) \right]^{N_z-c_{lz}} \times \; \prod_i^{n_z} p(L_{\textrm{J,obs},i}|\theta) \nonumber \\
& &\qquad C^{N_c}_{n_c} \; \prod_{l}^\mathcal{C} \left[1- A_l/A_\textrm{sky}\; p(I=1|\theta) \right]^{N_c-c_{lc}} \times \; \prod_i^{n_c} p(L_{\textrm{J,obs},i}|\theta)
\EEA
where $N_z$ is the number of high-$z$ sources in the Universe given the intrinsic luminosity function and $N_c$ is the number of objects which would potentially contaminate the luminosity function. The corresponding values for the observed sample are $n_z$ and $n_c$ where the total number of objects in the galaxy sample is given by $n_t = n_z + n_c$.
If $\mathcal{C}$ individual fields are included in the sample this is accounted for by taking the product over the $\mathcal{C}$ fields where $c_l$ corresponds to the number of objects in the $l$'th field such that $n_t = \sum_{l}^\mathcal{C} c_l$. 
The ratio $A_l/A_\textrm{sky}$ gives the fractional area the $l$'th field covers on the sky.
The $C^a_b$ terms are binomial coefficients which can be expressed in terms of the logarithm of the gamma function as
\BE
\ln(C^a_b) = \ln\Gamma(a+1) - \ln\Gamma(b+1) - \ln\Gamma(a-b+1) \; .
\EE
The number of contaminants is determined by the contamination fraction $f$.
By approximating the number of contaminants by its expectation value such that
\BE
N_c = \frac{f}{1-f} N_z  \quad\quad
c_{lc} = \frac{f}{1-f} c_{lz}  \quad\quad
n_c = fn_t \quad\quad
n_z = (1-f)n_t \nonumber
\EE
\Eq{eqn:postcontam} becomes
\BE\label{eqn:posterior2}
p(\theta \;|\; L_\textrm{J,obs},I_\textrm{V}=0) \propto \; p(\theta) \;
C^{N_z}_{(1-f)n_t} C^{\frac{f}{1-f} N_z}_{f n_t} \; \prod_{l}^\mathcal{C}
\left[1- A_l/A_\textrm{sky}\; p(I=1|\theta) \right]^{\frac{1}{1-f_l}(N_z-(1-f_l)c_{l})} \times \; \prod_i^{n_t} p(L_{\textrm{J,obs},i}|\theta)
\EE
Here the contamination $f$ varies between each field contributing to the final sample as indicated by the subscript $l$.
The $p(I=1|\theta)$ term on the right-hand-side represents the probability distribution of an object making it into the sample but is independent on the individual observations. However, it differs between the different observed fields (or surveys) in the sample as exposure times and hence depths differ from field to field. 
Expanding the detection probability for the considered sample we have that
\BE\label{eqn:ptheta}
p(I=1|\theta) =  \int_0^\infty p(I=1|L_\textrm{J,obs})  p(L_\textrm{J,obs}|\theta)  \;dL_\textrm{J,obs} \; .
\EE
Here $p(I=1|L_\textrm{J,obs})$ is the selection function for the $l$'th field accounted for completeness. 

The last term on the right-hand-side in  \Eq{eqn:posterior2} also appearing in \Eq{eqn:ptheta} represents the likelihood of the $i$'th object in the sample which can be expressed as 
\BEA
p(L_\textrm{J,obs}|\theta) &=& \int_0^\infty p(L_\textrm{J,obs} | L_\textrm{J,true}) \; p(L_\textrm{J,true} | \theta )\; dL_\textrm{J,true}\\
&\propto& \int_0^\infty \mathcal{N}(L_\textrm{J,obs} | L_\textrm{J,true}, \delta L_\textrm{J,field})\; \textrm{gamma}(L_\textrm{J,true} | k,L^\star) \; dL_\textrm{J,true} \; . \nonumber
\EEA
Where we use $p(L)\propto \frac{\Phi(L)}{\phi^{\star}}$ \citep[see Equation~(1) of][]{Kelly:2008p29070} and that gamma$(L_\textrm{J,true} | k,L^\star)$ is related to the Schechter luminosity function $\Phi(L_\textrm{J,true})$ as shown in \Eq{eqn:gamma}. 
\BE
\mathcal{N}(L_\textrm{J,obs} | L_\textrm{J,true}, \delta L_\textrm{J,field}) =  \frac{1}{\delta L_\textrm{J,field}\sqrt{2\pi}} \exp\left(-\frac{(L_\textrm{J,true}-L_\textrm{J,obs})^2}{2\; \delta L_\textrm{J,field}^2} \right) \label{eqn:errgauss}
\EE
represents the true luminosity inferred from the observations assuming a Gaussian measurement error with $\delta L_\textrm{J,field}$ being the median photometric error in the J-band in the given field.

By denoting the selection function with $\mathcal{S}(L_\textrm{J,obs})$ and defining
\BE
\mathcal{F}(L_\textrm{J,true}) =  \mathcal{N}(L_\textrm{J,obs} |L_\textrm{J,true}, \delta L_\textrm{J,field})\; \textrm{gamma}(L_\textrm{J,true} | k,L^\star) 
\EE
we can express \Eq{eqn:posterior2} as
\BEA\label{eqn:margpost}
p(\theta &|&  L_\textrm{J,obs},I_\textrm{V}=0) \propto \; p(\theta)\; \nonumber \\
&\times&  C^{N_z}_{(1-f)n_t} C^{\frac{f}{1-f} N_z}_{f n_t} \; \prod_{l}^\mathcal{C}
 \left[1- A_l/A_\textrm{sky}\; \int_0^\infty \mathcal{S}(L_\textrm{J,obs})  \int_0^\infty \mathcal{F}(L_\textrm{J,true}) \; dL_\textrm{J,true}  \;dL_\textrm{J,obs} \right]^{\frac{1}{1-f_l}(N_z-(1-f_l)c_{l})} \nonumber \\
&\times& \; \prod_i^{n_t} \int_0^\infty \mathcal{F}(L_\textrm{J,true}) \;dL_\textrm{J,true}
\EEA
This is the posterior probability distribution for a sample of $n_t$ objects assumed to be binomially distributed and to have an intrinsic Schechter luminosity function on the form shown in \Eq{eqn:schechter}, where each observed luminosity is related to the true luminosity through an assumed Gaussian error distribution and where the prior on $\log_{10} N_z$ is uniform.

Estimating $p(\theta \;|\; L_\textrm{J,obs},I_\textrm{V}=0)$ for a certain $\theta$ corresponds to estimating the probability that the luminosity function with the parameters $(k,L^\star,N_z)$ is the intrinsic luminosity function of the analyzed galaxy sample. 
The intrinsic luminosity function can for instance be obtained from this expression by sampling $\theta$ via a Markov Chain Monte Carlo approach. 
As described in Section~\ref{sec:LF} we used such an approach to infer the intrinsic
luminosity function at redshift 8 for the BoRG13 sample.

Having obtained the three-dimensional posterior distribution for $\theta$ the number density of high redshift galaxies, $\phi^\star$, and the luminosity density, $\epsilon$, can be determined using \Eqs{eqn:Nuni}{eqn:epsilon}. 
The current framework assumes that the fraction of contaminants is independent of luminosity for the BoRG sample.

Above we have partially followed \cite{Kelly:2008p29070} and refer to this work for further details. 

\end{appendix}

\end{document}